\documentclass[longauth,traditabstract]{aa} 
 \usepackage{graphicx}
 \usepackage{txfonts}
 \usepackage{natbib}

 \newcommand{\FunitAREA}{10$^{-18}$ erg~s$^{-1}$~cm$^{-2}$~arcsec$^{-2}$}
 \newcommand{\Funits}{10$^{-16}$ erg~s$^{-1}$~cm$^{-2}$}
 \newcommand{\FunitsSS}{10$^{-17}$ erg~s$^{-1}$~cm$^{-2}$}
 \newcommand{\FunitsSSAREA}{10$^{-17}$ erg~s$^{-1}$~cm$^{-2}$~arcsec$^{-2}$}
 \newcommand{\FunitsAREA}{10$^{-16}$ erg~s$^{-1}$~cm$^{-2}$~arcsec$^{-2}$}

 \newcommand{\degree}{\ensuremath{^\circ}}

\begin{document}

\title{CALIFA, the Calar Alto Legacy Integral Field Area survey:}

    \subtitle{I. Survey presentation\thanks{Based on observations collected at the Centro Astron\'omico
    Hispano Alem\'an (CAHA) at Calar Alto, operated jointly by the Max-Planck-Institut f\"ur Astronomie and the Instituto de Astrof\'isica de Andaluc\'ia
    (CSIC).}
    }

    \author{     
      S.F. S\'anchez\inst{1}
      \and
      R.C. Kennicutt\inst{2}
      \and
      A. Gil de Paz\inst{3}
      \and
      G. van de Ven\inst{4}
      \and
      J.M. V\'\i lchez\inst{5}
      \and
      L. Wisotzki\inst{6}
      \and
      C. J. Walcher\inst{6}
      \and
      D. Mast\inst{5,1}
      \and
      J. A. L. Aguerri\inst{9,28}
      \and
      S. Albiol-P\'erez\inst{20}
      \and
      A. Alonso-Herrero \inst{12}
      \and
      J. Alves\inst{22}
      \and
      J. Bakos\inst{9,28}
      \and
      T. Bart\'akov\'a\inst{34}
      \and
      J. Bland-Hawthorn\inst{7}
      \and
      A. Boselli\inst{19}
      \and
      D. J. Bomans\inst{25}
      \and
      A. Castillo-Morales\inst{3}
      \and
      C. Cortijo-Ferrero\inst{5}
      \and
      A. de Lorenzo-C\'aceres\inst{9,28}
      \and
      A. del Olmo\inst{5}
      \and
      R.-J. Dettmar\inst{25}
      \and
      A. D\'\i az\inst{10}
      \and
      S. Ellis\inst{8,7}
      \and
      J. Falc\'on-Barroso\inst{9,28}
      \and
      H. Flores\inst{31}
      \and
      A. Gallazzi\inst{14}
      \and
      B. Garc\'\i a-Lorenzo\inst{9,28}
      \and
      R. Gonz\'alez Delgado\inst{5}
      \and
      N. Gruel\inst{24}     
      \and
      T. Haines\inst{26}
      \and
      C. Hao\inst{32}
      \and
      B. Husemann\inst{6}
      \and
      J. Igl\'esias-P\'aramo\inst{5,1}
      \and
      K. Jahnke\inst{4}
      \and
      B. Johnson\inst{30}
      \and
      B. Jungwiert\inst{16,33}
      \and
      V. Kalinova\inst{4}
      \and
      C. Kehrig\inst{6}
      \and
      D. Kupko\inst{6}
      \and
      \'A. R. L\'opez-S\'anchez\inst{8,23}
      \and
      M. Lyubenova\inst{4}
      \and
      R.A. Marino\inst{1,3}
      \and
      E. M\'armol-Queralt\'o\inst{1,3}
      \and
      I. M\'arquez\inst{5}
      \and
      J. Masegosa\inst{5}
      \and
      S. Meidt\inst{4}
      \and
      J. Mendez-Abreu\inst{9,28} 
      \and
      A. Monreal-Ibero\inst{5}
      \and
      C. Montijo\inst{5}
      \and
      A. M. Mour\~ao\inst{17}
       \and 
      G. Palacios-Navarro\inst{21}
      \and
      P. Papaderos\inst{15}
      \and
      A. Pasquali\inst{29}
      \and
      R. Peletier \inst{11}
      \and
      E. P\'erez\inst{5}
      \and
      I. P\'erez\inst{27}
      \and
      A. Quirrenbach\inst{13}
      \and
      M. Rela\~no\inst{27}
      \and 
      F. F. Rosales-Ortega\inst{10,1}
      \and
      M. M. Roth\inst{6}
      \and
      T. Ruiz-Lara\inst{27}
      \and
      P. S\'anchez-Bl\'azquez\inst{10}
      \and
      C. Sengupta\inst{1,5}
      \and
      R. Singh\inst{4}
      \and
      V. Stanishev\inst{17}
      \and
      S.C. Trager \inst{11}
      \and
      A. Vazdekis\inst{9,28}
      \and
      K. Viironen\inst{1,24}
      \and
      V. Wild\inst{18}
      \and
      S. Zibetti\inst{14}
      \and
      B. Ziegler\inst{22}
            }

     \institute{
  Centro Astron\'omico Hispano Alem\'an, Calar Alto, (CSIC-MPG),
  C/Jes\'us Durb\'an Rem\'on 2-2, E-04004 Almeria, Spain \email{sanchez@caha.es}.
  \and
  Institute of Astronomy, University of Cambridge, Madingley Road, Cambridge CB3 0HA, UK.
  \and
  Departamento de Astrof\'{i}sica y CC. de la Atm\'{o}sfera, Universidad Complutense de Madrid, Madrid 28040, Spain.
  \and
  Max Planck Institute for Astronomy, K\"onigstuhl 17, D-69117 Heidelberg, Germany.
  \and
 Instituto de Astrof\'\i sica de Andaluc\'\i a (CSIC), Camino Bajo de Huetor,
  s/n,  E-18008, Granada.
  \and
 Leibniz-Institut f\"ur Astrophysik Potsdam (AIP), An der Sternwarte 16, D-14482 Potsdam, Germany.
  \and
 Sydney Institute for Astronomy,
 School of Physics A28, University of Sydney, NSW 2006, Australia .
 \and
 Australian Astronomical Observatory, PO BOX 296, Epping, NSW 1710, Australia
 \and 
 Instituto de Astrof\'\i sica de Canarias (IAC), Glorieta de la
 Astronom\'\i a S/N, La Laguna, S/C de Tenerife, Spain.
 \and
 Departamento de Fisica Teorica, Universidad Autonoma de Madrid, Cantoblanco, 28049, Madrid, Spain.
 \and
 Kapteyn Astronomical Institute, University of Groningen, NL-9700 AV, Groningen, Netherlands.
 \and
 Instituto de F\'\i sica de Cantabria, CSIC-UC, Avenida de los Castros S/N, 39005 Santander, Spain 
 \and
 ZAH, Landessternwarte, K\"onigstuhl 12, D-69117, Heidelberg, Germany.
 \and
 Dark Cosmology Centre, Niels Bohr Institute, University of Copenhagen, Juliane Maries Vej 30, DK-2100 Copenhagen, Denmark.
 \and
 Centro de Astrof\'\i sica da Universidade do Porto, Rua das Esteral, 4150-762, porto, Portugal.
 \and
 Astronomical Institute of the Academy of Sciences of the Czech Republic, v.v.i., Bocni II 1401, 14131 Prague, Czech Republic.
 \and
 CENTRA - Centro Multidisciplinar de Astrof\'isica, Instituto Superior T\'ecnico, Av. Rovisco Pais 1, 1049-001 Lisbon, Portugal.
 \and
 SUPA, Institute for Astronomy, University of Edinburgh, Royal Observatory, Blackford Hill, Edinburgh EH9 3HJ, UK.
 \and
 Laboratoire d'Astropysique de Marseille, UMR6110 CNRS, 38 rue F. Joliot-Curie, F-13388 Marseille, France.
 \and
 Department of Systems Engineering and Computing, University of Zaragoza, Teruel 44003, Spain.
 \and
 Department of Electronic Engineeing and Communications, University of Zaragoza, Teruel 44003, Spain.
 \and
 University of Vienna, T\"urkenschanzstrasse 17, 1180 Vienna, Austria.
 \and
 Department of Physics and Astronomy, Macquarie University, NSW 2109, Australia.
 \and
 Centro de Estudios de F\'\i sica del Cosmos de Arag\'on (CEFCA), C/ G. Pizarro, 1, Teruel, Spain.
 \and
 Astronomical Institute of the Ruhr-University Bochum Universitaetsstr. 150, 44801 Bochum, Germany.
 \and
 Department of Physics, University of Missouri-Kansas City, Kansas City, MO
 64110, USA.
 \and
 Dep. Física Teórica y del Cosmos, Campus de Fuentenueva, Universidad de Granada, 18071, Granada, Spain. 
 \and
 Depto. Astrofísica, Universidad de La Laguna (ULL), E-38206 La Laguna, Tenerife, Spain. 
 \and
 Astronomisches Rechen Institut, Zentrum fuer Astronomie der Universitaet Heidelberg, Moenchhofstrasse 12 - 14, 69120 Heidelberg, Germany.
 \and
 Institut d'Astrophysique de Paris, CNRS, 98 bis Boulevard Arago, 75014 Paris, France.
 \and
 Laboratoire GEPI, Observatoire de Paris, CNRS-UMR8111, Univ. Paris-Diderot 5 place Jules Janssen, 92195 Meudon, France
 \and 
 Tianjin Astrophysics Center, Tianjin Normal University, Tianjin
 300387, China.
\and 
Astronomical Institute, Faculty of Mathematics and Physics, Charles
University in Prague, Ke Karlovu 3, CZ-121 16 Prague, Czech Republic
\and
Dept. of Theoretical Physics and Astrophysics, Faculty of Science,
Masaryk University, Kotl\'a\v rsk\'a 2, CZ-611 37 Brno, Czech Republic
 }

     \date{Received ----- ; accepted ---- }

    \abstract {
     The final product of galaxy evolution through cosmic time 
     is the population of galaxies in the Local Universe. These 
     galaxies are also those that can be studied in most detail, thus 
     providing a stringent benchmark for our understanding of galaxy 
     evolution. Through the huge success of spectroscopic single-fiber, statistical surveys 
     of the Local Universe in the last decade, it has become clear, however, 
     that an authoritative observational description of galaxies 
     will involve measuring their spatially resolved
     properties over their full optical extent { (covering D25)} for a statistically 
     significant sample. We present here the Calar Alto Legacy Integral Field 
     Area (CALIFA) survey, which has been designed to 
     provide a first step in this direction. We summarize the survey goals and design, including 
     sample selection and observational strategy. We also showcase 
     the data taken during the first observing runs (June/July 2010) and outline 
     the reduction pipeline, quality control schemes and general 
     characteristics of the reduced data. 

     This survey is obtaining spatially resolved spectroscopic
     information of a diameter selected sample of $\sim$600 galaxies
     in the Local Universe (0.005$<z<$0.03). CALIFA has been designed
     to allow the building of two-dimensional maps of the following
     quantities: (a) stellar populations: ages and metallicities; (b)
     ionized gas: distribution, excitation mechanism and chemical
     abundances; and (c) kinematic properties: both from stellar and
     ionized gas components.  CALIFA uses the PPAK Integral Field Unit
     (IFU), with a hexagonal field-of-view of $\sim$1.3$\sq\arcmin$,
     with a 100\% covering factor by adopting a three-pointing
     dithering scheme. The optical wavelength range is covered from
     3700 to 7000 ~\AA{}, using two overlapping setups (V500 and
     V1200), with different resolutions: R$\sim$850 and R$\sim$1650,
     respectively. CALIFA is a legacy survey, intended for the
     community.  The reduced data will be released, once the quality
     has been guaranteed.

    The analyzed data fulfill { the expectations of the original observing proposal}, on the basis of a set
    of quality checks and exploratory analysis: (i) the final datacubes
    reach a 3$\sigma$ limiting surface brightness depth of $\sim$ 23.0
    mag/arcsec$^2$ for the V500 grating data ($\sim$22.8 mag/arcsec$^2$
    for V1200); (ii) about $\sim$70\% of the covered
    field-of-view is above this 3$\sigma$ limit; (iii) the data have a
    blue-to-red relative flux calibration within a few percent in most
    of the wavelength range; (iv) the absolute flux calibration is
    accurate within  $\sim$8\% { with respect to SDSS}; (v) the measured spectral resolution
    is $\sim$85 km~s$^{-1}$ for V1200 ($\sim$150 km~s$^{-1}$
    for V500); (vi) the estimated accuracy of the wavelength
    calibration is $\sim$5 km~s$^{-1}$ for the V1200 data ($\sim$10
    km~s$^{-1}$ for the V500 data); (vii) the aperture matched CALIFA
    and SDSS spectra are qualitatively and quantitatively
    similar. Finally, we show that we are able to carry out all
    measurements indicated above, recovering the properties of the
    stellar populations, the ionized gas and { the kinematics of both components}. { The associated maps illustrate the spatial variation of these parameters across the field, reemphasizing the redshift dependence of single aperture spectroscopic measurements.} We
    conclude from this first look at the data that CALIFA will be an
    important resource for archaeological studies of galaxies in the
    Local Universe.}


     \maketitle

  \section{Introduction}

Our understanding of the Universe and its constituents comes from
large surveys such as the 2dFGRS \citep{folkes99}, SDSS \citep{york00},
GEMS \citep{rix04}, VVDS \citep{le-fevre04}, and COSMOS
\citep{scoville07}, to name but a few. Such surveys have not only
constrained the evolution of global quantities such as the cosmic star
formation rate, but also enabled us to link these with the properties
of individual galaxies -- morphological types, stellar masses,
metallicities, etc. Compared to previous approaches, the major
advantages of this recent generation of surveys are: (1) the large
number of objects sampled, allowing for meaningful statistical
analysis to be performed on an unprecedented scale; (2) the
possibility to construct large comparison/control samples for each
subset of galaxies; (3) a broad coverage of galaxy subtypes and
environmental conditions, allowing for the derivation of universal
conclusions; and (4) the homogeneity of the data acquisition,
reduction and (in some cases) analysis.

 On the other hand, the cost of these surveys, in terms of telescope time,
 person-power, and time scales involved, is also unprecedented in astronomy. The
 user of such data products has not necessarily been involved in any step of
 designing or conducting the survey, but nevertheless takes advantage of the
 data by exploiting them according to her/his scientific interests.  This new
 approach to observational astronomy is also changing our perception of the
 scientific rationale behind a new survey: while it is clear that certain
 planned scientific applications are key determinants to the design of the
 observations and `drive' the survey, the survey data should, at the same time,
 allow for a broad range of scientific exploitation. This aspect is now often
 called a survey's \emph{legacy} value.

 Current technology generally leads to surveys either in the imaging or
 in the spectroscopic domain. While imaging surveys provide
 two-dimensional coverage, the information content of a photometric
 Spectral Energy Distribution (SED) is limited. This remains true for
 the new generation of multi-band photometric surveys such as COMBO-17
 \citep{wolf03}, ALHAMBRA \citep{moles08}, the planned PAU project
 \citep{benitez09}, COSMOS \citep{scovi07} or the LUS
 survey\footnote{http://www.inaoep.mx/$\sim$gtc-lus/}, which are geared
 towards better precision in redshift, mean ages and stellar masses,
 but are nevertheless unable to measure individual spectral lines and
 thus emission line ratios or internal radial velocity
 differences. Spectroscopic surveys such as SDSS or zCOSMOS
 \citep{lill07}, on the other hand, do provide more detailed
 astrophysical information, but they are generally limited to one
 spectrum per galaxy. One thus misses all information on the radial
 distribution of galaxy properties and on all details of the
 kinematics. Even when attempting to describe galaxies by their
 integrated properties only, this state of affairs also leads to
 aperture losses that are difficult to control.  For example, the $3''$
 diameter of the fiber used in the SDSS corresponds to different
 physical scales at different redshifts, with limited possibilities to
 correct for these aperture effects \citep[e.g.,][]{kewley05, elli05}.  The
 most popular method is to compare results of a fit to the photometry
 of the whole galaxy with the photometric fit corresponding to the area
 of the fiber only \citep{gomez03, brinchmann04}. Even more severely,
 \cite{zibetti09} have recently shown that spatially resolved stellar
 population analysis may lead to corrections of up to 40\% for the
 stellar mass of a galaxy, when compared to integrated light studies.

 An observational technique combining the advantages of imaging and
 spectroscopy (albeit with usually quite small field of view) is
 Integral Field Spectroscopy (IFS). This technique allows us to study
 both the integrated and spatially resolved spectroscopic properties of
 galaxies. IFS has the potential to provide observational evidence to
 constrain many outstanding questions of baryonic physics which are key
 to our understanding of galaxy evolution and, therefore, cosmology.
 Some of these are (1) the importance and consequences of merging,
 major and minor; (2) internal dynamical processes, such as bars,
 spiral arms, stellar migration; (3) environmental effects, such as
 tidal forces, stripping; (4) AGN feedback; (5) occurrence, spatial and
 temporal extent and trigger of star formation. Spatially resolved
 spectroscopic properties of a statistical sample of nearby galaxies is
 the dataset required to address these questions.

 However, so far, IFS has rarely been used in a `survey mode' to
 investigate sizeable samples. Among the few exceptions there is, most
 notably, the SAURON survey \citep{de-zeeuw02}, focused on the study
 of the central regions of 72 nearby early-type galaxies and bulges of
 spirals, and its extension ATLAS$^{\rm 3D}$ \citep[260 { early}-type
 galaxies at $z<0.01$;][]{cappellari10}.  Others are the on-going
 PINGS project \citep{rosales-ortega10} at the CAHA 3.5m of a dozen of
 very nearby galaxies ($\sim$10 Mpc) and the currently ongoing study
 of 70 (U)LIRGS at $z<0.26$ using different IFUs
 \citep{arribas08}. Finally, the VIRUS-P instrument is currently used
 to carry out two small IFS surveys, namely VENGA \citep[30 spiral
 galaxies, ][]{blanc10} and
 VIXENS\footnote{http://www.as.utexas.edu/$\sim$alh/vixens.html} (15
 starbursts). {\bf All these datasets are clearly focused on specific
   science questions, adopting correspondingly optimized sample
   selection criteria and also observing strategies. For example, at
   the redshifts of the galaxies in the ATLAS$^{\mathrm{3D}}$ sample
   SAURON has a field of view of $30'' \times 40''$ , or $< 7 \times
   9$~kpc, thus does not cover the outer parts of these galaxies. }



 On completion, CALIFA will be the largest and the most comprehensive
 wide-field IFU survey of galaxies carried out to date. It will thus
 provide an invaluable bridge between large single aperture surveys and
 more detailed studies of individual galaxies. With CALIFA we will fix
 observational properties of galaxies in the Local Universe, which
 will have a potential impact in the interpretation of observed
 properties at higher redshift \cite[e.g.][]{epi2010}.

 { CALIFA is an ongoing survey, which has been granted with 210 dark
   nights by the Calar Alto Executive Committee, spanning through 6
   semesters. The gathering of the data started in June/July 2010, and
   after a technical problem with the telescope, it was resumed in
   March 2011. {\bf This technical problem, fully repaired, has not
     affected the quality of the CALIFA data.} About 20 galaxies are
   observed per month. As mentioned above, there will be consecutive
   data releases of the fully reduced datasets, once certain
   milestones/number of observed galaxies are reached. The first one
   is planned for late 2012, when a total number of 100 galaxies is
   completed (ie., observed, reduced and quality of the data
   tested). The current status of data acquisition can be obtained
   from the CALIFA webpage\footnote{http://www.caha.es/CALIFA/}. }

 In this article we present the main characteristics of the survey,
 starting from the design requirements in Section
 \ref{sec:design}. Section \ref{sec:sample} describes the sample
 selection criteria, and the main characteristics of the CALIFA mother
 sample. In Section \ref{sec:obs} we describe the observing strategy,
 in particular the observations performed during the first CALIFA
 runs. The data reduction is described in detail in Section
 \ref{sec:redu}, and some of the first quality tests performed on the
 data are presented in Section \ref{quality}. The exploratory analysis
 performed on these first datasets, { obtained in 2010,} to verify that we will be able to
 reach our science goals are presented in Section \ref{explore}. A
 summary and conclusion of the results is presented in Section
 \ref{sum}. 

 \section{Design drivers} 
 \label{sec:design}

 CALIFA has been designed to increase our knowledge of the baryonic physics 
 of galaxy evolution. We intend to characterize observationally the local galaxy 
 population with the following key points in mind: 

 \begin{itemize}

 \item Sample covering a substantial fraction of the galaxy luminosity function. 

 \item Large enough sample to allow statistically significant conclusions for all 
 classes of galaxies represented in the survey.

 \item Characterization of galaxies over their \emph{full} spatial extent {\bf (covering in most cases the R25 radius) }, i.e. avoiding 
 aperture biases and harnessing the additional power of 2D resolution (gradients, 
 sub-structures: bars, spiral arms...). 

 \item Measurement of gas ionization mechanisms: star formation, shocks, AGN.

 \item Measurement of ionized gas oxygen and nitrogen abundances. 

 \item Measurement of stellar population properties: ages,
   mass-to-light ratios, metallicities, and
   (to a limited extent) abundance patterns.

 \item Measurement of galaxy kinematics in gas and stars, i.e. velocity fields for all 
 galaxies and velocity dispersions for the more massive ones. 

 \end{itemize}

 A careful assessment of these, partially competing, drivers and of the
 practical limits imposed by the instrument, the observatory and the
 timescale has led to the following key characteristics of the survey:
 (i) Sample: $\sim$600 galaxies of any kind in the Local Universe, covering the
 full color-magnitude diagram down to M$_B\sim-$18 mag, selected from
 the SDSS to allow good a-priori characterization of the targets; (ii)
 Instrument: PPAK IFU of the Potsdam Multi-Aperture Spectrograph (PMAS)
 instrument at the 3.5m telescope of CAHA, with one of the largest
 fields-of-view for this kind of instruments ($>$1 arcmin$^2$); (iii)
 Grating setups: two overlapping setups, one in the red (3750--7000
 ~\AA, R$\sim$850, for ionized gas measurements and stellar populations)
 and one in the blue (3700--4700 ~\AA, R$\sim$1650, for detailed stellar
 populations and for stellar and gas kinematics); and (iii) Exposure times: 1800s in the
 blue and 900s in the red.

\section{The CALIFA galaxy sample}
 \label{sec:sample}

{

We now briefly describe how the sample has been constructed. We limit this section to some
fundamental considerations that are important for understanding the potential
and limitations of the survey. In a later paper we will present a detailed
characterization of the sample and the distribution of galaxy properties,
including comparisons to other studies and the galaxy population as a whole.

The guiding principle for building the CALIFA galaxy sample followed from the
broad set of scientific aims outlined in the previous section, i.e.\ the sample
should cover a wide range of important galaxy properties such as morphological
types, luminosities, stellar masses, and colours. Further constraints were added
due to technical and observational boundary conditions, such as observing all
galaxies with the same spectroscopic setup to make scheduling as flexible as
possible. As detailed below, a small set of selection criteria was employed to
define a sample of 937 possible target galaxies, the so-called ``CALIFA mother
sample''. From this mother sample, the galaxies are drawn as actual targets only
according to their visibility, i.e.\ in a quasi-random fashion, so that the
final catalog of CALIFA galaxies will be a slightly sparsely sampled ($\sim$2/3)
subset of the mother sample.

 \begin{figure}
 \includegraphics[width=10cm,angle=0,clip=true,trim=0 0 0 100]{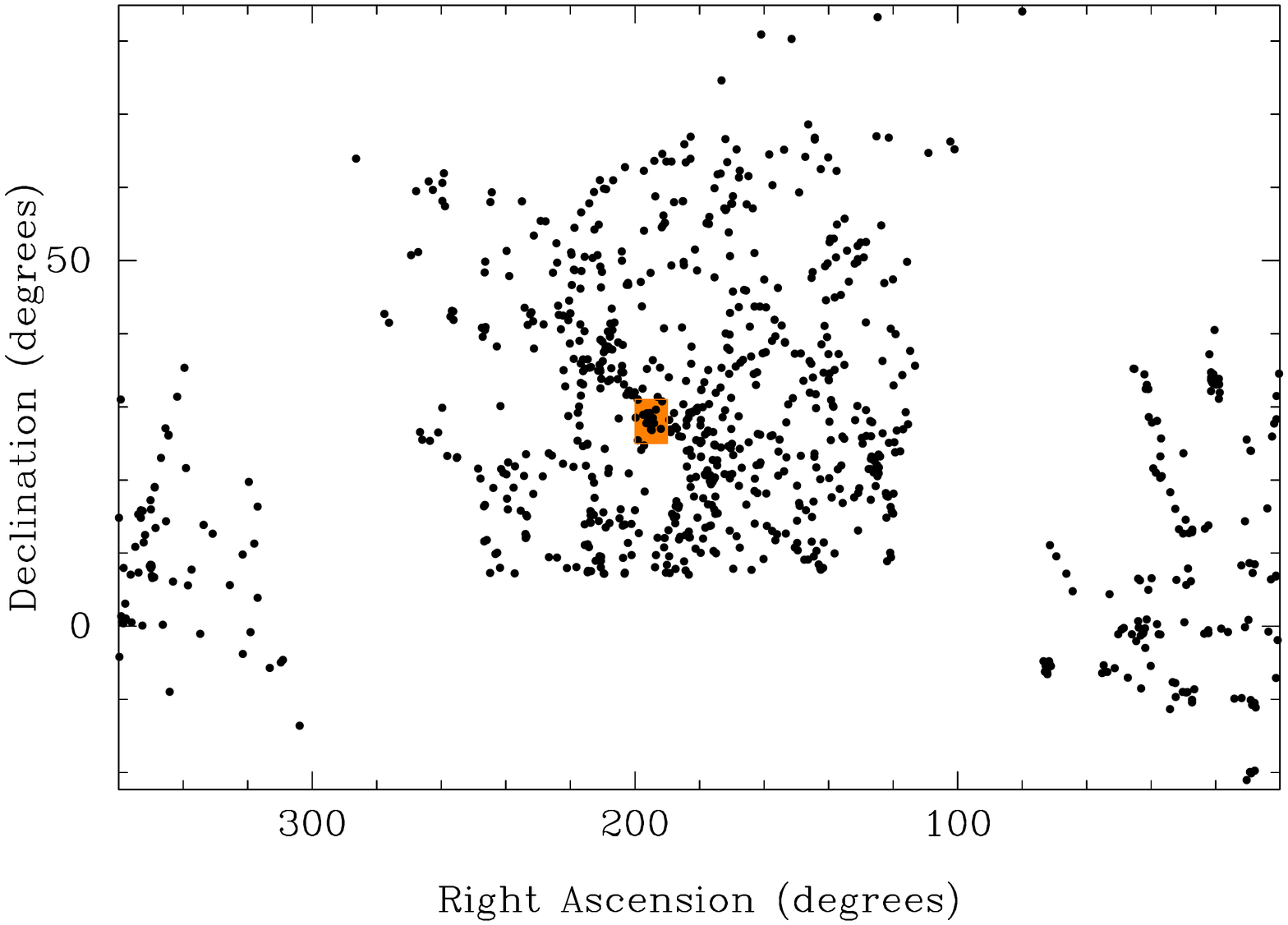}
 \caption{Distribution of the targets in the CALIFA mother sample in equatorial  coordinates. The distribution over the two Galactic caps is obvious. Note  that the north galactic plane part of the sample has been limited to $\delta > +7^\circ$  while the south galactic plane part does not have that limitation, to counterbalance the much lower number of SDSS objects in the south galactic plane region. { The orange square at the center of the figure indicates the location of the Coma cluster.}\label{fig:radec} }
 \end{figure}

The CALIFA mother sample was initially selected from the SDSS DR7 photometric
catalog (Abazajian et al. 2009), which ensures the availability of good quality
multi-band photometry, and in many cases (but not all, cf.\ below) nuclear
spectra. The CALIFA footprint is thus largely identical to that of the SDSS imaging
survey, modified only by an additional restriction of $\delta > 7\degr$ for
galaxies in the North Galactic hemisphere to ensure good visibility from the
observatory. Note that this criterion was not applied to objects at southern
Galactic latitudes, in order to at least partly counterbalance the much smaller
sky coverage of SDSS data in this region and improve on the observability of
CALIFA targets throughout the year. Figure~\ref{fig:radec} shows the resulting
distribution of the CALIFA mother sample in the sky.

The defining selection criteria of the CALIFA mother sample comprise a
combination of angular isophotal diameter selection with lower and
upper redshift limits ($0.005 < z < 0.03$). We further justify and
discuss the consequences of these selection criteria later in this
section. Note that while no explicit cut in apparent magnitude is
involved, the adopted criteria implicitly ensure that only relatively
bright galaxies enter the sample, with a roughly Gaussian distribution
of the total $r$-band magnitudes centred on a mean of $\overline{r} =
13.3$ and with a $1\sigma$ dispersion of 0.8\,mag. No CALIFA galaxy is
fainter than $r = 16$. In Fig.~\ref{fig:magz} we show the distribution
of the mother sample in the fundamental observed properties, total
$r$-band magnitudes and redshifts.

The choice of redshift range was driven by two requirements: (1) Objects within
the luminosity range of interest should have apparent sizes well matched to the
PPAK field-of-view (FoV). (2) All relevant emission lines in all galaxies should
be covered with a single spectral setup. As measure of the apparent galaxy size
we adopted the ``isoA\_r'' values (isophotal diameters $D_{25}$ in the SDSS
$r$-band) and selected only galaxies with $45'' < D_{25} < 80''$. We demonstrate
below in Sect.~\ref{datadepth}  that
this strategy indeed allows for a very efficient usage of the instrument, in the
sense that a large fraction of the FoV provides useful data. The upper limit in
the apparent diameters furthermore ensures that the contribution of light
from the galaxy at the position of the sky fibers is negligible (typically $>
27.5$ mag arcsec$^{−2}$ in the $r$-band). We thus avoid objects larger than the
FoV, which would require a mosaicing strategy plus separate sky exposures
leading to a dramatic reduction of the observing efficiency in terms of the number
of galaxies observable within a given time.

Not all galaxies in the SDSS photometric catalog obeying our isophotal
diameter criterion have spectra -- and therefore redshifts -- in the
SDSS spectroscopic database, which is known to become increasingly
incomplete for total magnitudes brighter that $r \sim 14$. In order to
(as much as possible) overcome such an undesirable bias \emph{against}
bright galaxies, we supplemented the SDSS redshifts with information
accessed through the SIMBAD database at CDS, which in turn is a
compilation of a large variety of observations and redshift
catalogs. Hence, there are SDSS-based spectra and redshifts for
$\sim$60\% of the CALIFA galaxies, while for the remainder we have
only the redshift information provided by SIMBAD. As the latter is
also not
100\% complete, there could be a few galaxies within the
CALIFA footprint without redshift measurements and, therefore, outside
of the CALIFA mother sample. It is difficult to quantify this
incompleteness, but it is unlikely to be more than a few percent, and
probably much less.

Our decision to construct a diameter-selected sample has several practical
advantages, besides the obvious benefit of efficiently using the instrumental
field-of-view. Another advantage has already been mentioned: for the adopted
redshift range, the distribution of apparent galaxy magnitudes
naturally favours relatively bright systems, and in fact there was no need to
define an additional faint flux limit to the survey (see
Fig.~\ref{fig:magz}). Furthermore, the range in \emph{absolute} magnitudes is
considerably broadened due to the factor of 6 between lowest and highest
redshifts, so that the CALIFA sample encompasses an interval of $> 7$~mag in
intrinsic luminosities. In fact, the low-redshift cutoff was mainly introduced
in order to limit the luminosity range and avoid swamping the sample with dwarf
galaxies, which -- given the limitation to 600 galaxies in total -- were
considered to be outside the main scientific interest of the CALIFA project.


Together with the wide range in luminosities comes a broad coverage of galaxy
colours, and the CALIFA sample includes substantial numbers of galaxies in all
populated areas of the colour-magnitude diagram, from the red sequence through
the green valley to the blue cloud (which is of course effectively truncated at
its faint end due to the low-redshift limit). This broad colour distribution is
illustrated in Fig.~\ref{fig:sample_cmd}, where we compare the $u-z$ vs.\ $M_z$
relation of galaxies in the CALIFA mother sample with the corresponding
distribution in the SDSS-NYU catalog \citep[e.g.][]{blanton05}. 
While there are (and should be) differences in the details of the
distribution -- which can be quantified, see below --, it is immediately clear
that CALIFA at least qualitatively represents a wide range of
galaxy types. Figure~\ref{fig:sample_cmd} also provides the number of CALIFA objects
per bin in $M_z$ and $u-z$ (recall, however, that the observed sample
size will be smaller by a factor $\sim 2/3$). These numbers show that there will
be sufficient statistics in several bins to make robust statements about
\emph{typical} galaxy properties, for early-type as well as late-type
galaxies. In fact, these were the numbers that drove the overall time request for
CALIFA to enable a total sample size of 600 galaxies.

The broad representation of galaxy properties in the CALIFA sample is also
reflected in the distribution of morphological types. While a thorough
characterization of the sample in terms of morphology and structural properties
will be the subject of a future paper, a qualitative impression can be obtained
already from rather simple diagnostics.  It has been demonstrated in the past
\citep{strat01} that bulge- and disk-dominated systems can be
reasonably well distinguished by their \emph{concentration indices}, defined as
the ratio $C$ of the $r_{90}$ and $r_{50}$ Petrosian radii provided by the SDSS
photometric catalog. Typically, a value of $C \ga 2.8$ requires the presence of
a substantial bulge, whereas $C \la 2.3$ is indicative of an exponential disk.
In Fig.~\ref{fig:sample_cmd} we coded the symbols into three groups, including a
class of transition or uncertain objects with $2.3 < C < 2.8$. Their different
distributions in the colour-magnitude diagram is immediately apparent.

Clearly, the majority of CALIFA galaxies ($\sim 2/3$) have substantial disk
components, including irregulars and interacting galaxies, and are more or less
actively forming stars. The final sample of $\sim 400$ of such galaxies will
clearly exceed any previous IFU study by a large factor. Interestingly, the
ominous ``green valley'' intermediate to star-forming and passive galaxies is
well covered by the sample.

On the other hand, CALIFA will also provide IFU data for some 200
bulge-dominated, morphologically early-type galaxies, most of which --
as expected -- cluster very strongly along the red sequence. While the
successful ATLAS$^{\mathrm{3D}}$ project \citep{cappellari10} has
already observed an even somewhat larger number of early-type
galaxies, CALIFA will complement the insights from
ATLAS$^{\mathrm{3D}}$ due to its much larger spectral coverage ($\sim
6\times$) and FoV, which will allow the study of the outer regions of
early-type galaxies.

\begin{figure}
\includegraphics[width=11cm,angle=0,clip=true,trim=70 0 0 70]{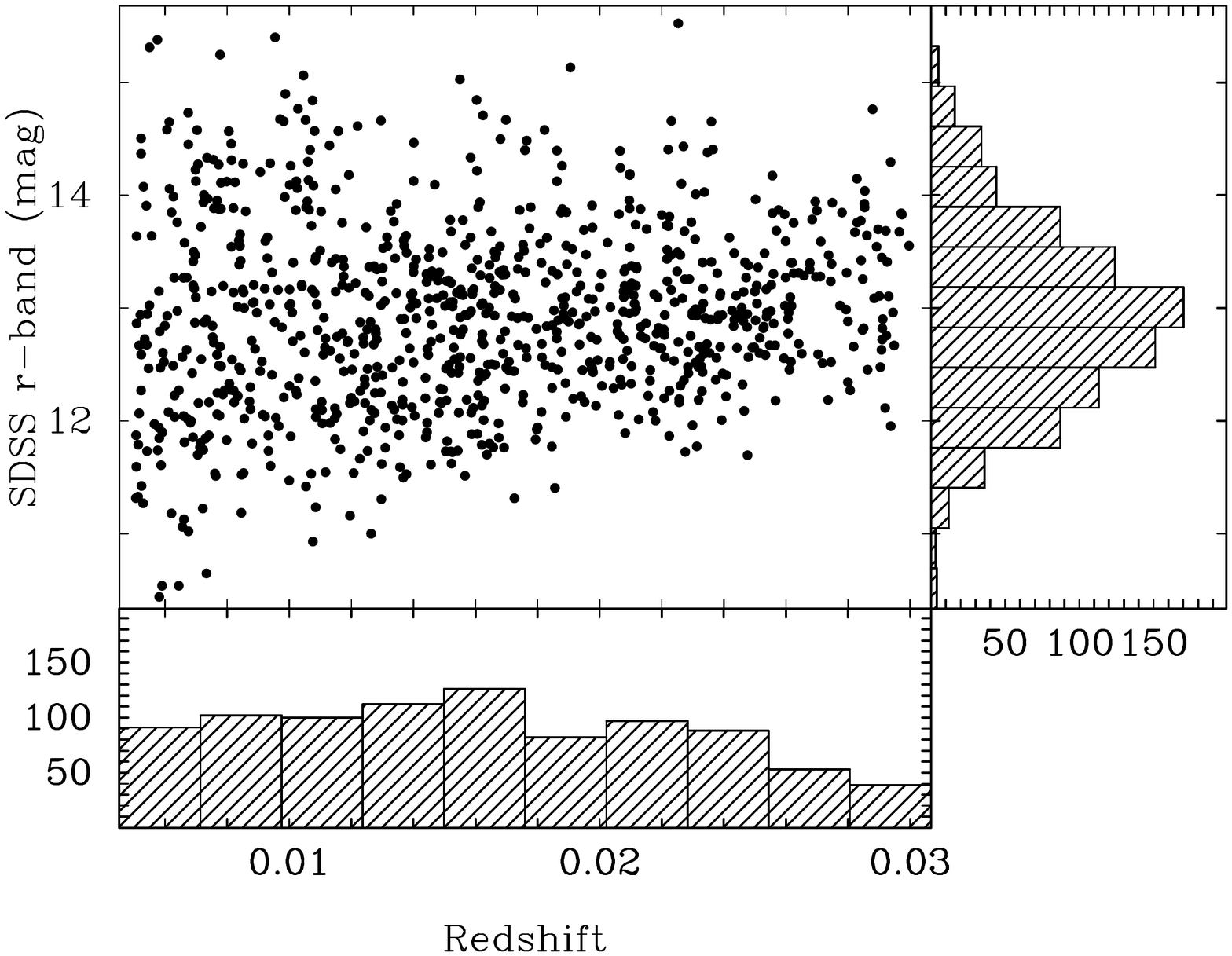}
\caption[]{ Apparent $r$-band magnitude and redshift distribution of the CALIFA mother  sample.}
\label{fig:magz}
\end{figure}

In terms of environment, the CALIFA sample is clearly dominated by field
galaxies. It will effectively include galaxy populations in groups, but much
denser environments will be poorly sampled. A rough estimate of the total number of
galaxy clusters in the sample can be obtained from the counts by \cite{bail11}, 
leading to a maximum of 10 clusters that will, at least partially,
be covered by CALIFA. Fortunately, the Coma cluster at $z\simeq 0.023$ is fully
covered by the CALIFA footprint and redshift range (cf.\ Fig.~\ref{fig:radec}). Therefore, there will be some (limited) ability to study the environmental
dependence of galaxy properties.

Any sample of galaxies faces the question about its ability to
represent, in a statistically well-defined way, the properties of the
galaxy population as a whole. In some cases, e.g.\ in the
ATLAS$^{\mathrm{3D}}$ project, it has been possible to construct and
observe strictly volume-limited samples for which this relation can be
directly made, but this is not a viable approach for a more generic
survey covering a wide range of luminosities. Many surveys are simply
flux-limited (such as the SDSS main galaxy sample), possibly
supplemented by an additional volume limit imposed by redshift cuts,
and statistical relations are then obtained by applying volume
corrections to the apparent trends.  One of the advantages of working
with a diameter-limited sample as in CALIFA is the fact that volume
corrections can then be applied just as easily as for 
flux-limited samples \citep[see,][, for
  extensive discussions of diameter-limited samples and volume
  correction]{davi90,dejo96}. We will always use such corrections for
CALIFA when deriving statistical trends within the galaxy population.

Of course, our ability to perform volume corrections does not imply that the
CALIFA sample is free from selection biases. Such biases may even occur
directly as a consequence of the selection procedure. For example, the
low-redshift limit of $z>0.005$ causes low-luminosity galaxies to be missing
from the CALIFA sample, with incompleteness setting in roughly around $M_r \sim
-20$ and becoming severe for $M_r \ga -18$. This type of bias is unavoidable
when applying explicit redshift cuts. The degree of incompleteness can then be
estimated by comparison with samples selected by other criteria. Nevertheless,
there may also be more subtle biases affecting our sample that we are currently
not aware of, and we will investigate this issue further.  We also continuously
check that the selection of CALIFA galaxies for observation -- supposed
to be essentially random, since it is based only on visibility during a given
night -- does not introduce spurious trends incompatible with the full sample.

As already mentioned, a more detailed characterization of the CALIFA mother sample
will be presented in a separate article that is currently being prepared. That
paper will provide a detailed analysis of the distribution of morphological and
multicolour properties, estimates of the survey selection function for several
parameters, and in particular a compilation of archival multi-wavelength
information.

}

\begin{figure}[t]
 \centering
 \includegraphics[width=11cm,angle=0,clip=true,trim=20 0 0 90]{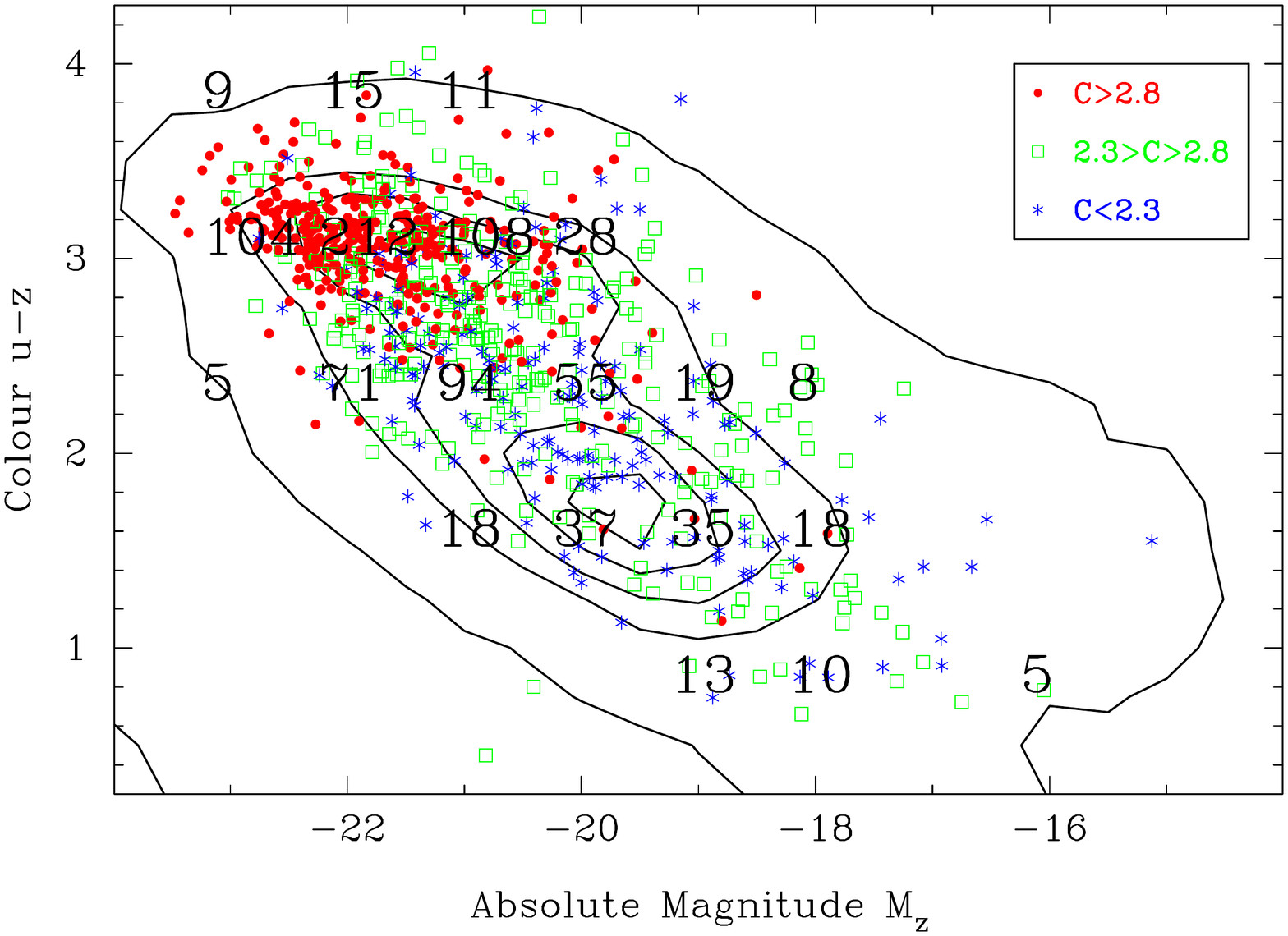}
 \caption{ Distribution of the CALIFA mother sample in the $u-z$ vs.\ $M_z$ colour-magnitude diagram. The overplotted numbers indicate the number of galaxies in bins of 1\,mag in $M_r$ and 0.75\,mag in $u-z$. Different colours and symbols represent a classification into bulge- and disk-dominated galaxies as well as intermediate cases, as suggested by the concentration index $C$. For comparison, the contours delineate the number density distribution of galaxies in the SDSS-NYU catalogue \textbf{(e.g., Blanton et al.\ 2005)}.}
  \label{fig:sample_cmd}
\end{figure}

 \section{Observing strategy}
 \label{sec:obs}

 The observations of the CALIFA survey officially started in July
 2010, being performed at the 3.5m telescope of the Calar Alto
 observatory with the Potsdam Multi Aperture Spectrograph, PMAS
 \citep{roth05} in the PPAK mode \citep{verheijen04,kelz06}. The PPAK
 fiber bundle consists of 382 fibers of 2.7 arcsec diameter each
 \citep[see fig.~5 in][]{kelz06}. Of these 382 fibers, 331 (the science
 fibers) are concentrated in a single hexagonal bundle covering a
 field-of-view of 74$\arcsec$$\times$64$\arcsec$, with a filling factor
 of $\sim$\,60\%. The sky background is sampled by 36 additional
 fibers, distributed in 6 bundles of 6 fibers each, along a circle
 $\sim$\,72 arcsec from the center of the instrument FoV. The
 sky-fibers are distributed among the science fibers within the
 pseudo-slit in order to have a good characterization of the sky,
 sampled with a similar distortion than the science fibers; the
 remaining 15 fibers are used for calibration purposes. Cross-talk
 between the adjacent spectra in the detector is estimated to be less than 5\% when using a
 pure aperture extraction \citep{sanchez06a}. Adjacent spectra on the CCD may be mapped to very different locations in the spatial plane \citep{kelz06}. However, it introduces an incoherent contamination, not
 important for the present study, that is minimized by adopting a
 more refined extraction procedure during the reduction process.

\begin{figure*}[tb]
 \begin{center}
 \includegraphics[height=6.5cm,angle=0,clip=true,clip=true,trim=60 0 70 80]{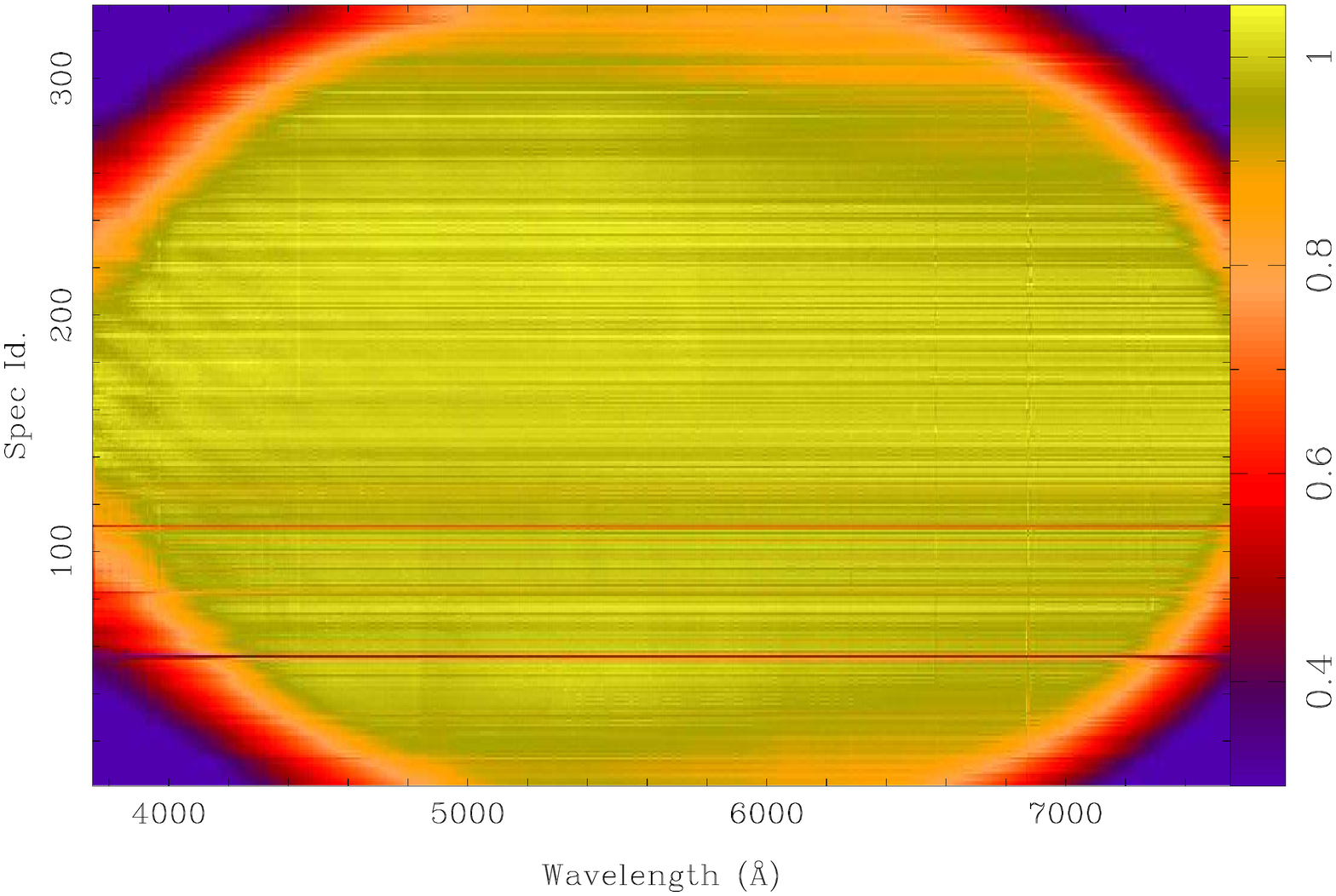}
 \includegraphics[height=6.5cm,angle=0,clip=true,clip=true,trim=100 0 70 80]{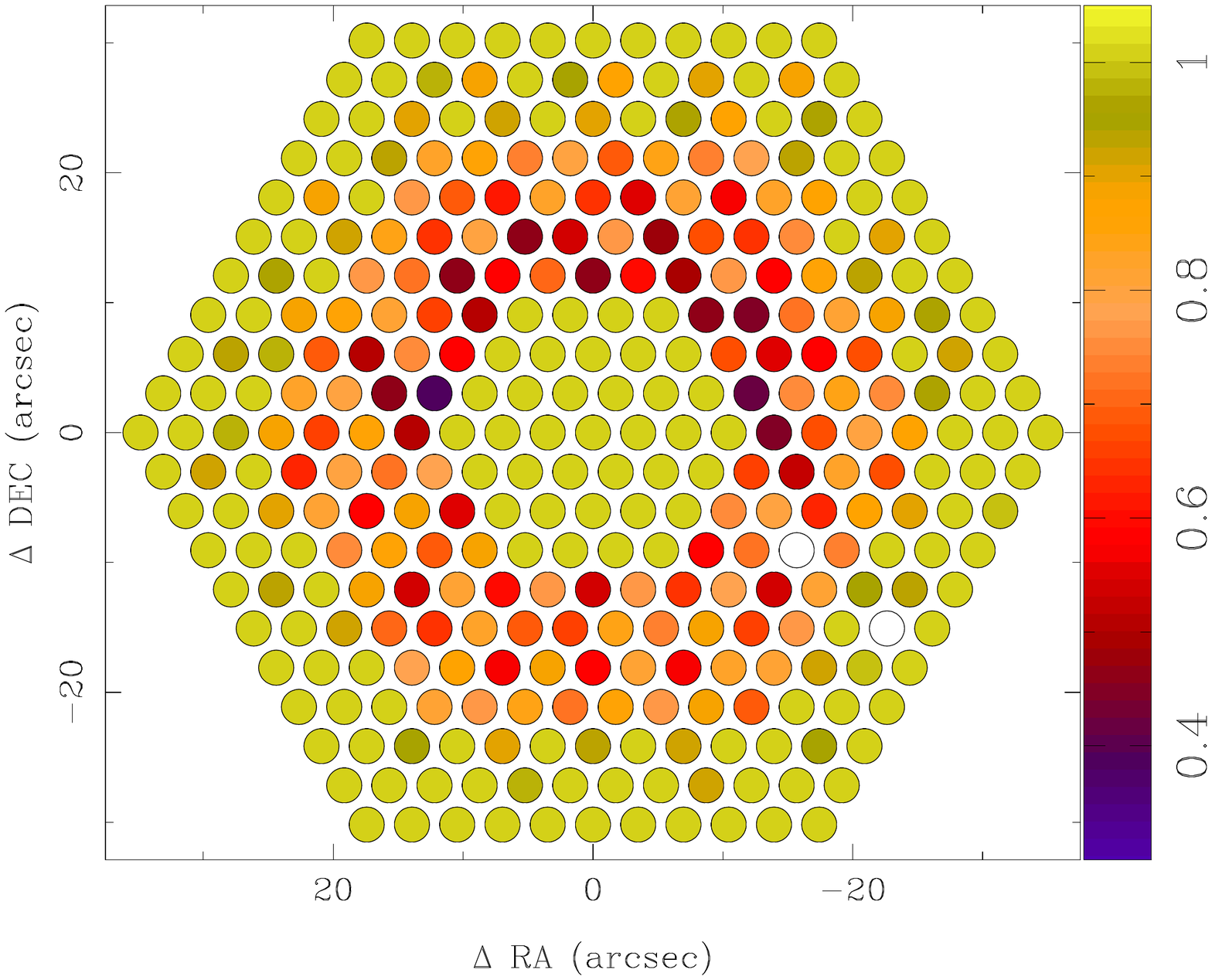}
 \caption{{\it Left panel}: Intensity map of the differential transmission
   fiber-to-fiber for the  V500-grating (known as fiber-flat), corresponding to
   the night of the 10th of June 2010. The effects of vignetting are clearly
   visible  with a significant drop of the
   transmission at the edges of the image.
  {\it Right panel}: Spatial distribution of the fraction of pixels
   free of vignetting, when adopting a conservative criterion of 75\% of
   the average transmission in the central fibers. Two fibers at the
   bottom/right show a transmission below this limit in the whole
   wavelength range.
   \label{fig:vig} }
 \end{center}
 \end{figure*}

 PMAS was upgraded with an E2V CCD231 4K$\times$4K in October 2009
 \citep{roth10}, which has increased nominally the wavelength range
 covered by a certain instrumental setup by a factor two, with respect
 to the values reported by \cite{roth05}.  However, this nominal
 increase is not fully met at the four corners of the CCD. As a result
 of a trade-off study for the PMAS fiber spectrograph, the optical
 system was deliberately designed to tolerate some vignetting at the
 edges of the full detector field-of-view, as reported by
 \cite{roth97}, and \cite{roth98}. Figure \ref{fig:vig} illustrates
 the effect of vignetting for the new 4Kx4K CCD, which was not
 noticeable in the previous 2Kx4K detector configuration.  The left
 panel shows a map of the differential fiber-to-fiber transmission for
 all the 331 fibers in the hexagonal fiber-bundle.  The effect of the
 vignetting is clearly identified as a drop in the transmission at the
 four edges of the map.  { 70\% of the spectra are free from
   vignetting .} However, 30\% of the spectra, namely those close to
 the edge of the detector, do suffer some loss of efficiency that
 gradually increases towards the corners of the CCD. The loss of
 throughput amounts to up to 30\% { in the best case} and 50\% {
   in the worst case}, for one half of those spectra, but more than
 50\% for the other half. It can be seen in Fig. \ref{fig:vig} (left
 panel), that where vignetting sets in, no more than a quarter of the
 full spectral range is affected. In summary the vignetting severely
 affects less than a 15\% of the fibers, in less than a 25\% of their
 wavelength range.

 The right panel illustrates the spatial distribution of the
 vignetting.  The mapping of the PPAK fibers from the telescope focal
 plane to the CCD implies that the areas affected by the vignetting
 correspond to an annular ring at about $\sim$15$\arcsec$ from the
 center of the FoV.  In addition, we report the existence of two broken
 fibers with a lower transmission than the average (fibers 65 and 127),
 not reported before.

 \begin{table*}
 \caption{Summary of the objects observed in the first CALIFA runs}             
 \label{tab:obj}      
 \begin{center}
 \begin{tabular}{lccrrrrl}        
 \hline\hline                 
 Name$^1$ & RA (J2000)$^2$ & DEC (J2000)$^2$ & Redshift$^2$ & g$_{SDSS}$$^2$ & u-g$^2$ & M$_{g}$ &Morph.$^1$ \\
  \hline                        
 NGC 5947  & 15:30:36.00 & +42:43:01.00 & 0.01965 & 14.20 & -20.41 & 1.57 & SBbc   \\
 UGC 09892 & 15:32:51.94 & +41:11:29.27 & 0.01893 & 14.83 & -19.70 & 1.66 & Sb     \\
 UGC 10693 & 17:04:53.01 & +41:51:55.76 & 0.02799 & 13.59 & -21.79 & 2.14 & E      \\
 UGC 10710 & 17:06:52.52 & +43:07:19.96 & 0.02795 & 14.52 & -20.86 & 1.98 & Sb     \\
 NGC 6394  & 17:30:21.42 & +59:38:23.62 & 0.02842 & 14.60 & -20.81 & 1.82 & SBb    \\
 NGC 6411  & 17:35:32.84 & +60:48:48.26 & 0.01227 & 12.99 & -20.59 & 2.37 & E      \\
 NGC 6497  & 17:51:17.96 & +59:28:15.15 & 0.02055 & 13.80 & -20.88 & 2.39 & SB(r)b \\
 NGC 6515  & 17:57:25.19 & +50:43:41.24 & 0.02284 & 13.72 & -21.22 & 2.03 & E      \\
 UGC 11228 & 18:24:46.26 & +41:29:33.84 & 0.01935 & 14.04 & -20.54 & 2.03 & SB0    \\
 UGC 11262 & 18:30:35.69 & +42:41:33.70 & 0.01862 & 15.16 & -19.33 & 1.82 & Sd     \\
 NGC 6762  & 19:05:37.09 & +63:56:02.79 & 0.00976 & 14.08 & -19.00 & 1.82 & S0/a   \\
 UGC 11649 & 20:55:27.62 & -01:13:30.87 & 0.01265 & 14.01 & -19.64 & 2.29 & SB(r)a \\
 UGC 11680 $^*$& 21:07:41.33 & +03:52:17.80 & 0.02599 & 14.56 & -20.66 & 2.89 & Sb \\
 NGC 7025  & 21:07:47.33 & +16:20:09.22 & 0.01657 & 13.17 & -21.07 & 2.28 & Sa     \\
 UGC 11694 & 21:11:52.02 & +11:16:34.11 & 0.01699 & 14.09 & -20.21 & 2.18 & S0     \\
 UGC 11717 & 21:18:35.41 & +19:43:07.39 & 0.02116 & 14.68 & -20.09 & 2.29 & S      \\
 UGC 11740 & 21:26:14.35 & +09:47:52.45 & 0.02151 & 14.70 & -20.11 & 1.90 & S?     \\
 NGC 7194  & 22:03:30.93 & +12:38:12.41 & 0.02718 & 13.87 & -21.44 & 2.02 & E      \\
 UGC 12127 & 22:38:29.41 & +35:19:46.89 & 0.02745 & 13.88 & -21.46 & 2.56 & E      \\
 UGC 12185 & 22:47:25.06 & +31:22:24.67 & 0.02217 & 14.35 & -20.52 & 2.03 & SBab   \\
 NGC 7549  & 23:15:17.26 & +19:02:30.43 & 0.01573 & 14.14 & -19.98 & 2.01 & Sb?    \\
 \hline                                   
 \end{tabular}
 \end{center}

 $(*)$ This object has accurate data with the V500 setup only.

 $^1$ Data obtained from the NED.

 $^2$ Data obtained from the NYU-SDSS catalogue.

 \end{table*}

 A dithering scheme with three pointings has been adopted in order to
 cover the complete FoV of the central bundle and to increase the
 spatial resolution of the data. This scheme was already used in
 previous studies using PPAK \citep{sanchez07b, castillo-morales10,
   perez-gallego10, rosales-ortega10}.  In most of these studies the
 offsets in RA and DEC of the different pointings, with respect to the
 nominal coordinates of the targets, were: 0$\arcsec$,0$\arcsec$ ;
 +1.56$\arcsec$,+0.78$\arcsec$, +1.56$\arcsec$,$-$0.78$\arcsec$. These
 offsets correspond to half the spacing between adjacent fibers, which
 allows to cover the holes between fibers in the central
 bundle. However, in order to correctly address the problems generated
 by the vignetting reported before, we adopted a different dithering
 scheme for CALIFA, with wider offsets in RA and DEC, i.e.:
 0$\arcsec$,0$\arcsec$ ; $-$5.22$\arcsec$,$-$4.84$\arcsec$ ;
 $-$5.22$\arcsec$, +4.84$\arcsec$. These offsets correspond to jumping
 one fiber to the adjacent hole. This procedure ensures that in the final dithered and rebinned dataset
 the throughput losses for any spectrum affected by vignetting are
 compensated through the spectra from at least two adjacent fibers,
 closer than 2$\arcsec$, which are unaffected by vignetting.
 A spatial recomposition of the three pointings is included in
 the standard data reduction scheme.

 All objects will be observed using two different and complementary
 setups. Firstly, the new grating, purchased for CALIFA and named V500, is used. It
 has a nominal resolution of $\lambda/\Delta\lambda\sim$850 at
 $\sim$5000~\AA\ (FWHM$\sim$6~\AA), to cover the widest possible
 wavelength range, nominally from [OII]$\lambda$3727 to
 [SII]$\lambda$6731 in the rest frame of all objects in the
 survey. Secondly, the V1200 grating \citep{roth05} is used,
 with a nominal resolution of $\lambda/\Delta\lambda\sim$1650 at
 $\sim$4500~\AA\ (FWHM$\sim$2.7~\AA). It covers the blue wavelength
 range, including [OII]$\lambda$3727, the Balmer break at
 $\sim$4000-4400~\AA, H$\delta$, H$\gamma$ and [OIII]$\lambda$4363. The
 main purpose of the first setup is to study the stellar populations
 and the properties of the ionized gas using the widest possible
 baseline of spectral features and emission line species. The
 observations with the second setup will provide accurate measurements
 of both the stellar and ionized gas kinematics, mostly through
 the H+K absorption features and the [OII]$\lambda$3727
 emission line, respectively.

 In order to avoid strong effects of the described vignetting on the
 final dataset, the wavelength range covered by each grating was
 selected in a complementary way. The range covered by the V500
 grating was selected such that the reddest emission line of
 interest ([SII]$\lambda$6731) was free of vignetting in all
 fibers. The range covered by the V1200 grating was selected such
 that the bluest emission line of interest ([OII]$\lambda$3727) was
 free of vignetting. This selection reduces the final wavelength
 range covered with each grating, but guarantees that all features of
 interest are unaffected by vignetting in at least one of the
 setups. A fixed setup was selected for each grating for the whole
 survey, to guarantee the homogeneity of the dataset. The final
 nominal wavelength range covered by each grating is 3745-7300~\AA~
 (V500) and 3400-4750~\AA~ (V1200), respectively. However, for the
 V1200 grating, the wavelength range bluer than 3700~\AA\ is of
 limited use, due to the sharp drop of the transmission of the PPAK
 IFU fore-optics and fibers in this wavelength range.

 The exposure time is fixed for all targeted objects. For the V500
 grating a single exposure of 900s per pointing of the dithering
 pattern is taken, while for the V1200 grating 3 exposures of 600s each are
 obtained per position.  The exposure times were selected based on our
 previous experience with the instrument \citep[in particular][ a
   feasibility study of the current survey]{marmol-queralto11}, and
 expectations about the performance of the V500 grating (which was
 unknown at the beginning of the survey).

 Data presented here were obtained during 16 nights (6 dark and 10 grey) 
 in June-July 2010. A total of 20 objects were observed in both setups
 (1 more was observed only with the V500 grating). Table
 \ref{tab:obj} lists the observed objects, including their coordinates,
 redshift, SDSS $g$-band { observed and absolute} magnitudes and $u-g$ color and their
 morphological classification (extracted from the NASA/IPAC Extragalactic Database\footnote{http://nedwww.ipac.caltech.edu/}). These objects
 are described here since we will use them to illustrate
 the current status of the data acquisition and its quality.

 \section{Data reduction}
 \label{sec:redu}

 The reduction of the CALIFA data is performed using a fully automatic
 pipeline developed ad-hoc. This pipeline operates without human
 intervention, producing both the scientifically useful frames and a
 set of quality control measurements that help to estimate the
 accuracy of the reduced data. The pipeline uses the routines included
 in the {\sc R3D} package \citep{sanchez06a} and the {\sc E3D}
 visualization tool \citep{sanchez04}, following the standard steps
 for fiber-based IFS \citep[e.g.,][]{sanchez06a,sandin10}. The
 pipeline will be upgraded and improved during this project on the
 basis of the quality tests performed on the acquired data. Here we
 present a summary of the current implementation of the data
 reduction.

 \subsection{Removal of electronic signatures and realignment of the frames}

  The CCD used for CALIFA is a new 4k$\times$4k E2V
 detector installed in the instrument in October 2009. In contrast to the
 old 2k$\times$4k one (Roth et al. 2005), the read-out software for the
 new CCD stores each frame in four different FITS files, corresponding
 to each of the four amplifiers of the detector, each one with
 different bias and gain levels. As a first step, the four files are
 combined into a single frame, after subtracting the corresponding bias
 level, and transforming from counts to electrons on the basis of the
 corresponding gain. A master bias frame is created by averaging all
 bias frames observed during the night, in order to cross-check the
 stability of the bias level. In contrast to the old CCD, the new CCD does
 not exhibit any detectable structure in the bias frame.

 The next step is clipping cosmic rays and combining different
 exposures taken at the same position on the sky. However, prior to
 combination, possible offsets between the projected spectra in the CCD
 due to flexure are estimated. As shown in \citet{kelz06}, flexure is a
 major issue when working with PPAK. In general, the amplitude of the
 flexure is stronger while targets are setting. However, in a large
 survey like CALIFA, it will not be feasible to avoid observing setting
 targets 100\%\ of the time. On the other hand, calibration frames are
 taken within a maximal range of 1.5 hours from the science
 frames. This ensures that in most cases the flexure pattern of the
 calibration and science frames is the same and, therefore, it is
 possible to trace the spectra projected in the CCD and extract them
 properly. However, in a few cases (e.g.~when the targets are observed
 when setting), there could be small linear offsets between the science
 and the calibration frames. To guarantee a correct handling of 
 flexure, these offsets are estimated based on the position of several
 ($\sim$50) spots in the raw data. They correspond to the Hg emission
 lines of the arc lamps on one calibration frame, and the night-sky
 light pollution Hg emission \citep{sanchez07a} on the science frames
 (including Hg$\lambda$4047, Hg$\lambda$4358 and Hg$\lambda$5461).  A
 linear shift is applied to the frame, when a significant offset
 ($>$0.1 pixel) is detected among the different exposures, and between
 them and their corresponding calibration frames (continuum and arc
 lamp exposures). This was the case in $\sim$2\% of the frames taken
 during the first observing runs.  Finally, a cosmic-ray cleaning
 algorithm is applied if only one single frame is taken at a certain
 position. The actual cleaning algorithm is based on the Laplacian Edge
 detection algorithm described in \cite{van-dokkum01}, recently
 implemented in {\sc R3D}.

 \begin{figure}
 \begin{center}
 \includegraphics[width=10cm,angle=0,clip=true,clip=true,trim=20 0 70 70]{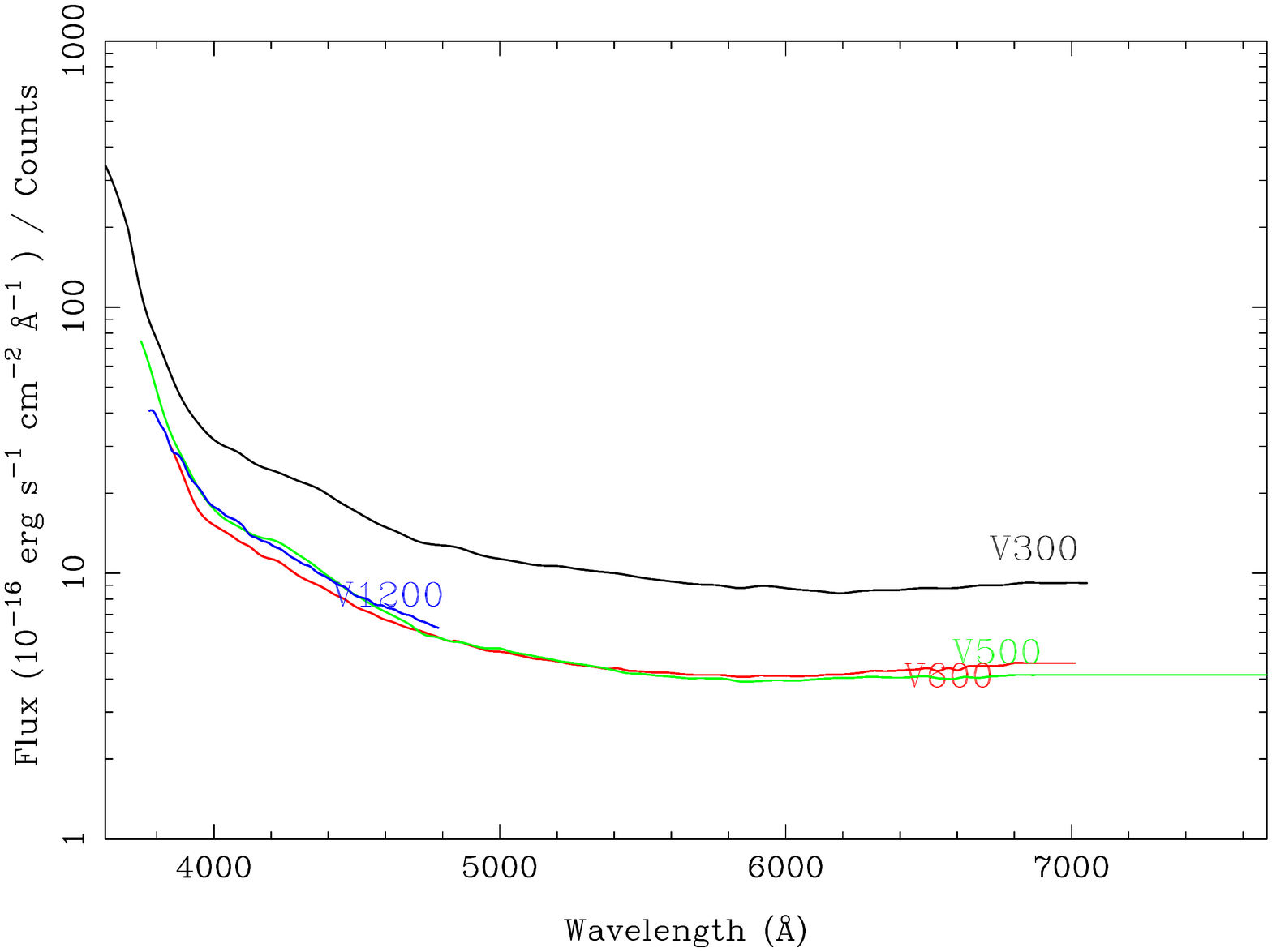}
 \caption{Flux density-to-counts ratio for the gratings used in the CALIFA survey, 
 derived for the most transparent night of those observed so far, as compared 
 to the most frequently used gratings for PMAS (V300 and V600). 
 \label{fig:effi}
 }
 \end{center}
 \end{figure}

 \begin{figure}[tb]
 \begin{center}
 \includegraphics[height=6.5cm,angle=0,clip=true,clip=true,trim=20 20 50 50]{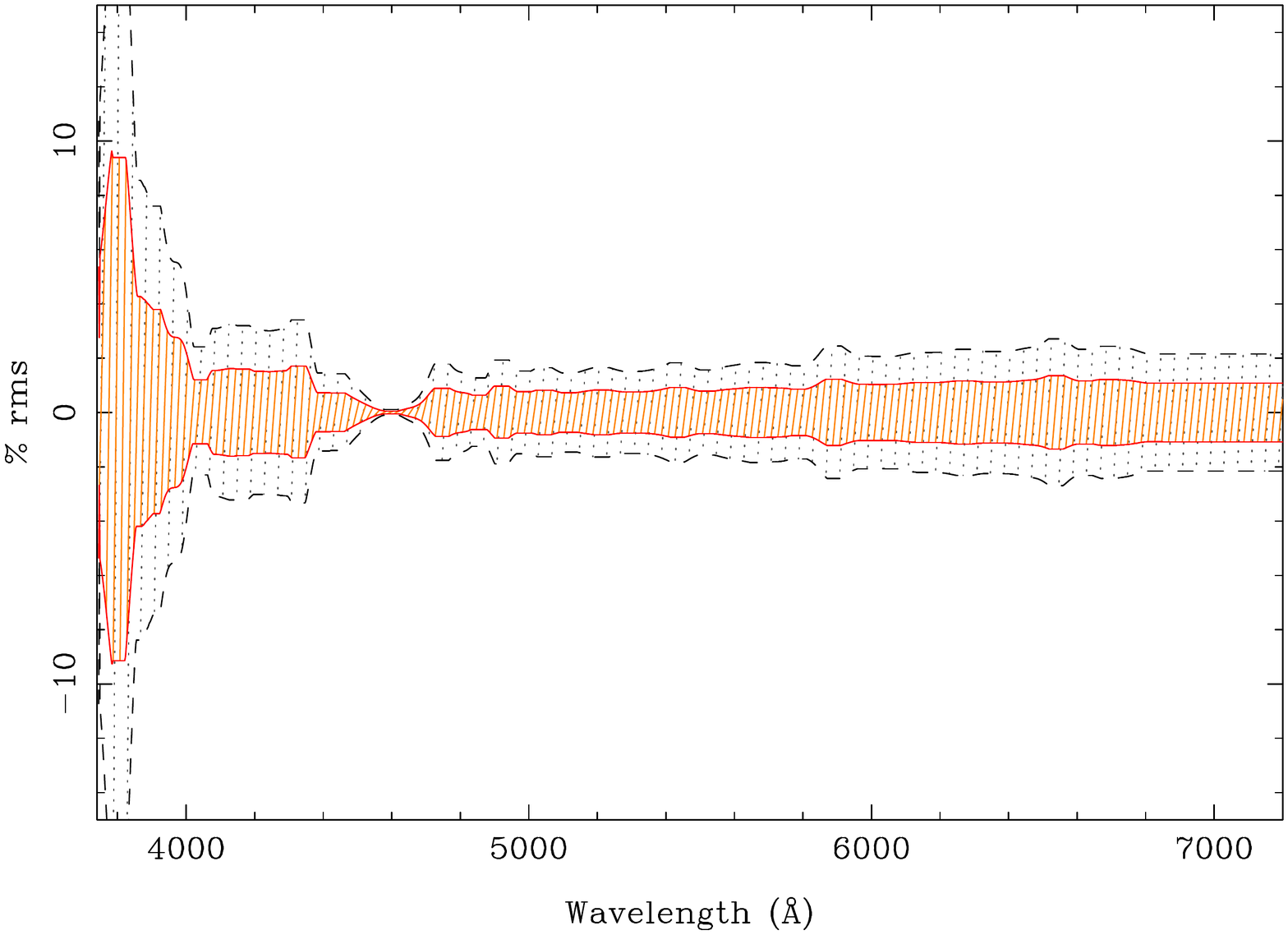}
 \includegraphics[height=6.5cm,angle=0,clip=true,clip=true,trim=20 20 50 50]{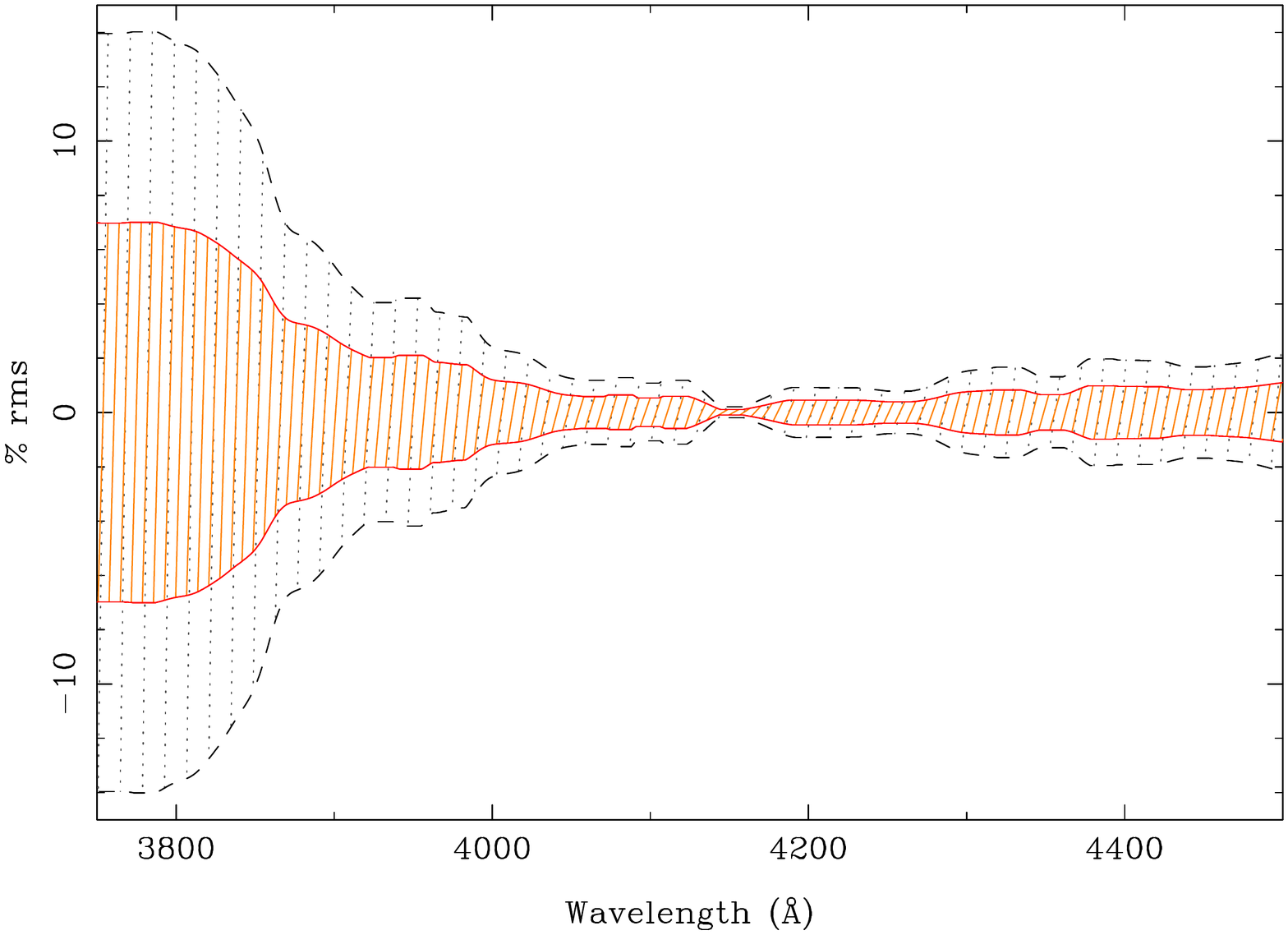}
 \caption{Relative differences, in percent, between the response curves found on different
   nights for the V500 grating dataset (top panel) and the V1200 grating (bottom
   panel), as a function of wavelength, normalized to unity at 4600 ~\AA\
   and 4150 ~\AA, respectively. The solid hashed region (and red
   solid line) shows the 1$\sigma$ error, while the dashed hashed region (and
   black dashed line) shows the 2$\sigma$ error.
   \label{fig:rms} }
 \end{center}
 \end{figure}

 \subsection{Spectral extraction, wavelength calibration and fiber transmission correction}

 The locations of the spectra on the CCD are determined using a continuum-illuminated 
 exposure taken before the science exposures. Each spectrum is
 then extracted from the science frames. In order to reduce the effects of 
 cross-talk we do not perform a simple aperture extraction, which would
 consist of co-adding the flux within a certain number of pixels of the location
 derived from the continuum-illuminated exposure. Rather, we adopt a modified version
 of the Gaussian suppression \citep{sanchez06a}, first used in the
 reduction of the PINGS data \citep{rosales-ortega10}, and described in
 \citet{sanchez11}. The cross-talk is reduced to less than 1\% when
 adopting this new method. This procedure relies on an accurate internal focus
 of the instrument. The quality of the instrumental focus, which depends on the
 internal temperature of the structure of the instrument, is controlled by
 fitting a Gaussian to a subset of the projected continuum illuminated spectra
 along the cross-dispersion axis, and comparing the results with the nominal
 value for the instrument ($\sim$2.5 pixels).

 The extracted flux, for each pixel in the dispersion direction, is stored in a
 row-stacked-spectrum file \citep{sanchez04}. Wavelength
 calibration is performed using HeHgCd lamp exposures obtained before and
 after each pointing. The number of identified emission lines used to perform
 the wavelength calibration ranges between 11 (for the V1200 grating) and 16
 (for the V500 grating), homogeneously distributed along the  wavelength range covered. 
 It was found that the adopted procedure yields an
 accuracy of about ten percent of the nominal pixel scale in the wavelength 
 calibration (i.e. , $\sim$\, 0.2 ~\AA\ for the V500 grating and $\sim$\, 0.1 ~\AA\
 for the V1200 one). Differences in the relative fiber-to-fiber transmission
 throughput are corrected by comparing the wavelength-calibrated row-stacked-spectra science
 frames with the corresponding frames derived from sky exposures taken during
 the twilight. These frames are also used to determine which pixels at the edge
 of the CCD are affected by the previously described vignetting effect.  Those
 pixels for which the transmission drops to $<75$\% are masked.

 \subsection{Sky subtraction}
 \label{subsecsky}

 PPAK is equipped with 36 fibers to sample the sky, distributed around the
 science fiber-bundle, in six small bundles of six fibers each, at a
 distance of $\sim$75$\arcsec$ from the center of the FoV (see Kelz et
 al. 2006).  By construction, the objects selected for the CALIFA final sample
 cover a substantial fraction of the FoV of the central PPAK bundle, with most
 of the sky fibers free from emission by the corresponding target. The
 procedure adopted to derive the night-sky spectrum is to combine the spectra
 corresponding to these fibers, performing a 2$\sigma$ clipping rejection to
 remove any possible contamination from low surface brightness regions of the
 galaxy and/or projected companions. Once obtained, the sky spectrum is
 subtracted from all spectra of the corresponding frame. The accuracy of the
 subtraction is controlled by comparing the equivalent width of the most
 prominent night-sky emission lines \citep{sanchez07a}, before and after
 the subtraction for each individual spectrum within the frame. In addition,
 each derived night-sky spectrum is used to derive the night-sky brightness at
 the location and time when the target is observed, by convolving the
 spectrum with the transmission curves of the Johnson $V$-band ($B$-band)
 filter for the V500 (V1200) setups. This value is used to control the
 actual conditions when the data are acquired.

 \begin{figure*}[tb]
 \begin{center}
 \includegraphics[height=5.2cm,angle=0,clip=true,trim=30 20 146 20]{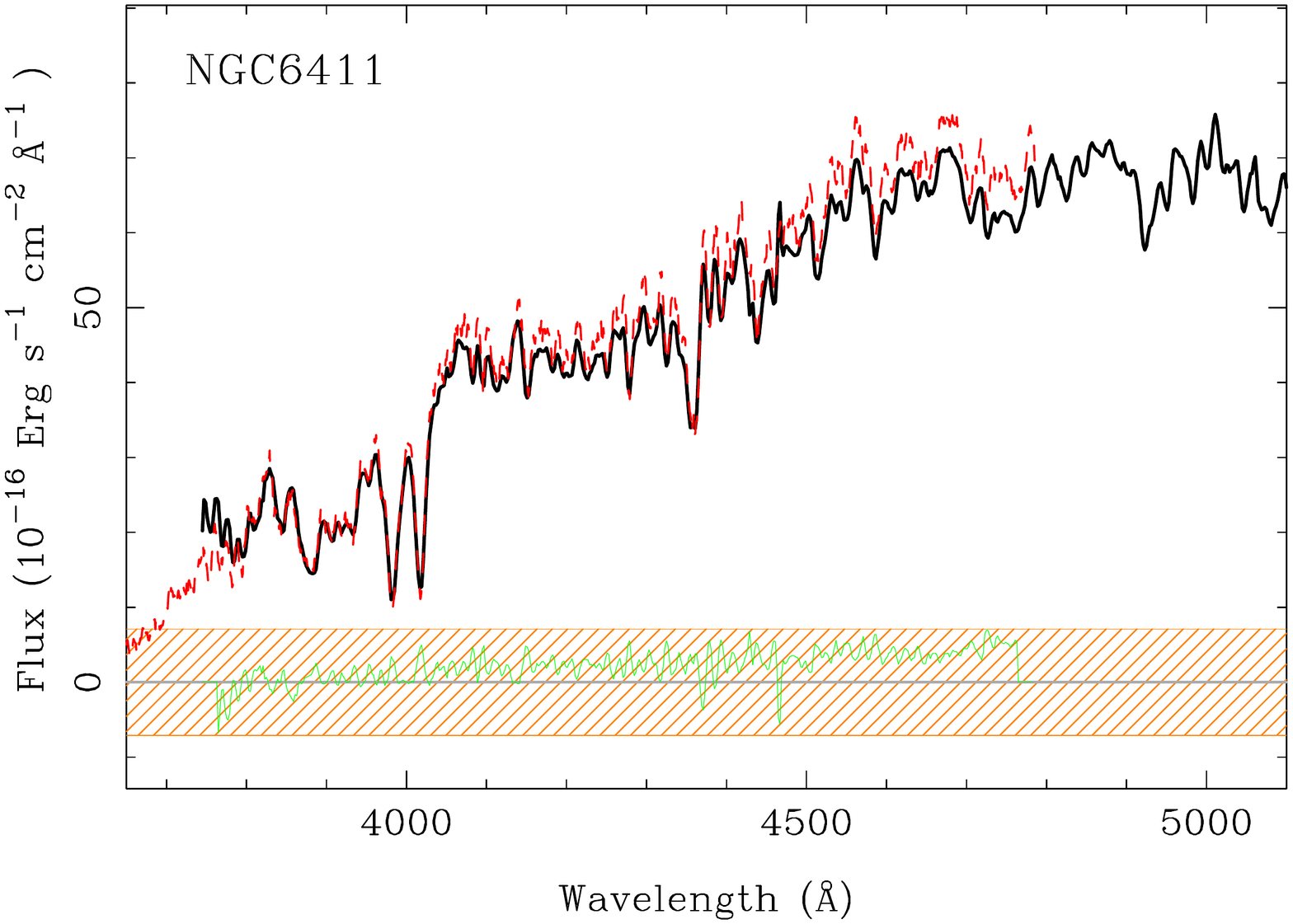}
 \includegraphics[height=5.2cm,angle=0,clip=true,trim=57 20 146 20]{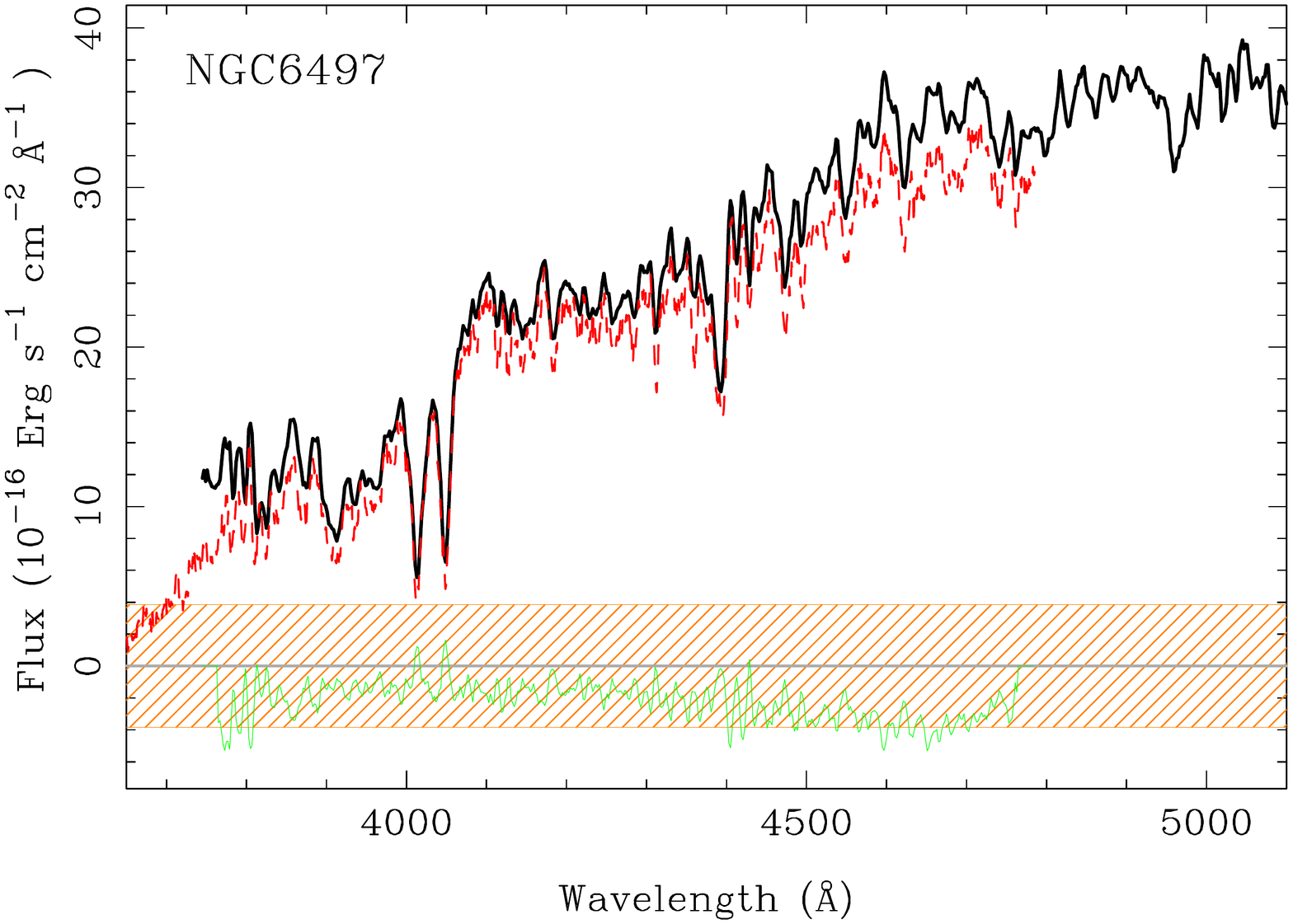}
 \includegraphics[height=5.2cm,angle=0,clip=true,trim=57 20 146 20]{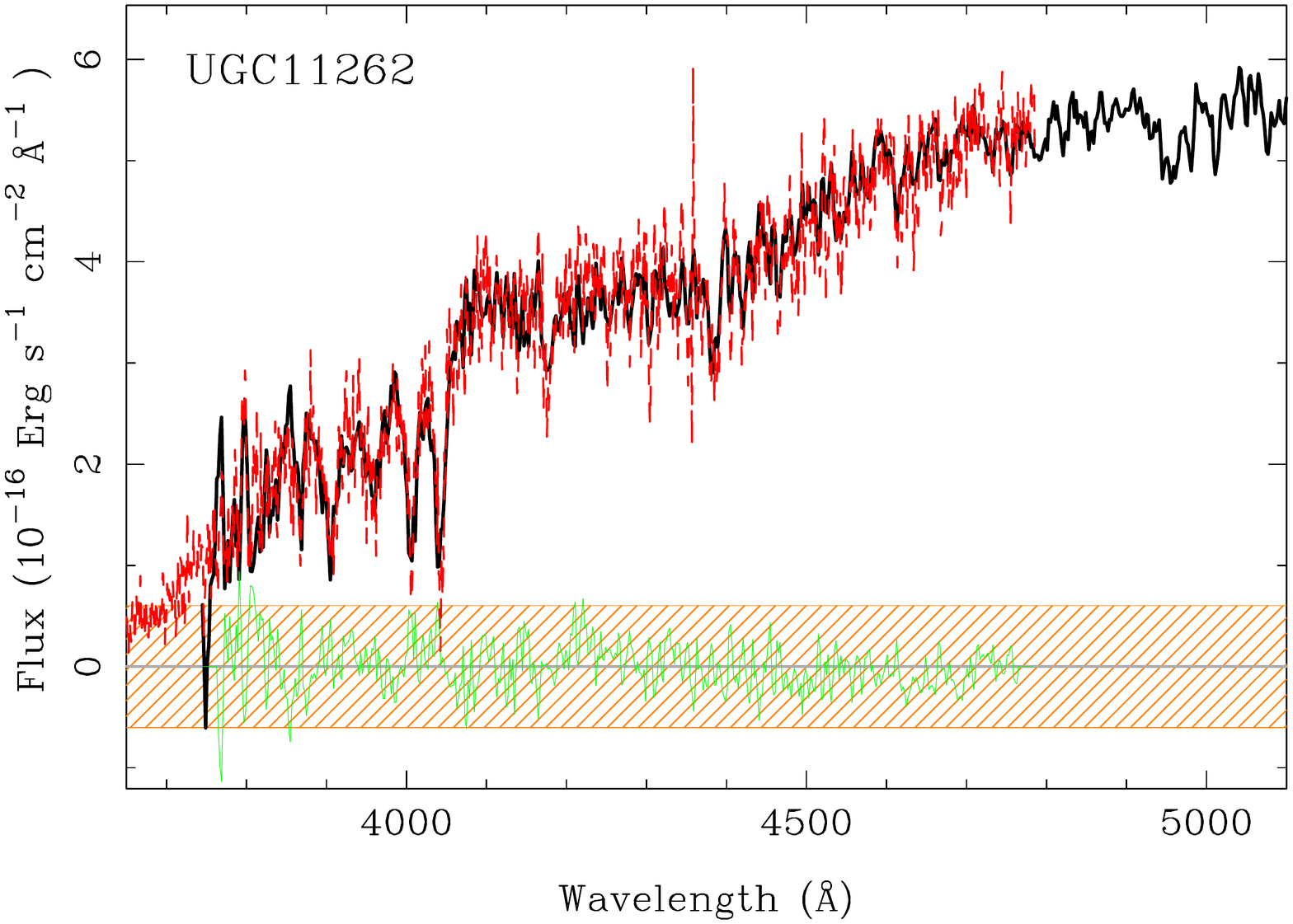}
 \caption{Comparison of the central 5$\arcsec$ spectrum extracted from the
   V500 (black, thick solid line) and V1200 (red, dashed line) grating datacubes for three typical objects
   (from left to right):  NGC 6411, the brightest object observed so far;
   NGC 6497, a spiral galaxy with an average brightness, and UGC11262, the
   faintest target observed so far. The green thin solid line shows
   the difference between both spectra. { The orange hashed region indicate the $\pm$10\%\ median intensity range of both spectra around the zero level, indicated with a grey continuous line.}   
   \label{fig:comp_r} }
 \end{center}
 \end{figure*}

 \subsection{Flux Calibration}

 Flux calibration is performed by comparing the extracted spectra of
 spectrophotometric standard stars from the Oke Catalogue
 \citep{oke90}, observed during each night, with their corresponding
 flux-calibrated spectra, publicly available on the webpage of the
 observatory\footnote{http://www.caha.es/pedraz/SSS/sss.html}.  For the
 currently observed nights, the stars were: BD+25d4655, BD+28d4211 and
 BD+33d2642 (although any suitable star from the list will be used
 in the course of the survey). When feasible, two different
 calibration stars are observed in the same night. Stars are not observed using the
 dithering pattern. A finite aperture of $r<$10$\arcsec$ was selected after several experiments,
 trying to maximize the covered flux and the signal-to-noise.
 This finite aperture, and the incomplete coverage of the FoV by the
 PPAK fiber bundle, produce a systematic offset of 15\% in the flux
 calibration, estimated on the basis of simulated extractions of
 stellar images with a typical seeing of $\sim$1$\arcsec$. This offset
 is well known, and it has already been reported by other authors
 (e.g.~PINGS, Rosales-Ortega et al. 2010). This systematic offset has
 been corrected prior to any further analysis. 

 R3D includes a procedure that finally provides the transformation
 function from observed counts to intensity, taking into account also the
 airmass and extinction of the observations of both the spectrophotometric
 standard stars and the science targets. To apply this procedure we adopted the
 extinction measured by the Calar Alto Vistual EXtinction monitor (CAVEX) at the moment of the
 observations, and the average extinction curve for the observatory \citep{sanchez07a}.
 The procedure ensures a good relative flux calibration from the
 blue to the red part of the spectra, if the weather conditions throughout the night
 are stable. However, an absolute offset between the derived and real fluxes
 is expected if the weather conditions varied during the observations. If more
 than one standard star is observed, the pipeline performs a comparison
 between the transformations derived using each of them, in order to estimate
 the photometric stability of the night considered.

 Figure \ref{fig:effi} shows the derived transformation function from counts to
 flux density for both the V1200 and V500 gratings (corrected for the offset in
 resolution), together with those for the most frequently used gratings for PMAS (the
 V300 and V600 ones). These transformation functions were derived by selecting
 the curves for the most transparent nights included in the current study,
 together with the ones derived in a previous study 
 \citep[][for the V300  and V600 gratings]{marmol-queralto11}. It should be
 noted that there was a major update in the instrument, with an exchange of
 the CCD, between the observations taken with the V300 grating and the other observations. The
 offset between the efficiency found for this grating and the remaining gratings
 can easily be explained by the improvement in efficiency of the new CCD. 
 Otherwise all efficiency curves are quite similar. This 
 basic conclusion would not change if instead of using the best curve, we had
 selected the mean or median transformation function. These results are in
 agreement those reported by \citet{roth05}, where it was stated that
 all the V-gratings show basically the same efficiency (after correcting for the
 resolution effects).

 \begin{figure*}[tb]
 \begin{center}
 \includegraphics[height=6.7cm,angle=0,clip=true,trim=30 20 157 60]{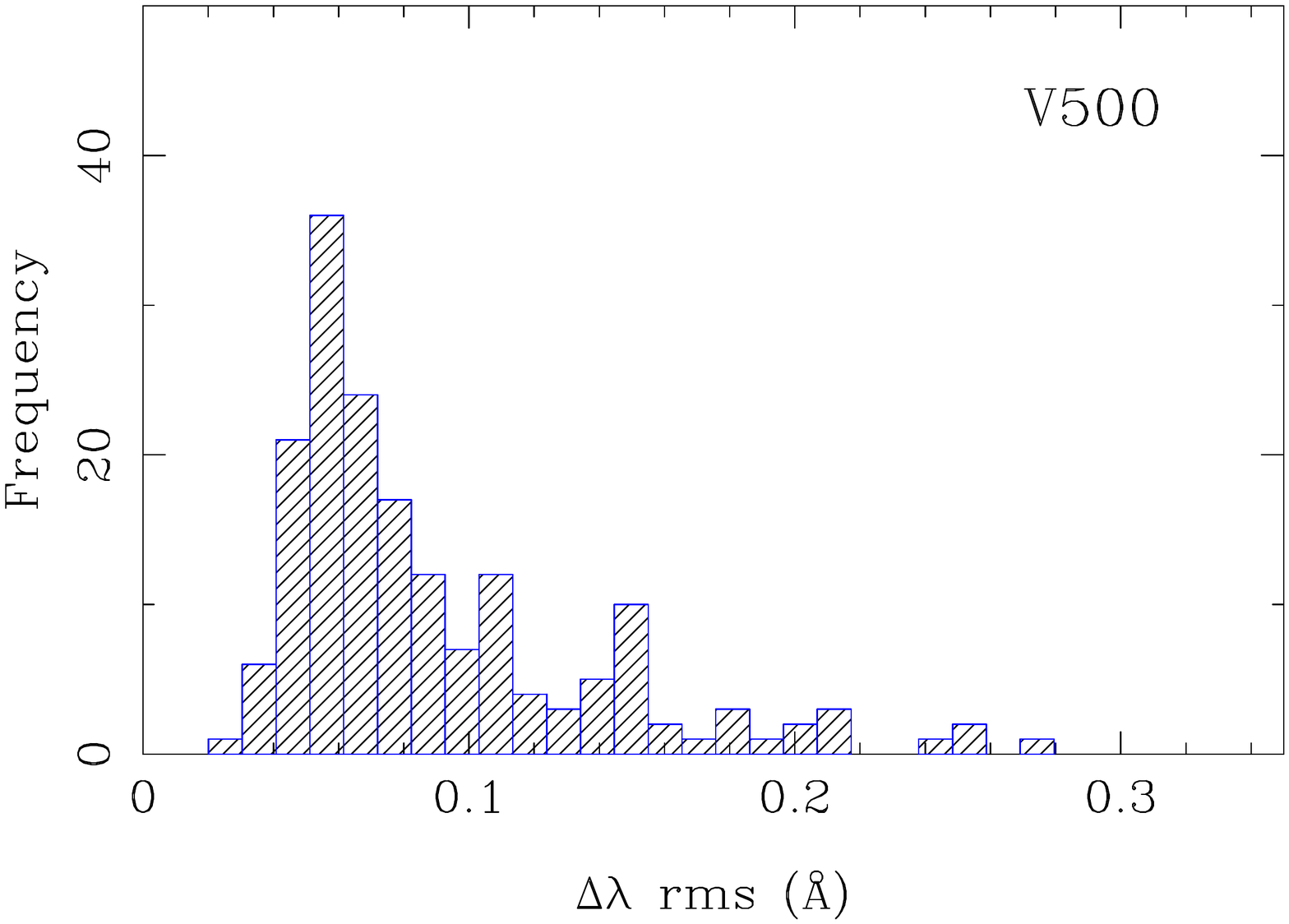} 
\includegraphics[height=6.7cm,angle=0,clip=true,trim=98 20 157 60]{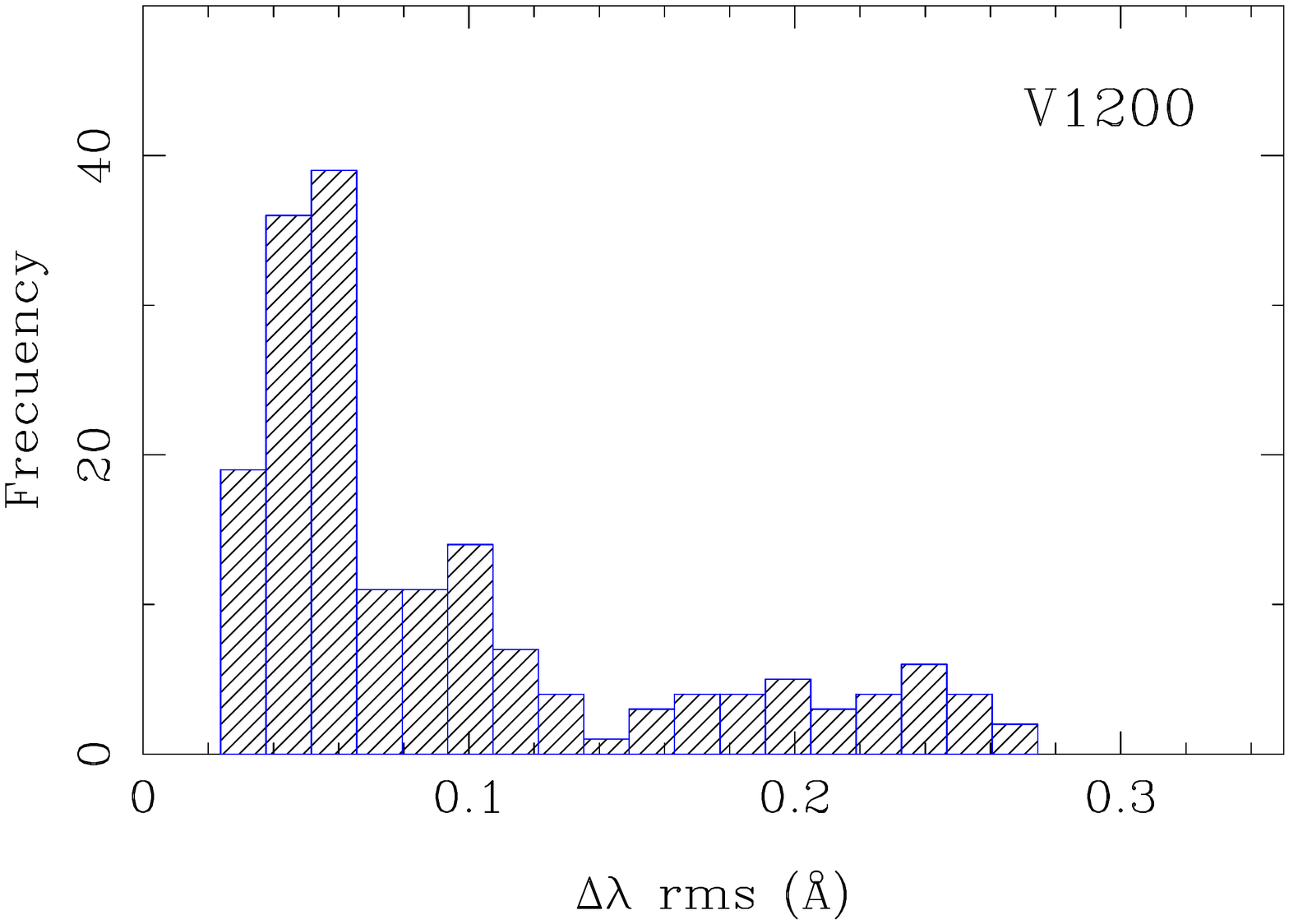}
 \caption{Distribution of the $rms$ of the difference between the nominal and
   measured wavelengths of the night-sky emission lines, derived from the statistical analysis
   of all the considered emission lines within the same science frame
   for the V500 (left panel) and V1200 (right panel)
   grating datasets. The mean shift is around zero.
   \label{fig:wave} }
 \end{center}
 \end{figure*}

 Figure \ref{fig:rms} shows the dispersion between the different response
 curves found for different calibration stars and/or nights as a function of
 the wavelength for the V500 and V1200 gratings,  normalized by the offset
 in the zeropoints at 4600~\AA\ and 4150~\AA, respectively. { The differences between one response curve and another as a function of the wavelength reflect the intrinsic dispersion of the flux calibration. The flux calibration obtained by applying this response curve is just a relative one. An absolute calibration is obtained by rescaling using a factor derived after the comparison with broad-band imaging, as described in forthcoming sections.}
On average, the
 relative flux calibration shows a dispersion of $\sim$2-3\% { ($\sim$0.025 mag)}, for wavelengths
 redder than of 3850~\AA. At shorter wavelengths, the error in the accuracy of the
 relative flux calibration becomes worse, being about $\sim$8\% { ($\sim$0.08 mag)} at
 $\sim$3700-3750~\AA. These values are similar to the ones found in classical
 slit spectroscopy. 

 \subsection{Spatial re-arranging and image reconstruction}

 Once the frames corresponding to each particular pointing are reduced,
 the science spectra corresponding to the three dithered exposures are
 combined in a single frame of 993 spectra. To do so the images are rescaled
 to a common intensity and response function by comparing the
 integrated spectra within an aperture of 30$\arcsec$/diameter. This
 procedure takes into account possible variations in the atmospheric
 transmission during the science exposures. In addition, the procedure
 produces a final position table, where the individual position tables
 (corresponding to the PPAK central bundle) are combined using
 the offsets provided for the dither. Finally, the data are
 spatially resampled to a datacube with a regular grid. This is done
 by adopting a flux-conserving variation of Shepard's interpolation method
 \citep{shepard1968}. This method is a very simple and robust
 tessellation-free interpolation of scattered data, where the intensity
 at each interpolated point is the sum of a weighted average of the
 intensities corresponding to $n$ adjacent scattered points (the
 original spectra) within a certain boundary distance ($r_{\rm lim}$):

 $$ F(i,j) = \sum_{k=1}^{k=n} w_{i,j}^k f_k \;\;\; r_{1...n} < r_{\rm lim} $$

\noindent where $F(i,j)$ is the reconstructed intensity in the pixel $(i,j)$ of
 the final datacube at a certain wavelength, $w_{i,j}^k$ is the weight
 at this pixel of the adjacent spectrum $k$, and $f_k$ is the intensity
 of this adjacent spectrum, at the considered wavelength.

 The weights are derived by a Gaussian function:

 $$ w = N \, {\rm exp}[-0.5 (r/\sigma)^2] $$  

\noindent where $N$ is a normalization parameter, $r$ is the distance between
 the pixel $(i,j)$ and the spectrum $k$, and $\sigma$ is the parameter
 that defines the width of the Gaussian function. The normalization
 parameter is derived for each interpolated pixel, being defined as the inverse
 of the sum of the different weights of the spectra contributing to
 this pixel:

 $$ N(i,j) = \frac{1}{\sum_{k=1}^{k=n} w_{i,j}^k}  \;\;\; r_{1...n} < r_{\rm lim} $$

 This normalization guarantees that the integrated flux is
 preserved. The adopted image reconstruction assumes a boundary limit
 of $r_{\rm lim}=$5$\arcsec$, $\sigma=$1$\arcsec$, and a final pixel
 scale of 1$\arcsec$/pixel for the resulting datacube. Prior to this
 interpolation, the flux corresponding to those spectral pixels masked
 due to the vignetting are replaced by the average of the fluxes at the
 two nearest fibers not affected by this effect. { Finally, the flux intensity is corrected by the difference in aperture between the original fibers ($\sim$2.7$\arcsec$/diameter) and the final pixels.}

 \subsection{Differential Atmospheric Refraction}
 \label{DAR}

 The differential atmospheric refraction \citep[DAR,][]{fili82} is
 corrected once the data are spatially resampled to a datacube with a
 regular grid. In the case of IFS data, the DAR can be corrected
 empirically, after the observations, without requiring to know the
 original orientation of the instrument and without the need of a
 compensator \citep{emse96,medi,roth04}. To do so, the reconstructed datacubes
 are thought of as
 a set of narrow-band images with a band-width equal to the spectral
 resolution.  These images can be recentered using the theoretical
 offset determined by the DAR formulae \citep{fili82}, or, as in 2D
 imaging, tracing the intensity peak of a reference object in the
 field-of-view (or a DAR reference observation) along the spectral
 range, and recentering it.  Note that this latter approach is
 basically unfeasible in slit spectroscopy, this being one of the
 fundamental differences between the two methods.


 We adopted this latter empirical correction for the CALIFA pipeline,
 using the corresponding tools included in {\tt R3D}. For doing so,
 { the centroid of the observed object (i.e., the galaxy) is   derived at each wavelength,} 
by determining its barycenter in an
 image slice extracted from the datacube, within a range of
 20\AA\ around the considered wavelength.  Then, the shifts along the
 wavelength are determined by comparing the corresponding coordinates
 to a common reference. A polynomial function of order 3 is fitted to
 each shift (X and Y) along the wavelength to increase the accuracy of
 the determined offset \cite[e.g.][]{wiso03}.  Then, the full datacube
 is shifted by resampling and shifting each image slice at each
 wavelength. {\bf In general, the empirical correction produces
   similar results to the theoretical one. However, the theoretical correction is not well defined when different pointings of the same object are observed during different nights, due to weather restrictions ($\sim$15\% of the observed targets). These observations may span over different dark-time periods. The use of the empirical correction guarantees the homogeneity of the treatment of DAR correction for all the datasets.}

 \begin{figure*}
 \begin{center}
 \includegraphics[height=6.5cm,angle=0,clip=true,clip=true,trim=60 0 70 80]{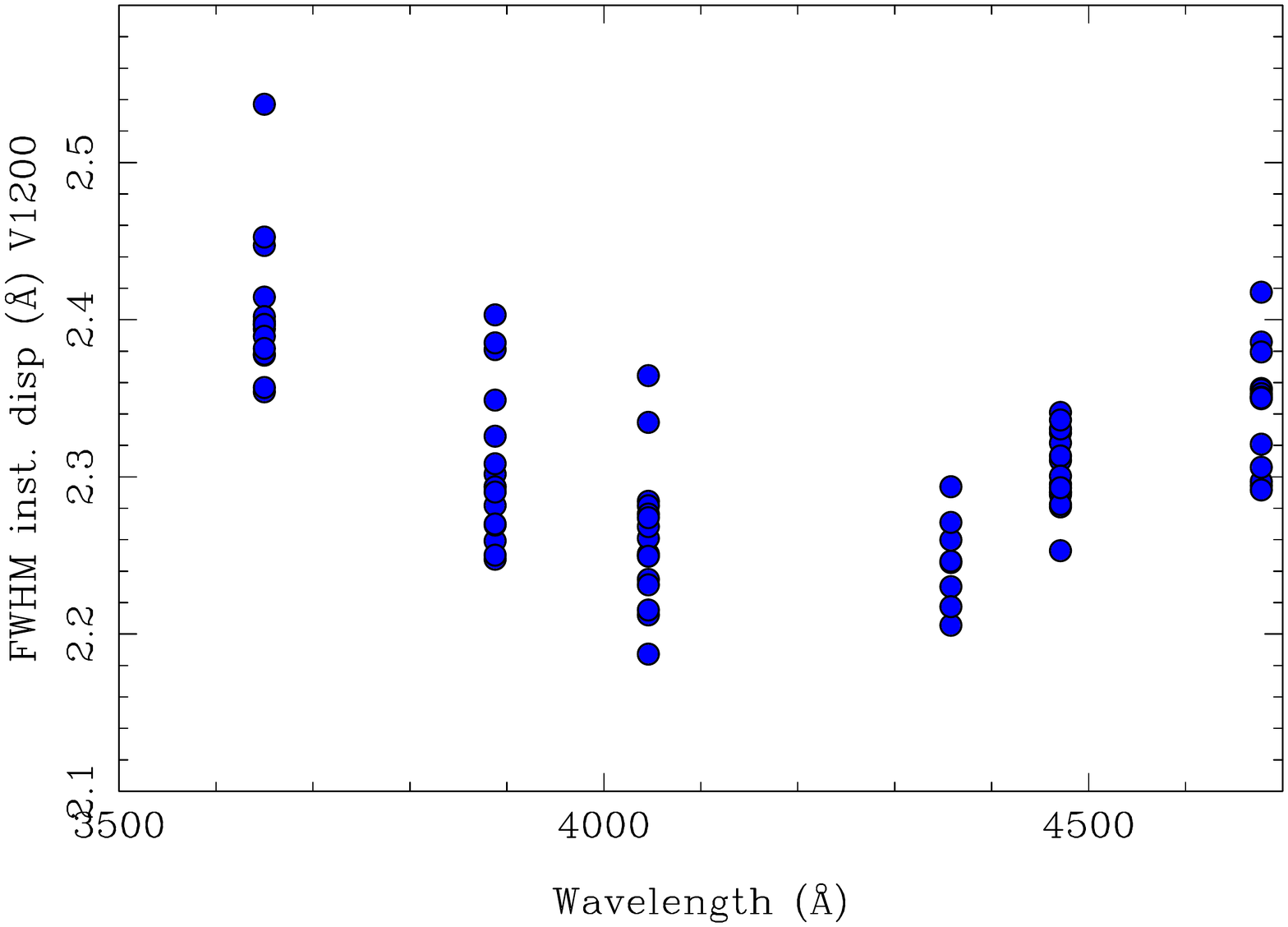}
 \includegraphics[height=6.5cm,angle=0,clip=true,clip=true,trim=100 0 70 80]{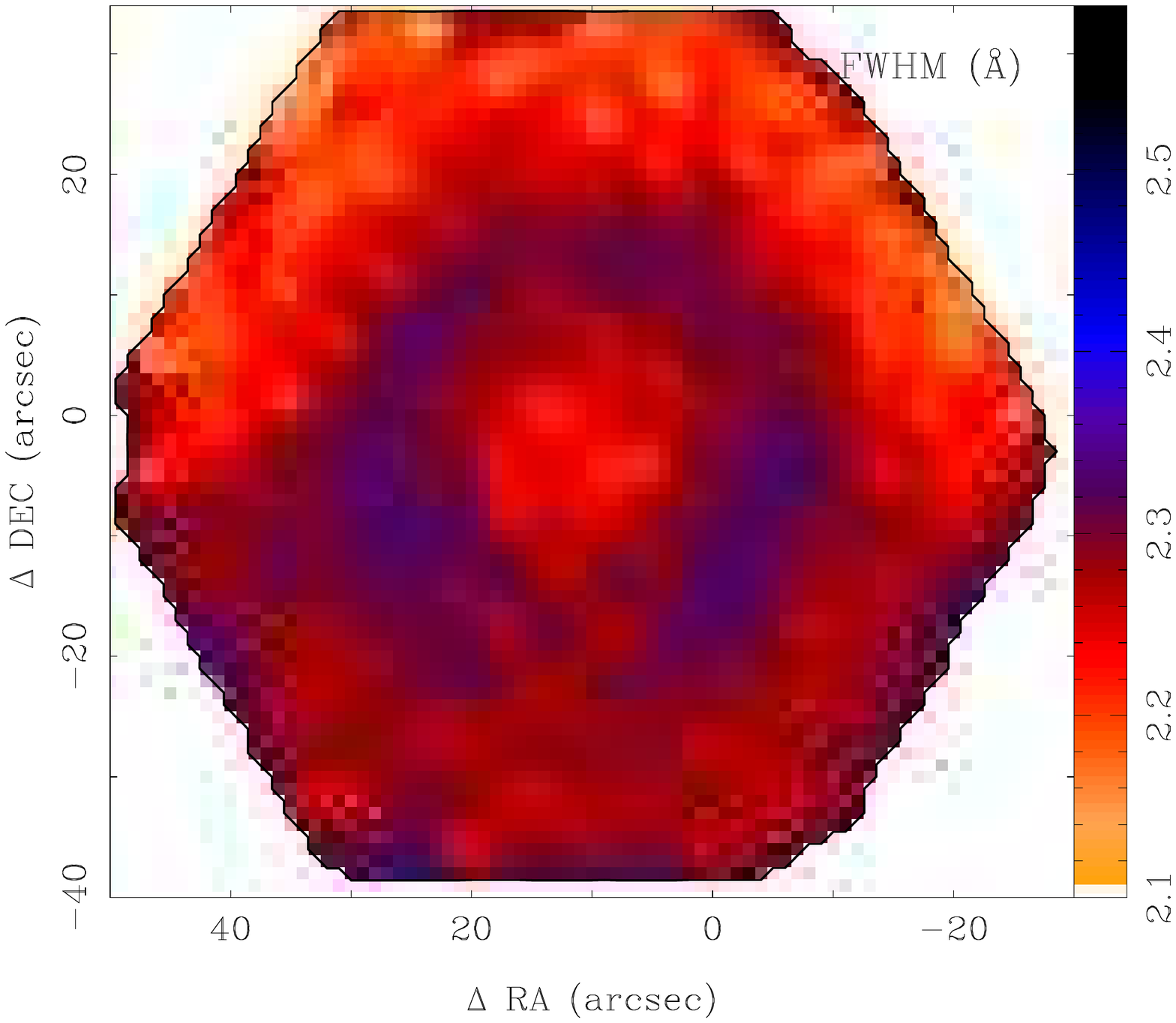}
 \caption{{\it Left panel}: The distribution of the FWHM of the instrumental dispersion derived from
   the analysis of the emission lines within the arc-lamp frames at different
   wavelengths and different positions within the FoV, for the V1200 data. {\it Right panel}: The spatial distribution
   of the average FWHM of the instrumental dispersion, derived from the same
   arc lines.
 \label{fig:res}}
 \end{center}
 \end{figure*}

 \subsection{Absolute flux re-calibration}
 \label{absflux}

 In order to get the best possible absolute flux calibration, we recalibrate
 our data based on SDSS photometry, available for all targets by
 construction of the sample. Of the five SDSS filters ($ugriz$), two
 namely, the $g$ ($\lambda_{\rm eff}=$4770~\AA) and $r$ ($\lambda_{\rm eff=}$6231~\AA) filters, are covered almost
 completely by the V500 grating data, while $u$ ($\lambda_{\rm eff}=$3594~\AA) is only
 partially covered by both the V500 and V1200 grating data. Therefore, we use
 the V500 grating data and the $g$ and $r$ band SDSS images to perform a
 primary flux recalibration. 

 To perform the primary recalibration, we measure the counts of each
 galaxy in the SDSS image inside a 30$\arcsec$ diameter aperture. These
 counts are converted to flux density following the counts-to-magnitude
 prescription in the SDSS
 documentation\footnote{http://www.sdss.org/dr6/algorithms/fluxcal.html}.
 The accuracy of this photometry was cross-checked by obtaining the
 magnitudes of any detected star in each field, using the same
 procedure, and comparing them with the values listed in the SDSS DR6
 photometric catalogue. We obtained less than 0.05 magnitude dispersion
 for the stars with $g <$ 17.5 mag. Once we ensure the accuracy of this
 photometry, we extract the spectrophotometry from the reduced
 datacubes corresponding to the V500 grating setup, coadding the flux
 of individual spectra inside a 30$\arcsec$ diameter aperture and
 convolving this spectrum with the SDSS $g$ and $r$ filter passbands
 \citep[ADPS database,][]{moro00}.  Using these two data pairs, a
 scaling solution is found, by adopting the average of the flux ratio
 in both bands.

 Once the V500 datacube has been recalibrated, a new scaling solution
 for the V1200 data is derived by comparing the 5$\arcsec$ aperture
 extracted spectra around the central position of each object from
 these cubes with those extracted from the original reduced V1200
 data. In this case, a low-order polynomical function is fitted to the
 ratio between both spectra within the common wavelength range
 ($\sim$3745-4770~\AA). The result of this polynomial fitting is then
 applied to the V1200 data, to match them to the V500 ones.  Figure
 \ref{fig:comp_r} shows the central aperture extracted spectra for
 both setups together with the difference between them
 (5$\arcsec$/diameter), for three typical targets. It illustrates the
 agreement between both datasets, once the recalibration procedure has
 been applied. The V1200 spectra have been convolved with a Gaussian
 function to compensate for the difference in resolution, for the
 purpose of this figure only. The typical difference between the
 spectra is $\sim$10\% of the average. The strongest differences are
 located at the wavelengths of the more significant spectral features
 (absorption and/or emission lines).\newline


\noindent We conclude that the current version of the pipeline, which operates
 automatically, is able to reduce the data, producing the required
 data cubes to perform early experiments and estimations of their quality.
 The pipeline will be continuously upgraded, producing different
 versions of the data, on the basis of the results of the different
 foreseen quality controls, prior to releasing the data to the community.
 { In the next section we describe the early tests implemented to estimate the quality of the data. In some cases these tests have already induced slight modifications in the pipeline, as we will describe below in detail.}

 \section{Quality of the data}
 \label{quality}


 In parallel to the reduction of the data, the pipeline performs a set of
 automatic tests, which are stored during the reduction process, creating a set
 of tables and figures. These allow the CALIFA team to check the 
 quality of the data and to identify possible problems in the data reduction. 
 In this section we present the basic results of the currently implemented 
 quality controls on the data acquired so far. We expect the quality control 
 to become more comprehensive with time.

 \subsection{Accuracy of the wavelength calibration}
 \label{s:wavecalib}


 In general, the wavelength solution is found with an accuracy of the order of
 10-15\% of the nominal pixel scale, i.e., the $rms$ of the fit with the low
 order polynomial function is $\sim$0.2-0.3~\AA\  for the V500 grating and
 $\sim$0.1-0.2~\AA\  for the V1200 grating. 

 In order to obtain an independent estimate of the accuracy of the
 wavelength calibration of the science frames, we compare the nominal
 and recovered wavelengths of the most prominent night-sky emission
 lines in each individual spectrum prior to the sky subtraction. The central wavelength of these
 lines was determined by fitting a Gaussian function to each line. This
 provides us with 331 estimations of the relative offsets between both
 values for each considered night-sky line.  For the V500 grating we use the
 HgI$\lambda$5461, [OI]$\lambda$5577, NaD$\lambda$5893 and
 [OI]$\lambda$6300 lines, while for the V1200 grating we use the
 HgI$\lambda$4046 and HgI$\lambda$4358 ones. Only measurements derived
 using emission lines with a signal-to-noise higher than 10 are
 considered. Finally, a few hundred measurements of the differences
 between the nominal and recovered wavelengths are retained for each
 frame corresponding to both gratings. A simple statistical analysis
 allows us to estimate the final accuracy of the wavelength
 calibration, deriving both the median offset (if any) and the $rms$ of
 each dataset.

 No mean shift was found between the nominal and measured wavelength
 for the different analyzed sky lines. Figure \ref{fig:wave} shows the
 distribution of the $rms$ of the difference between the nominal and
 measured wavelengths of the night-sky emission lines, found for each
 science frame, for both setups (left and central panel). In both cases
 the values range between 0.01~\AA\ and 0.35~\AA. The resulting median
 value is very similar for both distributions, being around
 $\sim$0.1~\AA, which corresponds to $\sim$8 km s$^{-1}$ at the
 wavelength of [OII]$\lambda$3727 and $\sim$3 km s$^{-1}$, at the
 wavelength of H$\alpha$.

 \begin{figure*}
 \begin{center}
 \includegraphics[height=6.7cm,angle=0,clip=true,trim=30 20 117 60]{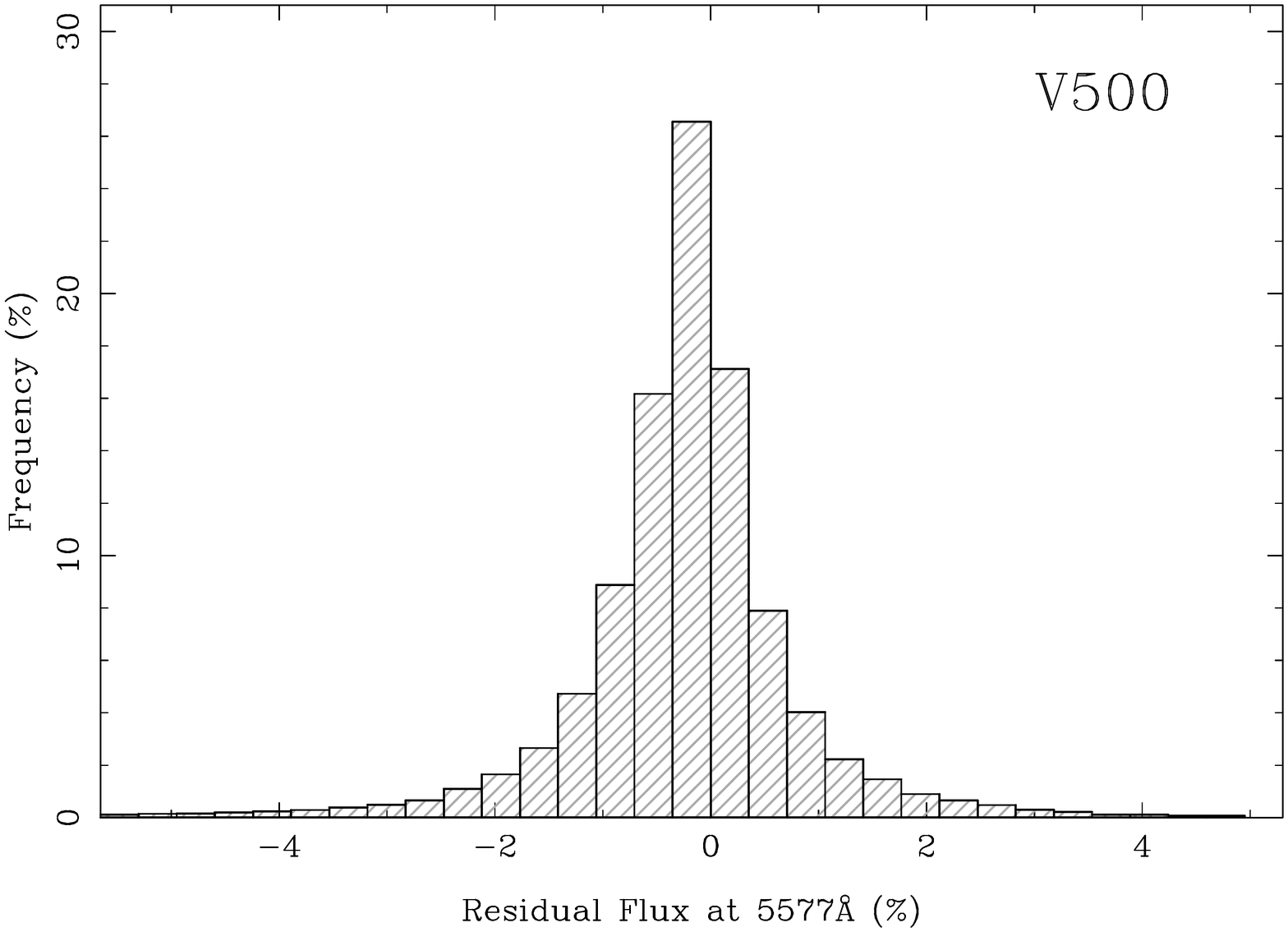} 
\includegraphics[height=6.7cm,angle=0,clip=true,trim=30 20 117 60]{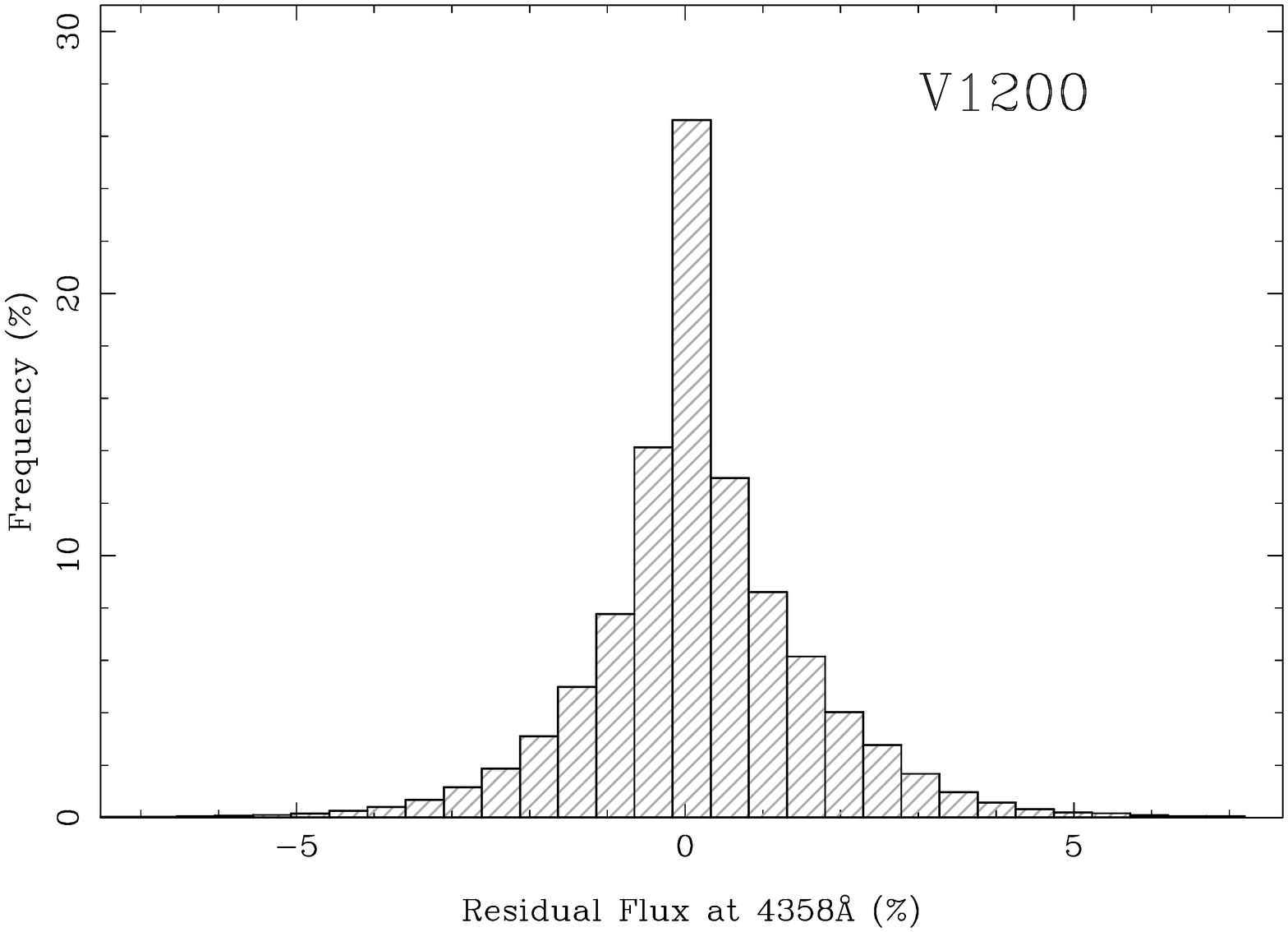}
 \caption{Distribution of the relative residual flux after subtracting the
   night-sky spectrum for the strongest emission lines on the corresponding
   wavelength range covered by each setup: [OI]$\lambda$5577 for the V500 grating (left
   panel) and  HgI$\lambda$4358 for the V1200 grating (right panel).
 \label{fig:sub}}
 \end{center}
 \end{figure*}

 \begin{figure}
 \begin{center}
 \includegraphics[width=9.5cm,angle=0,clip=true,trim=30 30 80 100]{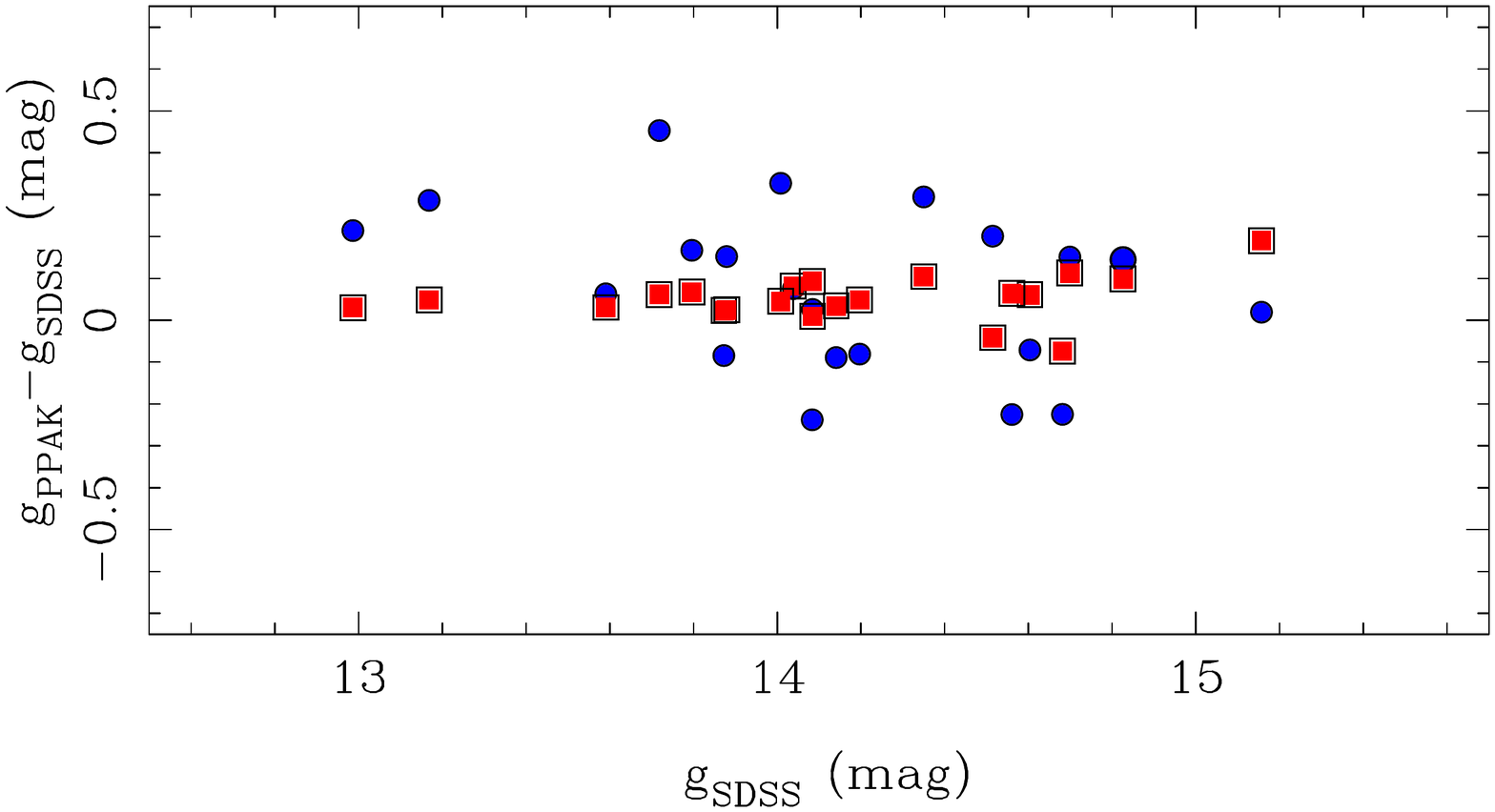}
 \caption{Comparison of the $g$-band magnitude derived from the
   integrated spectrum within an aperture of 30$\arcsec$ diameter extracted from
   the V500 datacubes and the corresponding magnitudes derived from the SDSS images,
   before (blue solid circles) and after (red solid squares) applying
   the flux recalibration procedure described in the text.
 \label{fig:phot}}
 \end{center}
 \end{figure}

 \subsection{Empirical spectral resolution}
 \label{emp_res}

 The spectral resolution of a dataset is limited at best to the nominal 
 instrumental resolution. However, different effects, such as problems with,
 or degradation in, the internal focus of the instrument, errors in the
 tracing/extraction process or in the wavelength calibration may 
 affect the final spectral resolution.

 In order to estimate the real spectral resolution for each science frame, we
 fit the strongest night-sky emission lines (see Section \ref{s:wavecalib}) 
 in each individual science spectrum and before subtracting the night-sky
 spectrum with a Gaussian function using FIT3D \citep{sanchez06b}. Night-sky 
 lines are unresolved at our resolution, thus their widths provide good
 estimates of the final resolution at their respective wavelengths. The median
 and standard deviation of the FWHMs are stored for each science frame.

 The spectral resolution found for the V1200 grating is very similar
 to the nominal value ($\sim$2.7~\AA), and is consistent with the
 values reported by Roth et al. (2005). For the V500 grating on the
 other hand, there is no published information about its nominal
 spectral resolution, which was estimated to be $\sim$6~\AA\ (Roth,
 private communications). The empirical resolution estimated from {
   night-sky lines} is $\sim$6.5~\AA. These instrumental spectral
 resolutions correspond to velocity resolutions of $\sigma\sim$85
 km~s$^{-1}$ and $\sigma\sim$150 km~s$^{-1}$ for the V1200 and V500
 datasets, fulfilling the requirements of the survey, for both setups.

 The same experiment can be repeated using the arc-lamp frames
 obtained during the night instead of the science frames. This
 procedure has the advantage that all selected lines have high
 signal-to-noise, and are evenly distributed over the full spectral
 range. However, the derived instrumental dispersion cannot be
 directly compared with the science data, since the arc-lamp exposures
 are an order of magnitude shorter in time and, therefore, they are
 not affected by the degradation in resolution due to the drift of the
 spectra on the CCD in long exposures (due to the change of the
 flexure pattern). For this experiment, a datacube for each reduced
 arc lamp frame was created following the prescriptions for science
 frames.  Then, for each spectrum of this datacube, we fit each
 emission line with a single Gaussian function, deriving the central
 wavelength, intensity and width (in FWHMs). These latter values are
 considered as an estimate of the instrumental dispersion. As expected
 the derived values are smaller (and more precise) than thos
 reported above, being 2.30$\pm$0.11~\AA\ for the V1200 (ranging
 between 2.2-2.4~\AA) and 5.65$\pm$0.21~\AA\ for the V500 (ranging
 between 5.3-6.1~\AA). { However, the final resolution achieved in the
   data is more similar to that estimated from the night-sky lines,
   than that estimated from the ARC-lamps, which are observed in
   shorter integration times.}

 Although not clearly appreciated in the measurements performed using
 the night-sky lines, there is a clear spatial and wavelength
 dependence of the instrumental resolution. Figure \ref{fig:res} shows
 both distributions, for the V1200 data (a similar result is found for
 the V500 data). The left panel shows the distribution with
 wavelength, which shows the larger difference, i.e. $\sim$13\%. At
 each wavelength we plot values found at different locations within
 the FoV, shown in the right-panel. The amplitude of the spatial
 pattern is $\sim$5\%. The pattern seen in the spatial distribution,
 which has a maximum in an annular ring at $\sim$15$\arcsec$ from the
 center, and an underlying gradient, is a consequence of the
 correspondence between the spatial distribution of the PPAK fibers
 within the FoV and their positions along the entrance pseudo-slit of
 the spectrograph. The fibers located in the described ring correspond
 to those at the edges of the pseudo-slit. This implies they also lie
 at the edges of the CCD, where the resolution is worse. The
 underlying gradient is possibly an effect of a slight tilt of the
 focal plane of the CCD with respect to the focal plane of the
 spectrograph camera.

 \begin{figure}
 \begin{center}
 \includegraphics[width=8.5cm,angle=0,clip=true,clip=true,trim=30 20 100 100]{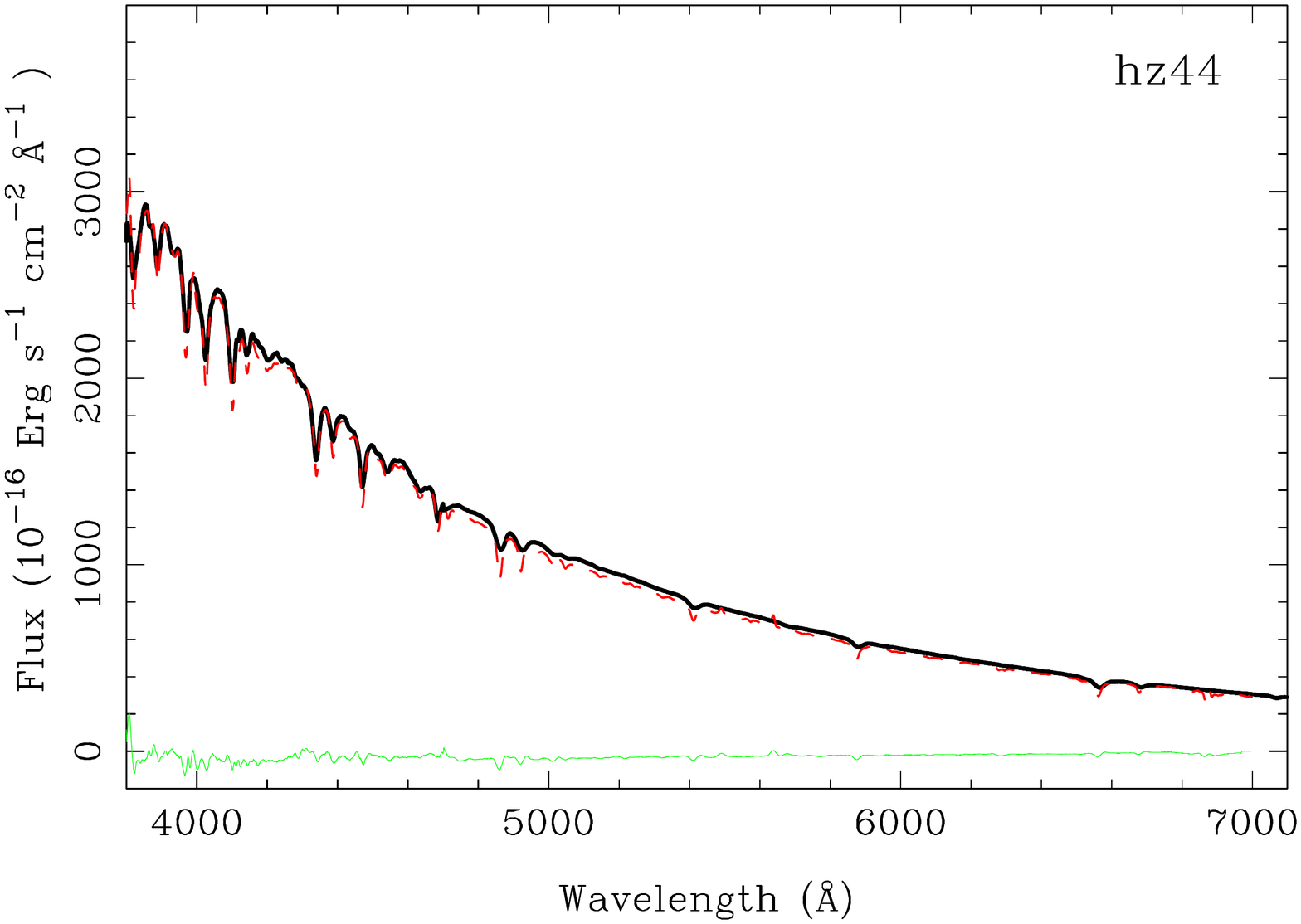}
 \includegraphics[width=8.5cm,angle=0,clip=true,clip=true,trim=30 20 100 100]{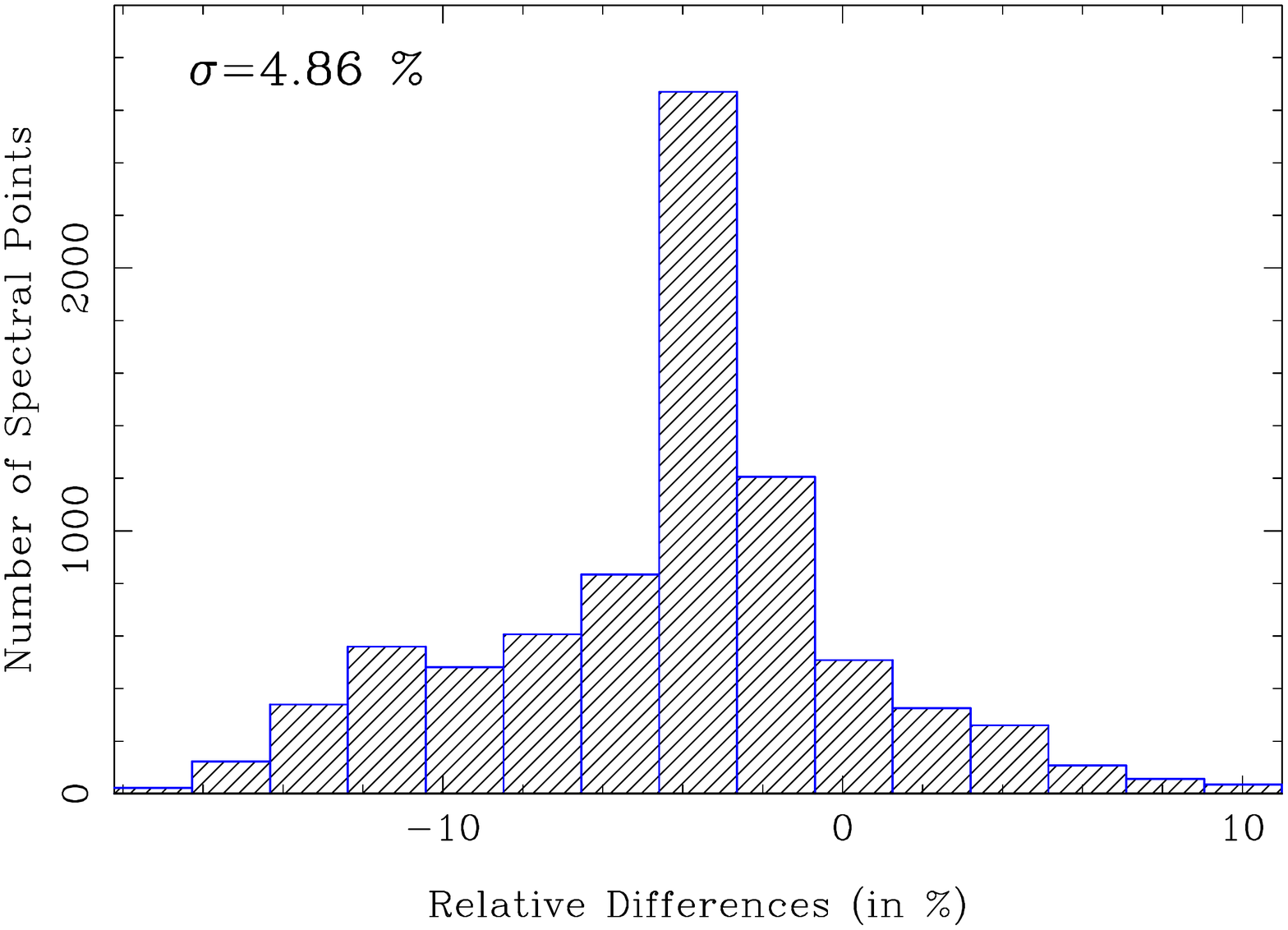}
 \caption{{ {\it Top panel: } Comparison of the published flux-calibrated spectrum of a calibration star (black, thick solid line), with the flux re-calibrated spectrum extracted from the V500 data (red, dashed line). The green line shows the residuals, once corrected by the different resolution of each spectra. {\it Bottom panel:} Distribution of the residuals in intensity, pixel-to-pixel, for all the calibration stars observed along the run. The standard deviation is lower than $\sim$5\% for the considered wavelength range (3800-7000\AA).  \label{fig:comp_stars} }}
 \end{center}
 \end{figure}

 \begin{figure}
 \begin{center}
 \includegraphics[height=8cm,angle=0,clip=true,clip=true,trim=110 20 50 70]{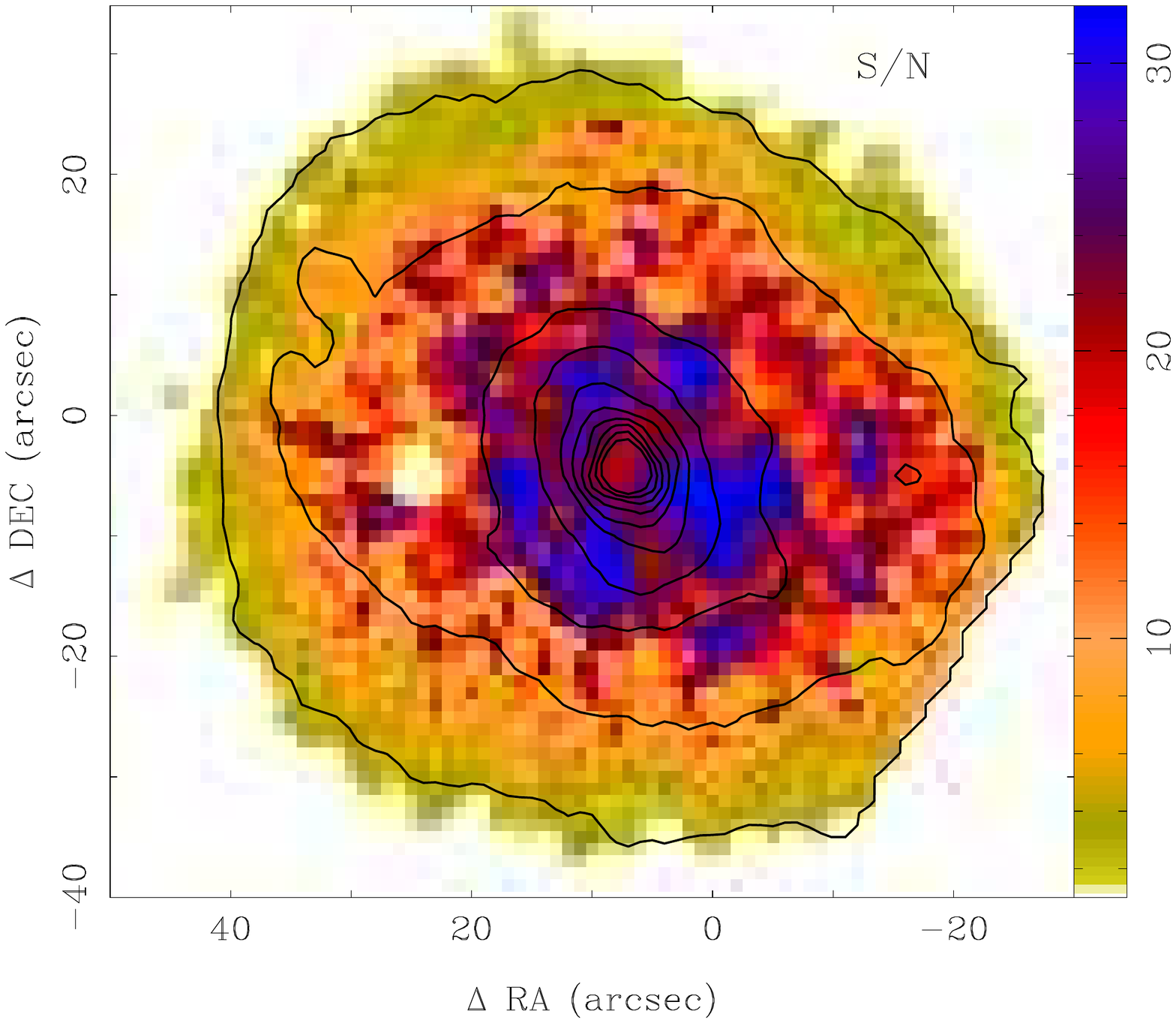}
 \caption{Color image of the signal-to-noise map derived for the V500
   data corresponding to the continuum emission of the NGC 5947 galaxy at the wavelength range between 4480 and 4520~\AA, based on the rough estimation
   described in the text. The contours shows the intensity of the $V$-band
   image synthesized from the datacube, with the 1st contour at 1.5\,
   {\FunitAREA} ($\sim$23.5 mag/arcsec$^2$) and consecutive ones with a step of 3\, {\FunitAREA}. 
 \label{fig:sn_map} }
 \end{center}
 \end{figure}


 { The correction of these instrumental signatures were not originally
   considered in the version of the pipeline described in the Section
   \ref{sec:redu}. These data allowed us to understand the changes in
   resolution along the wavelength and across the field. We found no
   clear dependencies of the instrumental resolution with the
   positioning of the telescope, at least for the range of airmasses
   at which we observe the targets ($<1.4$). There is a clear
   dependency with the temperature of the instrument, which affects
   its internal focus. This is corrected by the focus procedure
   adopted every night, and a careful  monitoring of this temperature
   (changes larger than $\sim$0.5$\degree$ C are rare, along a single
   night). However, both effects produce mostly a change in the global
   value, rather than a change in the described pattern, which seems
   to be rather stable. 

  Based on these results, we implemented a procedure in the pipeline
  to correct for the differences in the spectral resolution across the
  FoV and along the spectral range by fixing it (in terms of FWHM) to
  the worst value found for each dataset (based on the measurements of
  the FWHM in the ARC-lamp frames). For doing so, the described
  derivation was repeated for each arc-lamp exposure, associated with
  each science frame. Then a low order polynomial function is fitted
  to the derived values of the resolution, per spaxel and per
  pixel. Finally, the differential resolution is derived by obtaining
  the quadratic difference between the estimated value at each pixel
  and the globally worst value. The science spectra are then convolved
  pixel-to-pixel by this differential resolution, normalizing it along
  the spectral range and across the FoV. The final resolution achieved
  is slightly worse than the one described above, due to
  flexure-related shifts in wavelength: $\sim$2.7\AA\ for the V1200
  and $\sim$6.5\AA\ for the V500, as derived from the measurements of
  the night-sky emission lines width. However, with the adopted
  procedure the resolution is normalized along the spectral range and
  across the field. 

  Any further analysis described in this article is based on these spectra, with a homogeneous spectral resolution (in terms of FWHM).
 }

 \begin{figure*}
 \begin{center}
 \includegraphics[width=17cm,clip=true,trim=0 380 0 140]{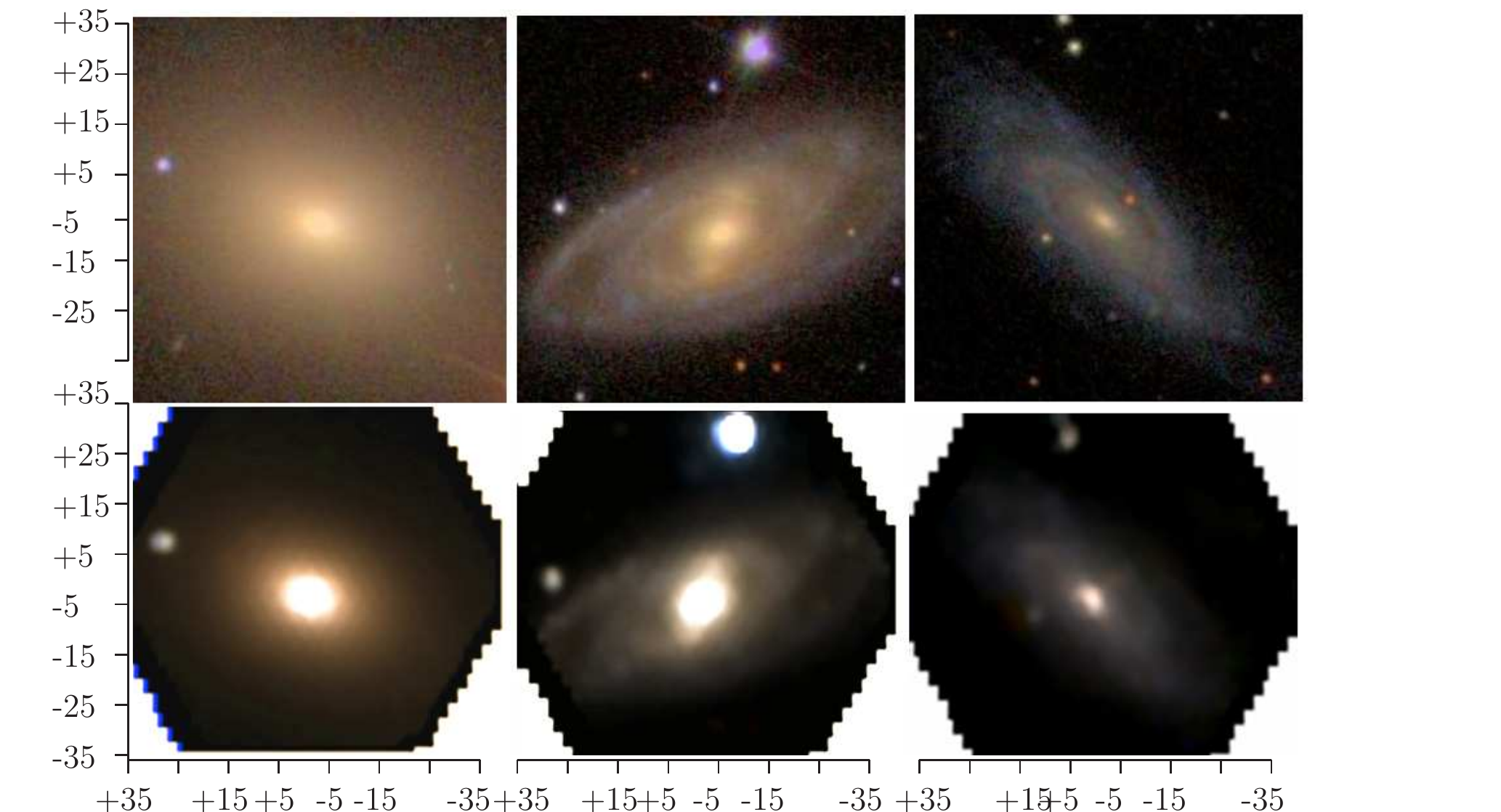}
 \includegraphics[width=16cm,trim=0 670 0 25,clip=true]{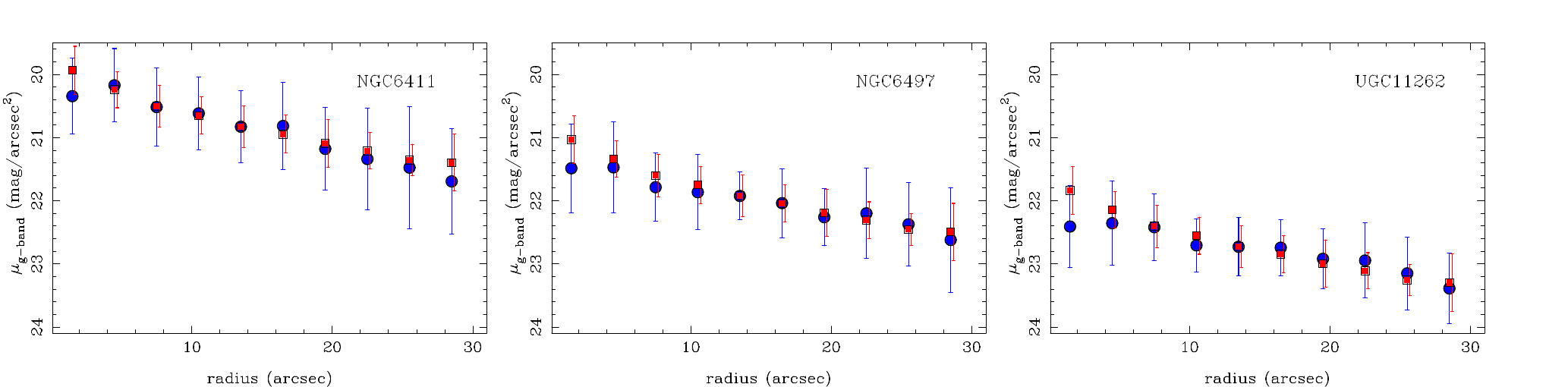}
 \includegraphics[width=16cm,trim=0 660 0 25,clip=true]{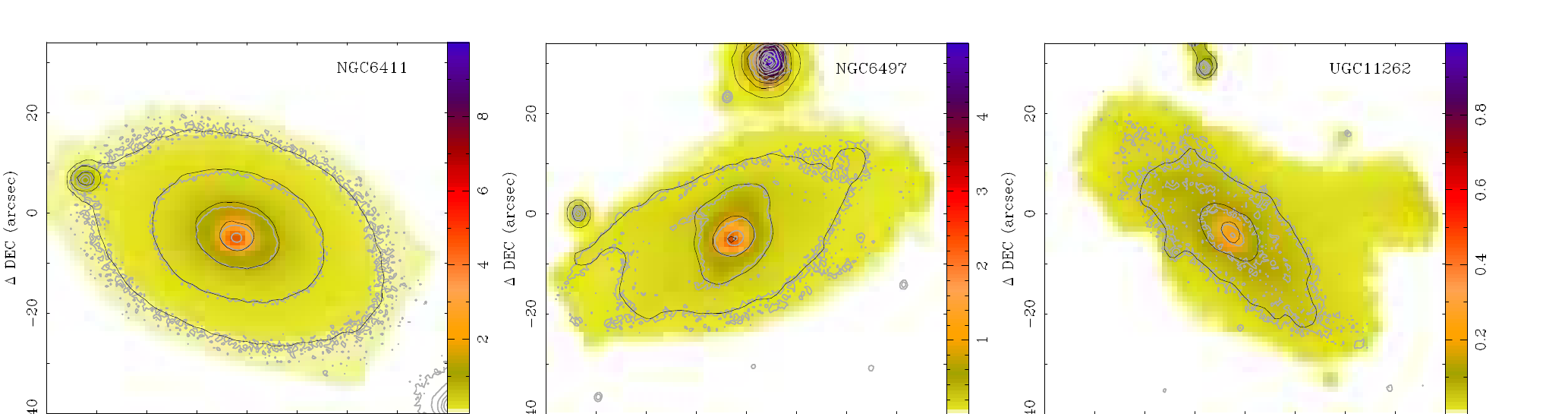}
 \caption{Three-color images from the SDSS imaging survey (first panel),
   vs. those obtained from the V500 CALIFA data (second panel), for
   the same objects shown in Figure \ref{fig:comp_r}: NGC 6411 (left),
   NGC 6497 (middle) and UGC 11262 (right). For the CALIFA data,  $u$,
   $g$ and $r$-band images were synthesized from the corresponding
   datacubes. { The third panel shows a comparison between the
     azimuthal averaged surface brightness profiles at the $g$-band,
     derived from synthesized images created from the V500 datacubes
     (blue circles) and the SDSS ones (red squares).} {\bf The bottom
     panel shows the synthesized $g$-band image, color-scaled,
     together with two logarithm-scaled counter plots: one
     corresponding to the SDSS $g$-band image, in grey color, and the other corresponding to the synthesized image, in black }  \label{fig:img_sdss}}
 \end{center}
 \end{figure*}

 \begin{figure*}[tb]
 \begin{center}
 \includegraphics[height=4.5cm,angle=0,clip=true,trim=30 10 120 40]{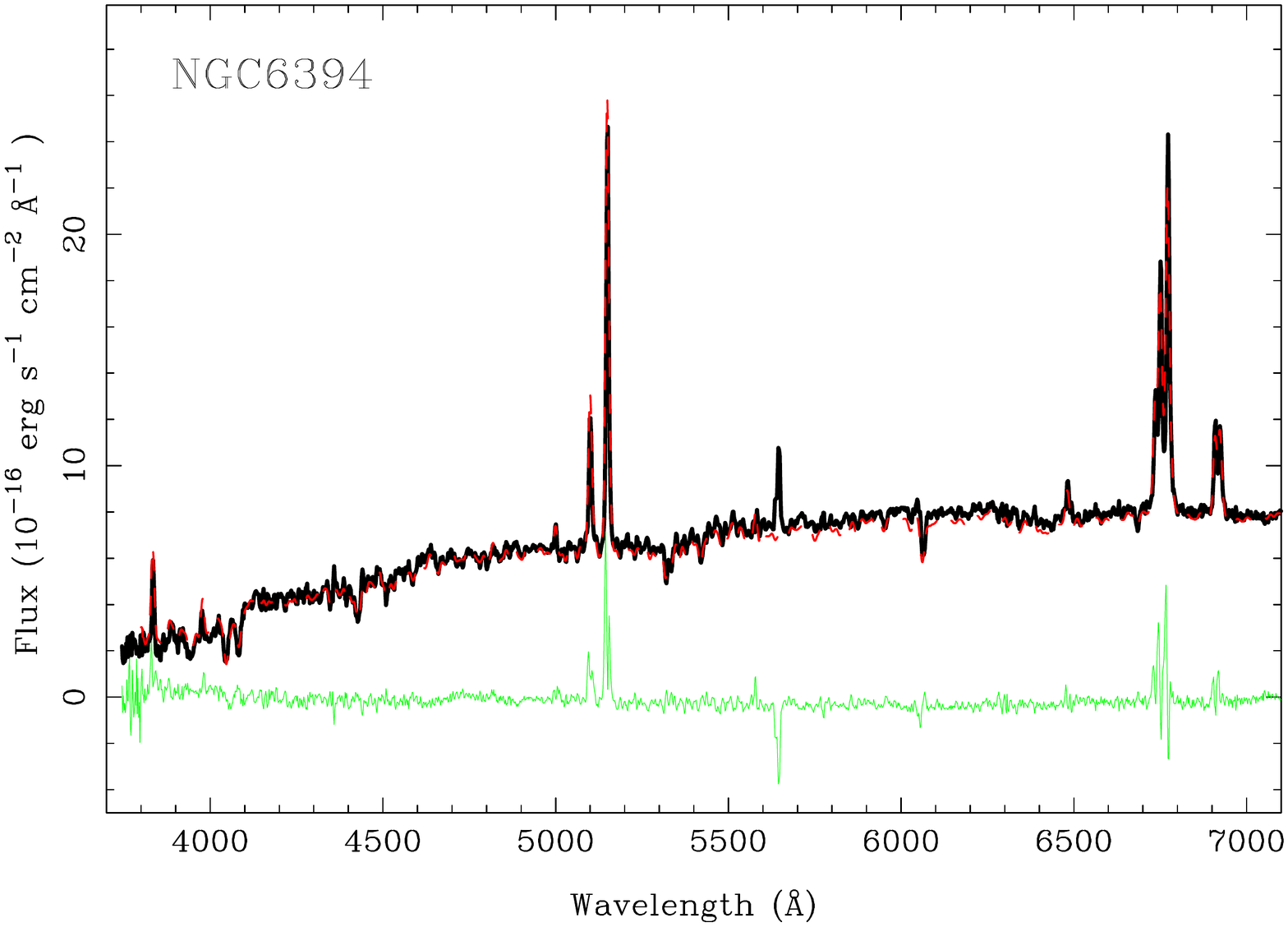}
 \includegraphics[height=4.5cm,angle=0,clip=true,trim=30 10 120 40]{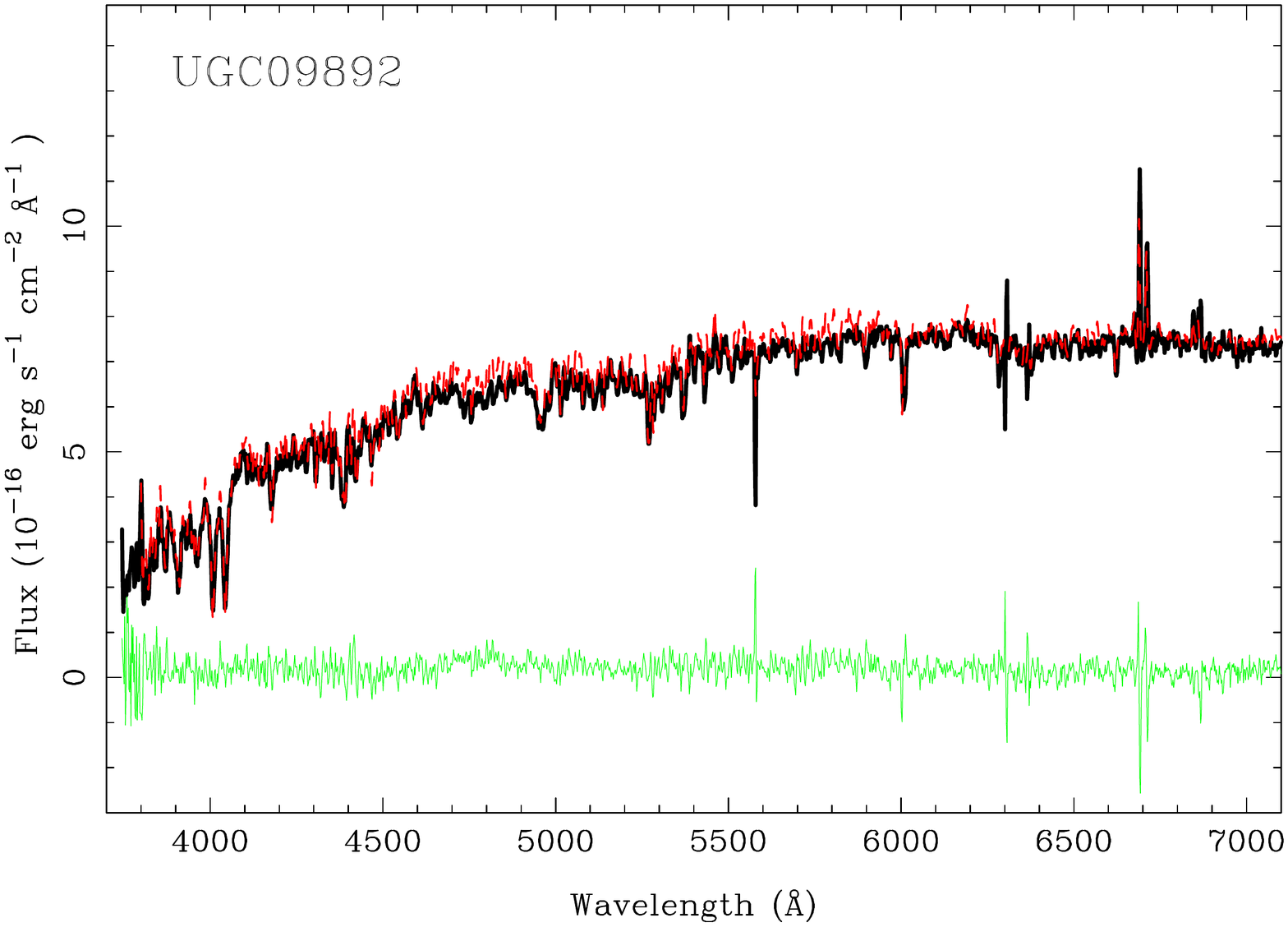}
 \includegraphics[height=4.5cm,angle=0,clip=true,trim=30 10 120 40]{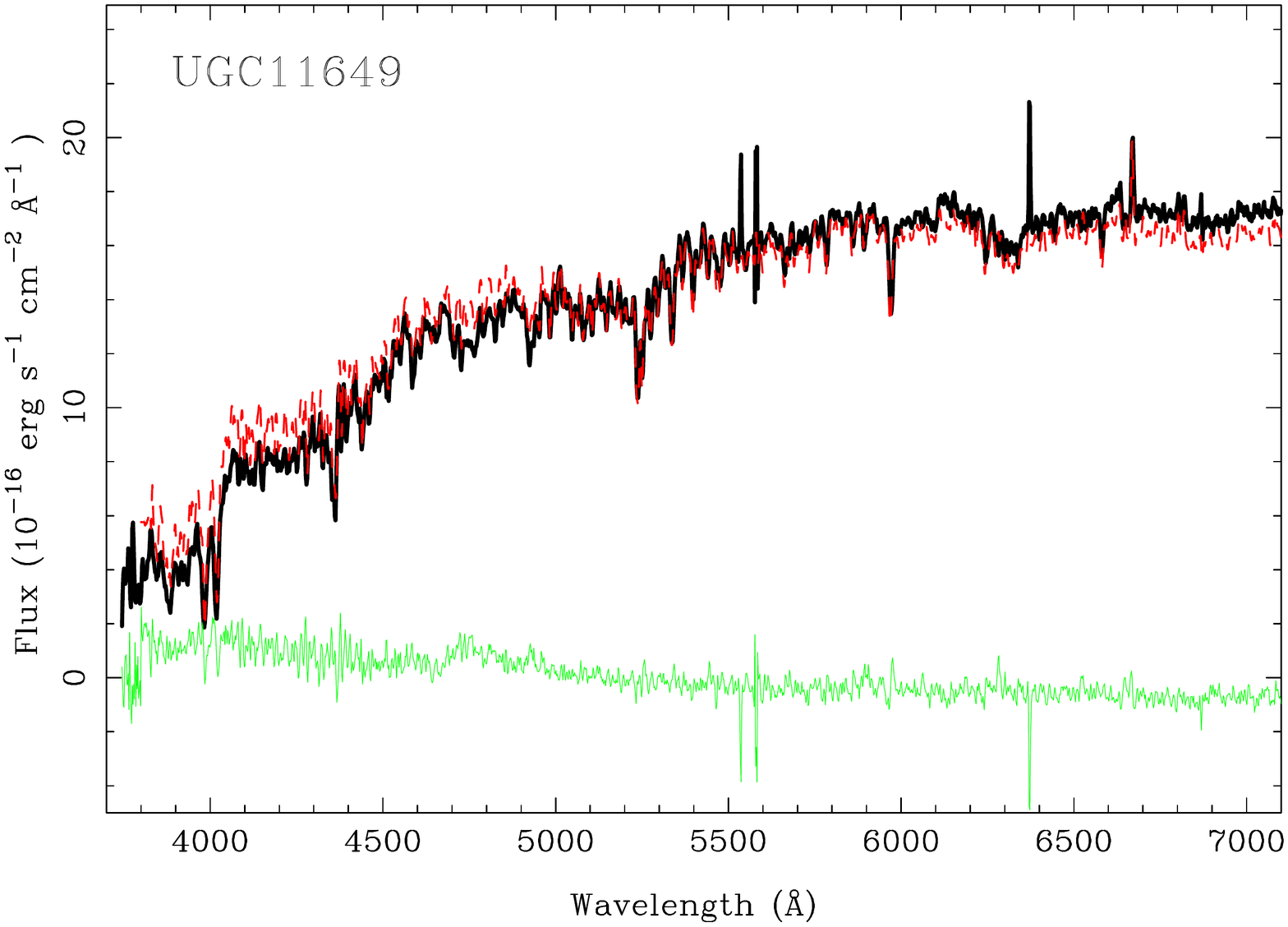}
 \caption{Comparison of the central 5$\arcsec$ spectrum extracted from the
   V500 (thick back solid-line) grating datacubes with the corresponding spectrum 
   obtained by the SDSS (thick red dashed-line),
   and the difference between both spectra (thin green solid-line). There are the only three galaxies in the 
   presented sample with SDSS spectra: NGC 6394, UGC 09892
   and UGC 11649 (from left to right)  \label{fig:comp_sdss_spec} }
 \end{center}
 \end{figure*}

 \subsection{Accuracy of the sky subtraction}

 The determination of the night-sky spectrum and its subtraction from
 the data is an issue in many IFUs, mostly due to the small FoV and
 the resulting lack of a pure night-sky spectrum. By design the CALIFA
 sample galaxies have an optical size that fits within the FoV of the
 central bundle of the PPAK IFU, while most of the 36 sky fibers are
 free of any contamination from galaxy light. The procedure adopted to
 subtract the sky is therefore rather simple, as explained above
 (Section \ref{subsecsky}).

 In order to estimate the accuracy of this procedure, we measured the flux
 corresponding to the strongest night-sky lines before and after subtracting
 the estimated night-sky spectrum. These lines are [OI]$\lambda$5577~\AA\ \, for
 the V500 grating and Hg$\lambda$4358~\AA\ \, for the V1200 grating. Their typical intensity
 ranges between 1-6 \FunitsSS, measured from adding the flux within a range of
 15 pixels around their nominal wavelengths, and subtracting the flux in two 
 adjacent continua of the same width, at shorter and longer wavelengths. After
 subtracting the night-sky spectra the typical residual value ranges between
 $\pm$0.5~~\Funits ~ for 99\% of the spectra, i.e., the sky is subtracted with an
 accuracy of 1-5\%, with a typical residual of less than 2\%. Figure
 \ref{fig:sub} shows the distribution of intensities obtained for each
 considered line once the night-sky spectra have been subtracted.

\begin{table*}
 \caption{Comparison of the PPAK central spectrum with the SDSS single-fiber spectrum.}         
 \label{tab:comp_sdss} 
 \centering                       
 \begin{tabular}{lccrrccrr} 
 \hline\hline                
 NAME &   \multicolumn{4}{c}{PPAK central spectrum} & \multicolumn{4}{c}{SDSS spectrum} \\
      & log$_{10}$ (Age) Gyr & $Z$ & A$_V$ & $\sigma_v$  km s$^{-1}$  &
 log$_{10}$ (Age) Gyr & $Z$ & A$_V$ & $\sigma_v$  km s$^{-1}$  \\
 \hline                        
 NGC 6394   &1.1  &0.016  &0.7  &139 &1.3  &0.016  &0.5  &141\\
 UGC 09892  &0.7  &0.019  &0.4  &116 &0.8  &0.021  &0.3  &144\\
 UGC 11649  &1.3  &0.024  &0.2  &228 &1.0  &0.024  &0.2  &215\\
 \hline 
     & H$\alpha$/H$\beta$ & [OIII]/H$\beta$ & [NII]/H$\alpha$ &  [SII]$\lambda$$\frac{6717}{6731}$ 
 & H$\alpha$/H$\beta$ & [OIII]/H$\beta$ & [NII]/H$\alpha$ &  [SII]$\lambda$$\frac{6717}{6731}$  \\
 \hline 
 NGC 6394  & 6.9$\pm$ 2.0   & 8.2$\pm$ 2.2   & 1.1$\pm$0.1 & 1.2$\pm$0.8
          & 6.6$\pm$ 1.7   &10.3$\pm$ 2.4   & 1.3$\pm$0.1 & 1.0$\pm$0.1\\ 
 UGC 09892 & 8.0$\pm$ 3.5   & 0.7$\pm$ 0.6   & 0.5$\pm$0.1 & 0.7$\pm$0.1
          &     $---$      &     $---$      & 0.5$\pm$0.1 & 1.5$\pm$0.7\\ 
 UGC 11649 &     $---$      &     $---$      & 1.1$\pm$1.1 & 1.3$\pm$0.2 
          & 3.8$\pm$ 4.1   & 2.2$\pm$ 2.8   & 1.1$\pm$0.5 & 1.5$\pm$0.6 \\
 \hline                        
 \end{tabular}
 \end{table*}

 The reduction pipeline stores the actual night-sky spectrum subtracted
 from each science frame, to allow inspection in more detail if
 needed. In particular, the night-sky surface brightness is measured by
 the pipeline and stored as a parameter to evaluate the quality of the
 data. To do so, each night-sky spectrum is convolved with the
 $B$-band (V1200) and $V$-band (V500) filter response curves obtained
 from the ADPS database (Moro \& Munari 2000), and then the
 corresponding surface brightness was estimated, applying the zero
 points listed in Fukugita et al. (1995). For the currently observed
 data the median values of the derived night-sky surface brightness are
 within $\sim$0.5 mag of the typical values for a dark night at Calar Alto
 (\citealp{sanchez07a}).

 \subsection{Accuracy of the flux calibration}

 The selected observational and reduction strategy uses the
 spectrophotometric calibration stars observed during the nights to
 perform an initial flux calibration. The main aim of this flux
 calibration is to correct for the relative transmission of the system
 from blue to red, and provide a preliminary absolute zero point. As
 indicated in Sec. \ref{sec:redu}, the accuracy of the relative
 transmission is better than a few percent in most of the covered
 wavelength range, which is sufficient for the science goals of this
 survey. On the other hand, the accuracy of the absolute
 spectrophotometric calibration was $\sim$24\%, on average, for the
 clear/photometric nights, and it can be worse for data taken under
 non-photometric conditions. Therefore a flux recalibration procedure
 is implemented in the pipeline, anchoring our photometry to that of
 the SDSS data (Sec. \ref{absflux}). It is expected that the final
 absolute spectrophotometric accuracy improves to
 $\sim$8\%\ photometric error, on the basis of our previous experience
 with the PINGS data \citep{rosales-ortega10}.

 We cross-check the improvement in the spectrophotometric calibration by
 extracting  from the datacubes the integrated spectra within an aperture of 
 30$\arcsec$ diameter around the peak intensity of the object, before and after
 the flux recalibration. The corresponding $g$ and $r$-band magnitudes were
 derived from these spectra, following the procedures described in Sec. \ref{sec:redu}. The magnitudes derived from the spectroscopy were then compared 
 with the corresponding ones obtained from the SDSS images, adopting a similar 
 aperture, and following the same procedure as used to recalibrate the data. 

 \begin{figure*}
 \begin{center}
{ \includegraphics[angle=0,width=5.5cm,clip=true,trim=30 10 120 40]{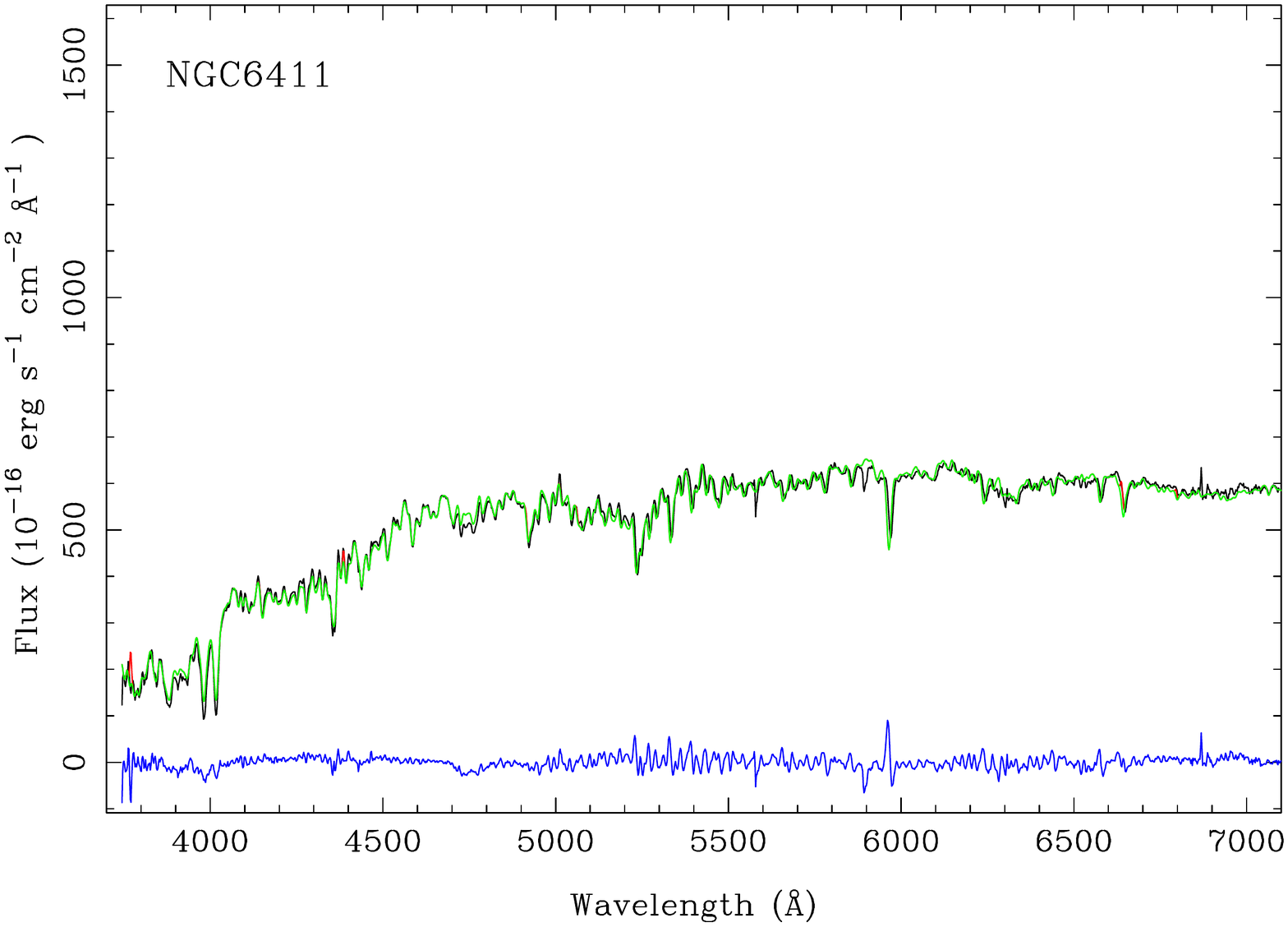}
 \includegraphics[angle=0,width=5.5cm,clip=true,trim=30 10 120 40]{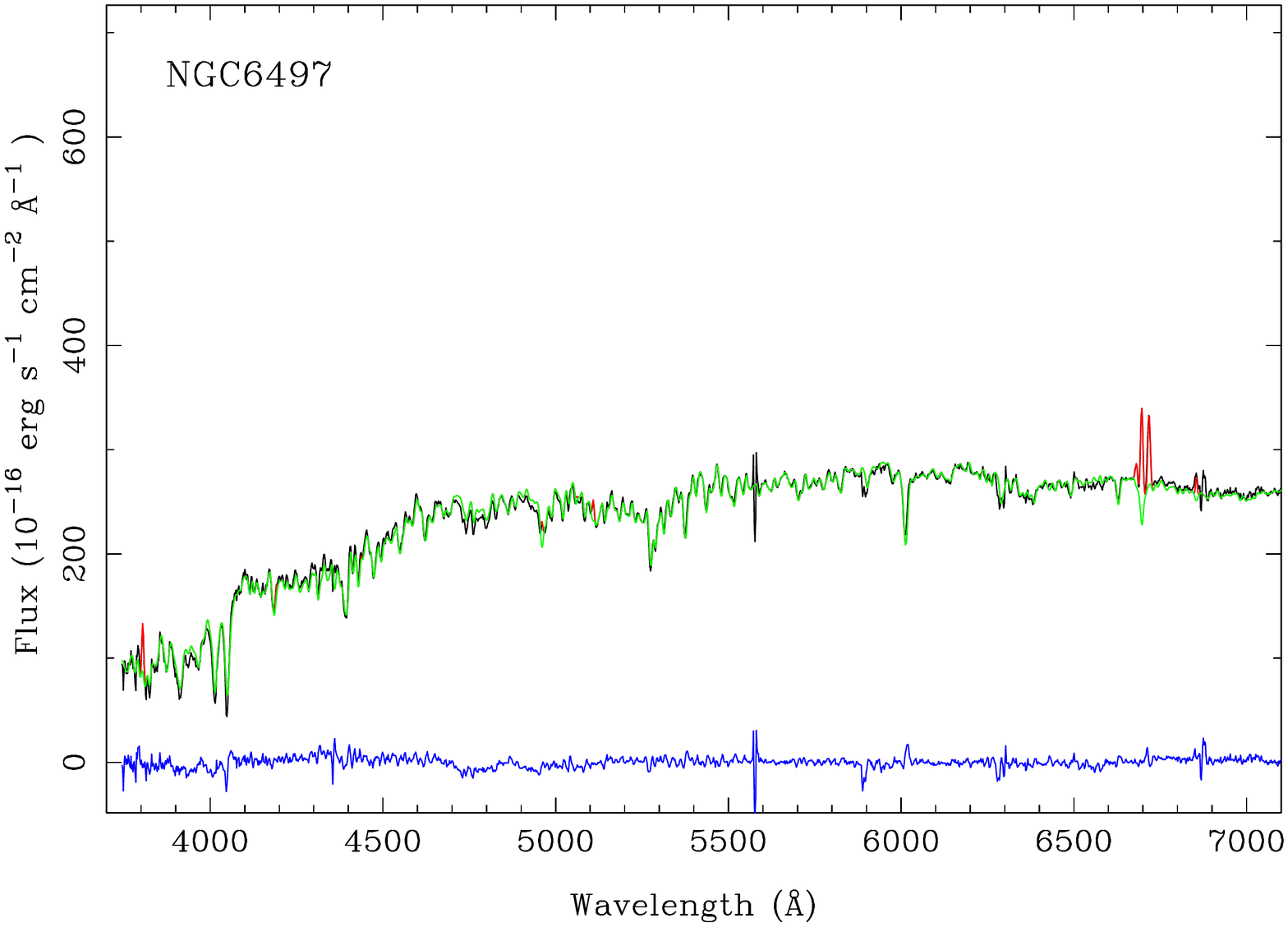}
 \includegraphics[angle=0,width=5.5cm,clip=true,trim=30 10 120 40]{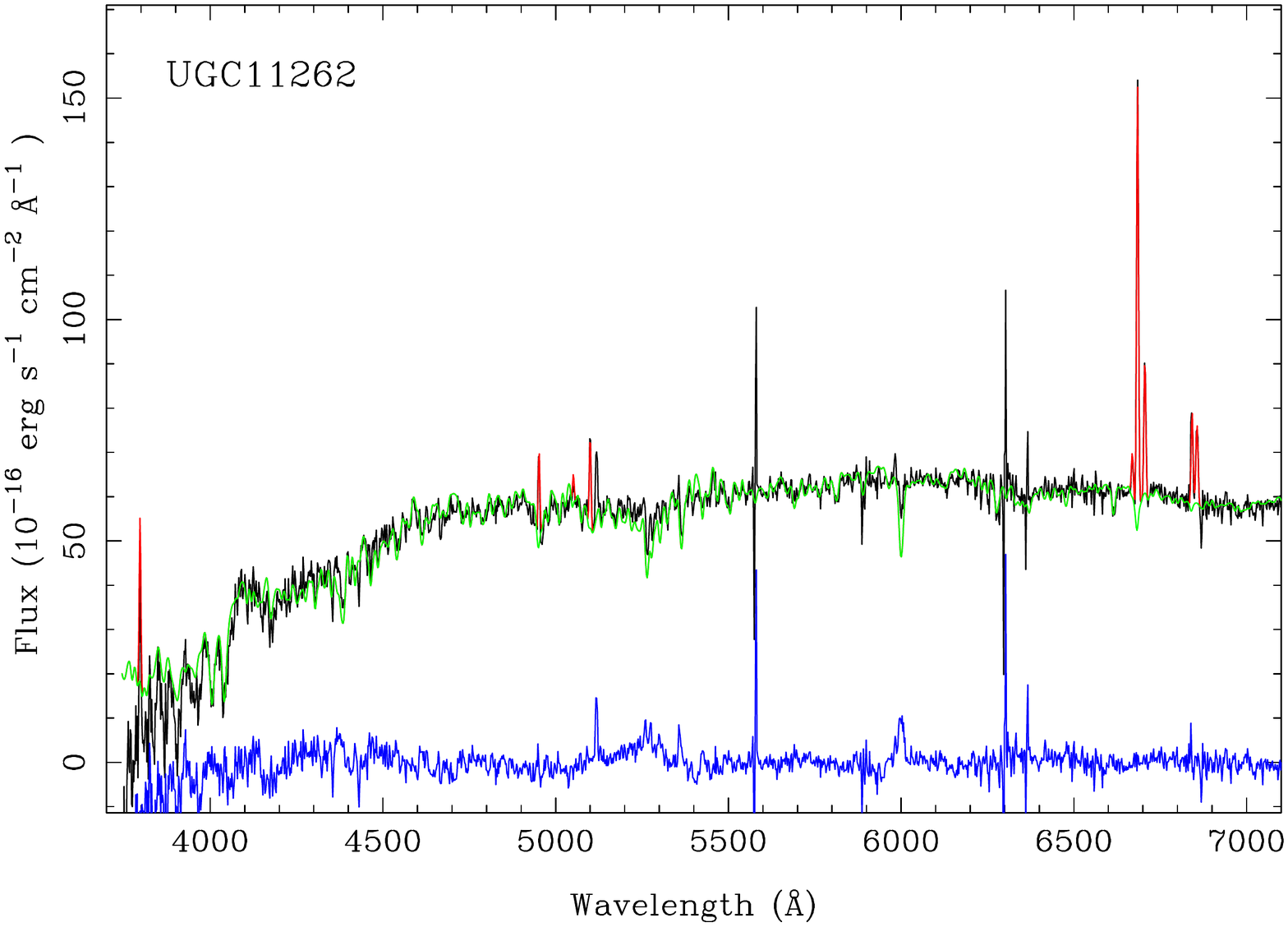}
}
\vspace*{5mm}
{\includegraphics[angle=0,width=5.5cm,clip=true,trim=30 10 120 40]{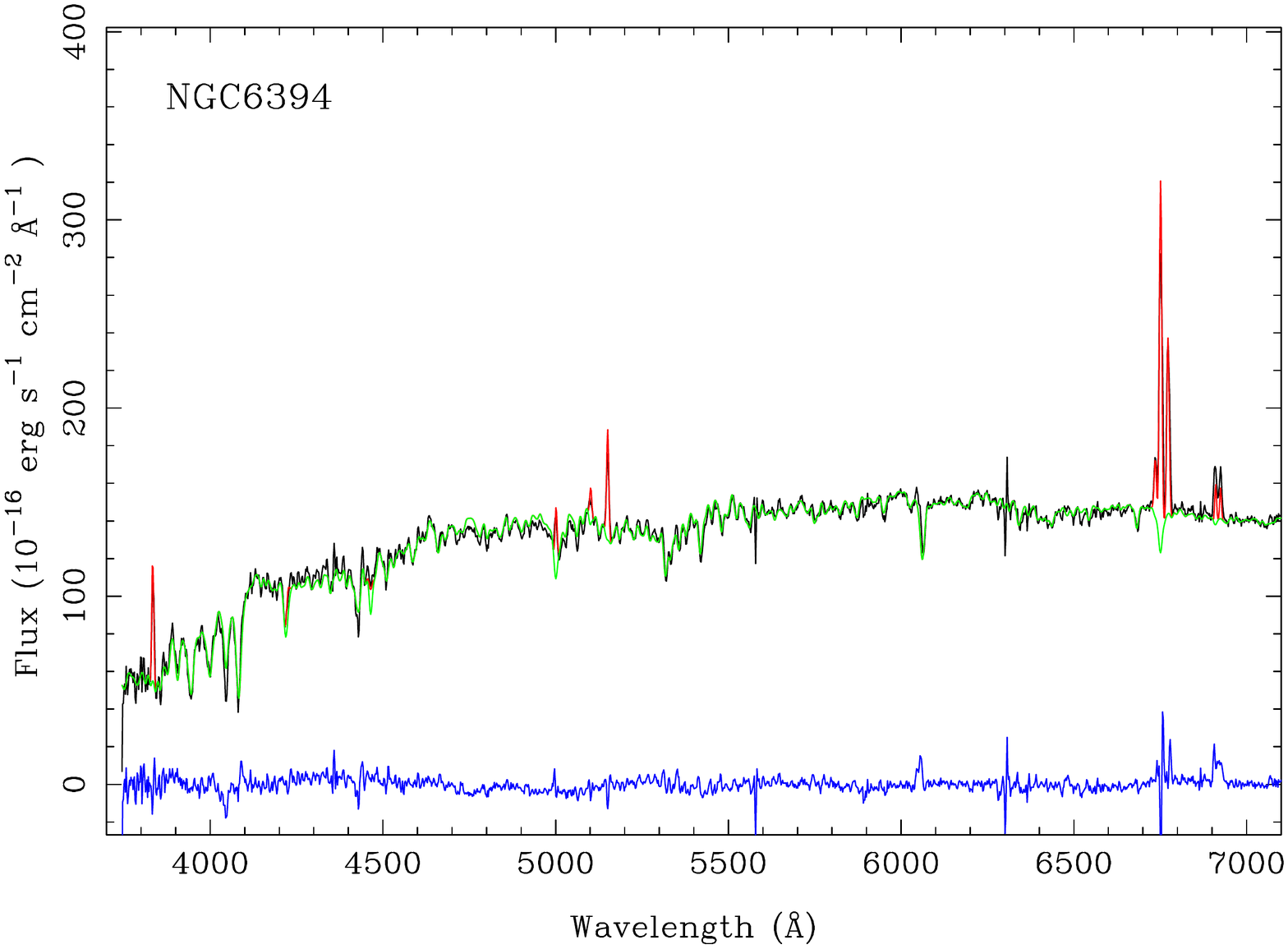}
 \includegraphics[angle=0,width=5.5cm,clip=true,trim=30 10 120 40]{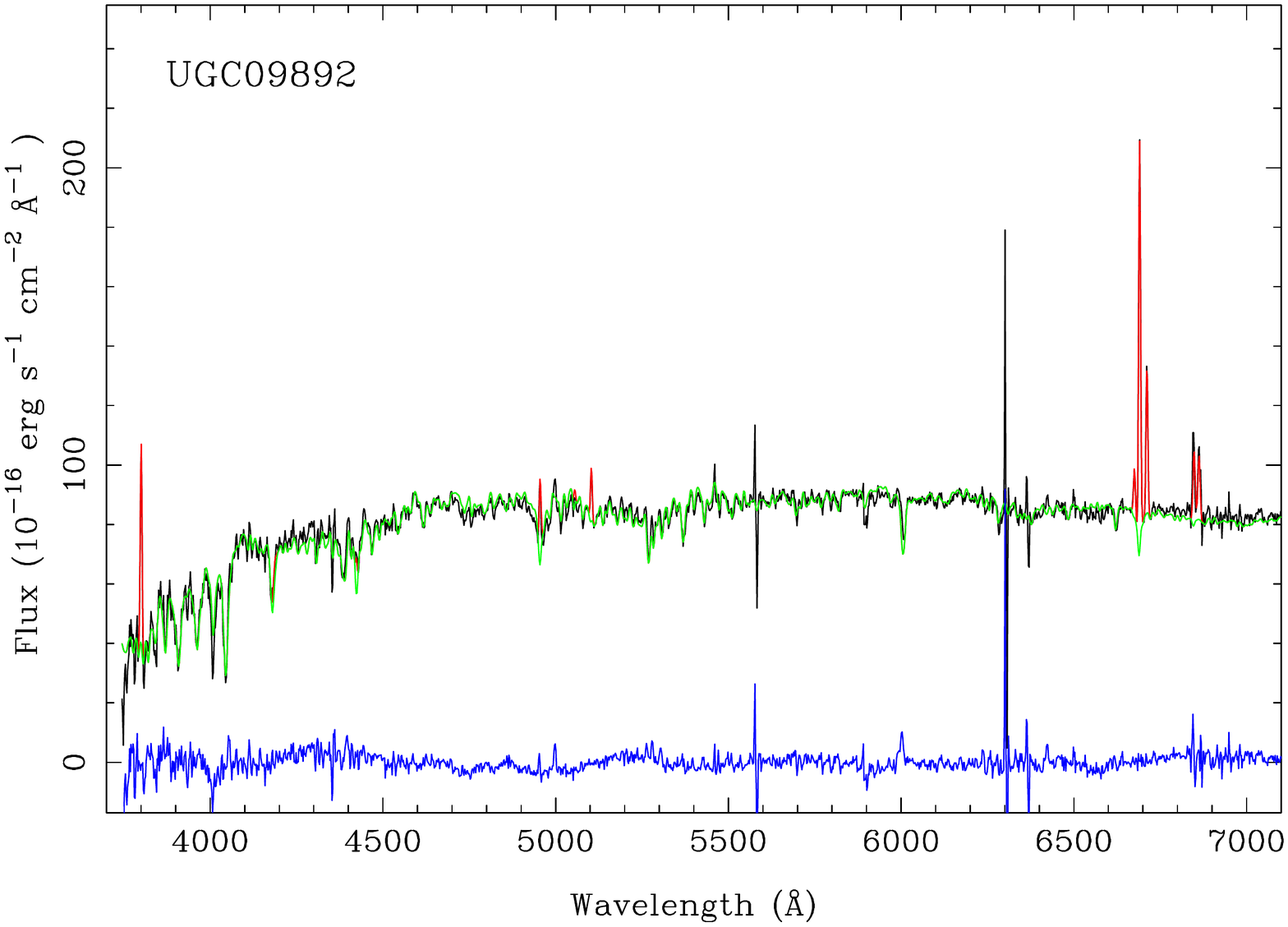}
 \includegraphics[angle=0,width=5.5cm,clip=true,trim=30 10 120 40]{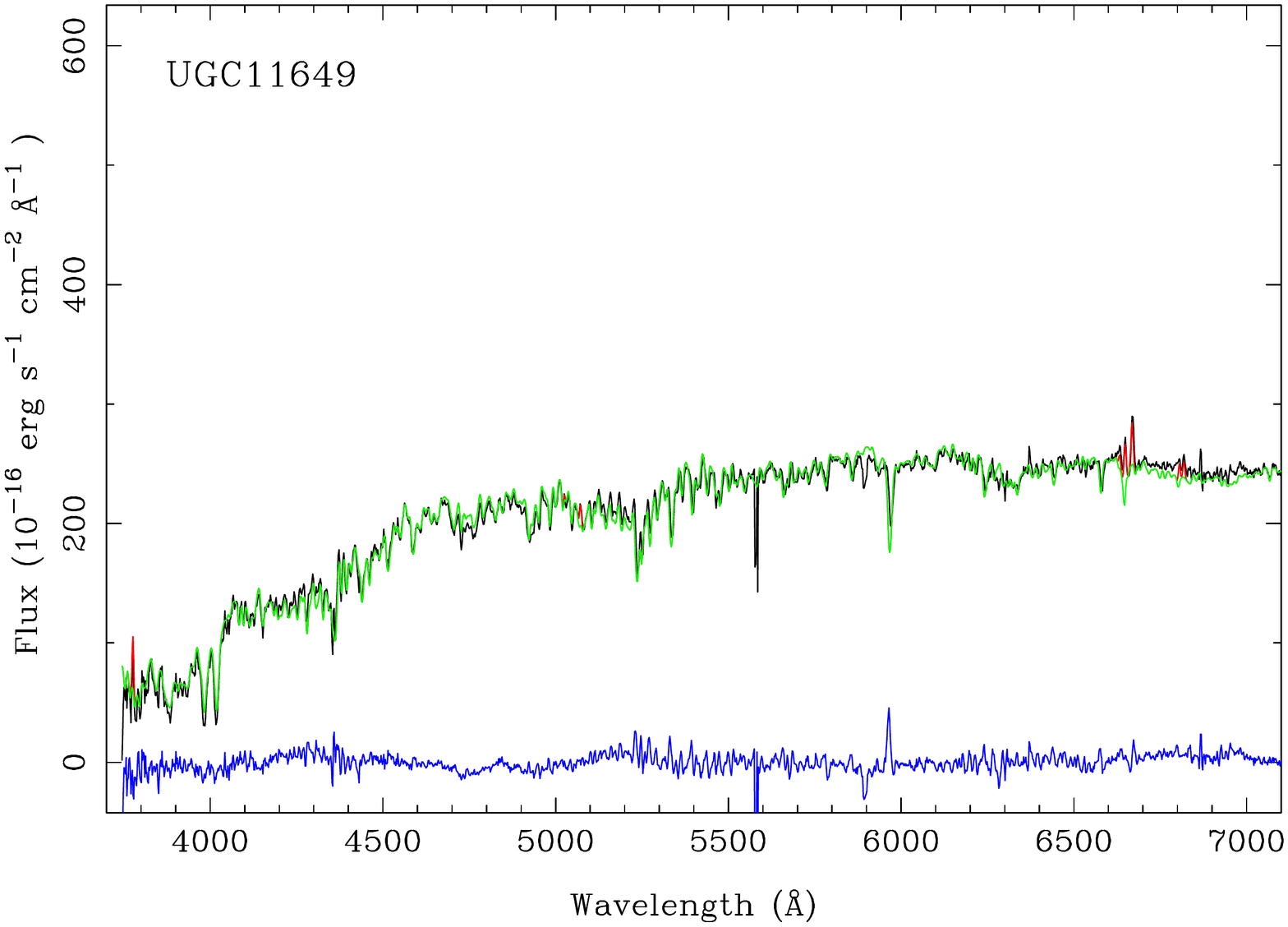}
}
 \caption{\label{fig:ssp} Results of the fitting procedure applied to each single
 spectrum contained in the dataset, described in the text, for the six galaxies
 shown in Figures \ref{fig:comp_r} and \ref{fig:comp_sdss_spec}. Each panel
 shows the integrated spectrum of the corresponding galaxy (solid-black line) derived by coadding
 the spectra within a 30$\arcsec$ diameter aperture centered on the peak
 emission in the $V$-band. The green line shows the best-fit multi-SSP 
 stellar population, while the red line shows the best model derived for the
 emission of ionized gas. The blue line is the residual from the fit.
 }
 \end{center}
 \end{figure*}

 Figure \ref{fig:phot} shows the distribution of the $g$-band
 magnitudes derived from the datacubes and the SDSS images before (blue
 solid circles) and after (red solid squares) the recalibration
 procedure. As expected, there is a significant improvement in the
 photometric agreement. Prior to the recalibration, both photometric
 datasets match within $\sim$20\%. This dispersion is similar to that
 found in the zero points of the flux calibration derived
 night-by-night for the V500 data. Once the recalibration was applied,
 both datasets match each other within $\sim$0.06 mag, i.e., $\sim$7\%
 (Fig. \ref{fig:phot}, red squares).  Similar results are found for the
 $r$-band.

 { To reinforce the validity of our flux re-calibration procedure, we
   repeated it for the flux-calibration standard stars observed during
   the run. The frames corresponding to the calibration stars were
   reduced using the same procedure as used for the science
   objects. Instead of using the transmission curve derived for the
   corresponding night, we adopted the master curve shown in Fig
   \ref{fig:effi}, and the extinction information provided by the
   extinction monitor. With this procedure we try to simulate as closely
   as possible the reduction conditions of the science frames. Next,
   to apply the same re-calibration procedure outlined before, the $g$
   and $r$-band magnitudes were synthesized from the published
   flux-calibrated spectra. The top panel of Figure \ref{fig:comp_stars}
   shows a typical example of how the flux-recalibrated spectrum
   corresponding to a calibration star agrees with the published
   flux-calibrated one. The bottom panel shows the distribution of the
   relative differences in intensity, pixel-to-pixel, between the
   published and re-calibrated spectra, for all the calibration stars
   observed along the run. The standard deviation is below
   $\sim$5\% for the considered wavelength range (3800-7000\AA)}

 \subsection{Signal-to-noise and depth of the data}
\label{datadepth}

 The pipeline performs a rough estimation of the S/N in each
 individual spectrum within the final reduced datacube. The median and
 the standard deviation of the intensity values are computed over a
 wavelength range free from strong spectral features. The wavelength
 range selected to perform this analysis was 4480-4520~\AA\ for both
 gratings, in order to simplify the comparison between the derived
 values. Assuming that the scatter is entirely due to noise, which is a
 rather conservative assumption, the signal-to-noise per spatial pixel
 is estimated as the ratio between the standard deviation and the
 median flux.  Figure \ref{fig:sn_map} illustrates the result of this
 procedure for the V500 datacube of NGC 5947. A typical patchy structure
 results for most of the objects, due to the intrinsic differences in
 the signal-to-noise map in dithered exposures.

 Once the signal-to-noise map was obtained, we determined those pixels
 for which S/N ranges between 3 and 4. The average flux
 corresponding to these spaxels ($\sim$100-200 spaxels) is considered a
 rough estimation of the 3$\sigma$ detection limit of the considered
 datacube. This flux is transformed to the corresponding AB magnitude
 using the standard equations, and stored for further quality control. 
 The depth of the datacubes depends on
 several factors, for each considered setup. It is mostly related to
 the night-sky brightness and the transparency and photometric
 stability of the night. Other factors, like the effects of
 the atmospheric seeing, have a much reduced effect due to the large
 size of the PPAK fibers ($\sim$2.7$\arcsec$ diameter).

 \begin{figure*}
 \begin{center}
{
 \includegraphics[width=6cm,angle=0,clip=true,clip=true,trim=110 20 170 70]{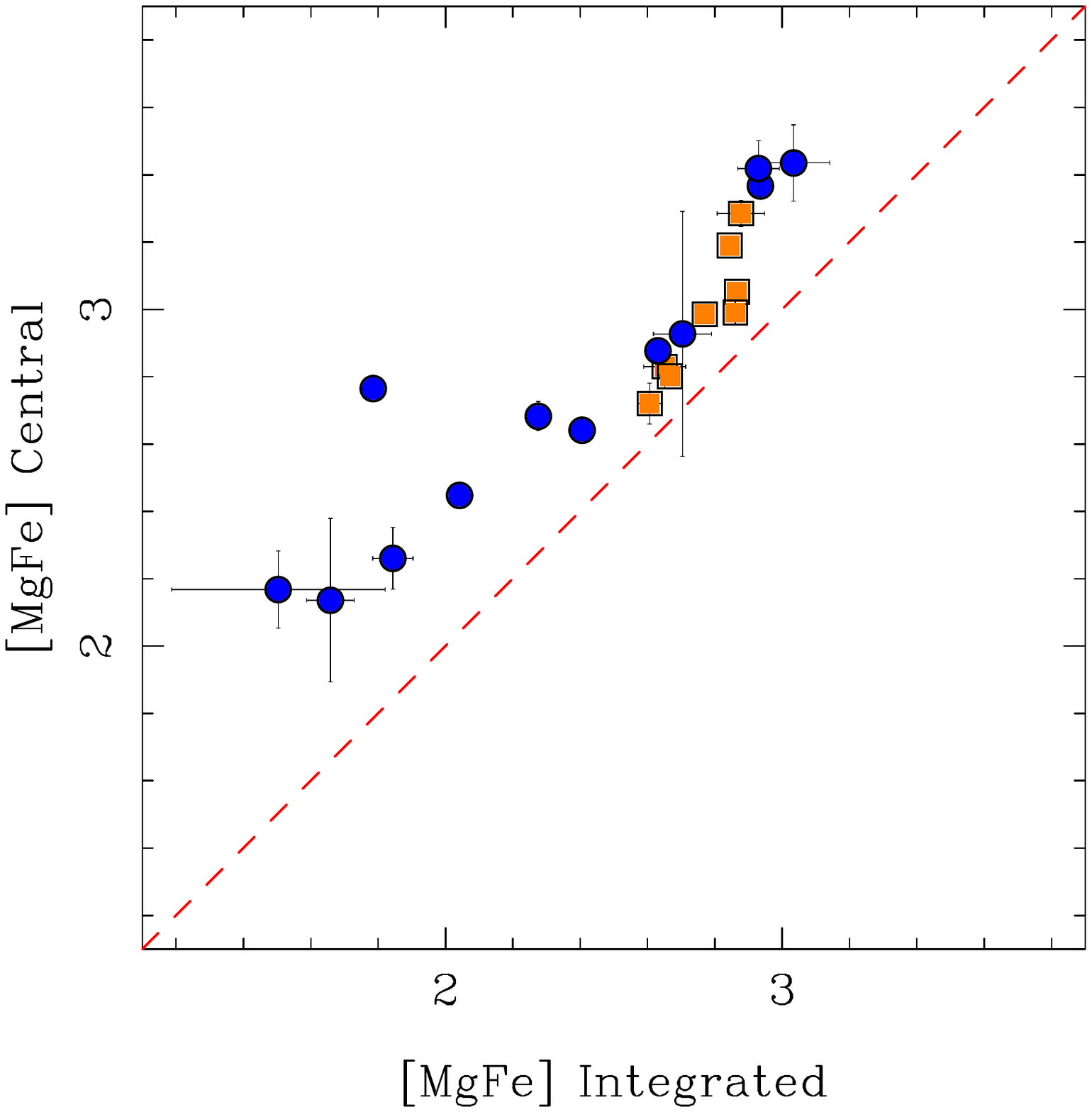}
 \includegraphics[width=6cm,angle=0,clip=true,clip=true,trim=110 20 170 70]{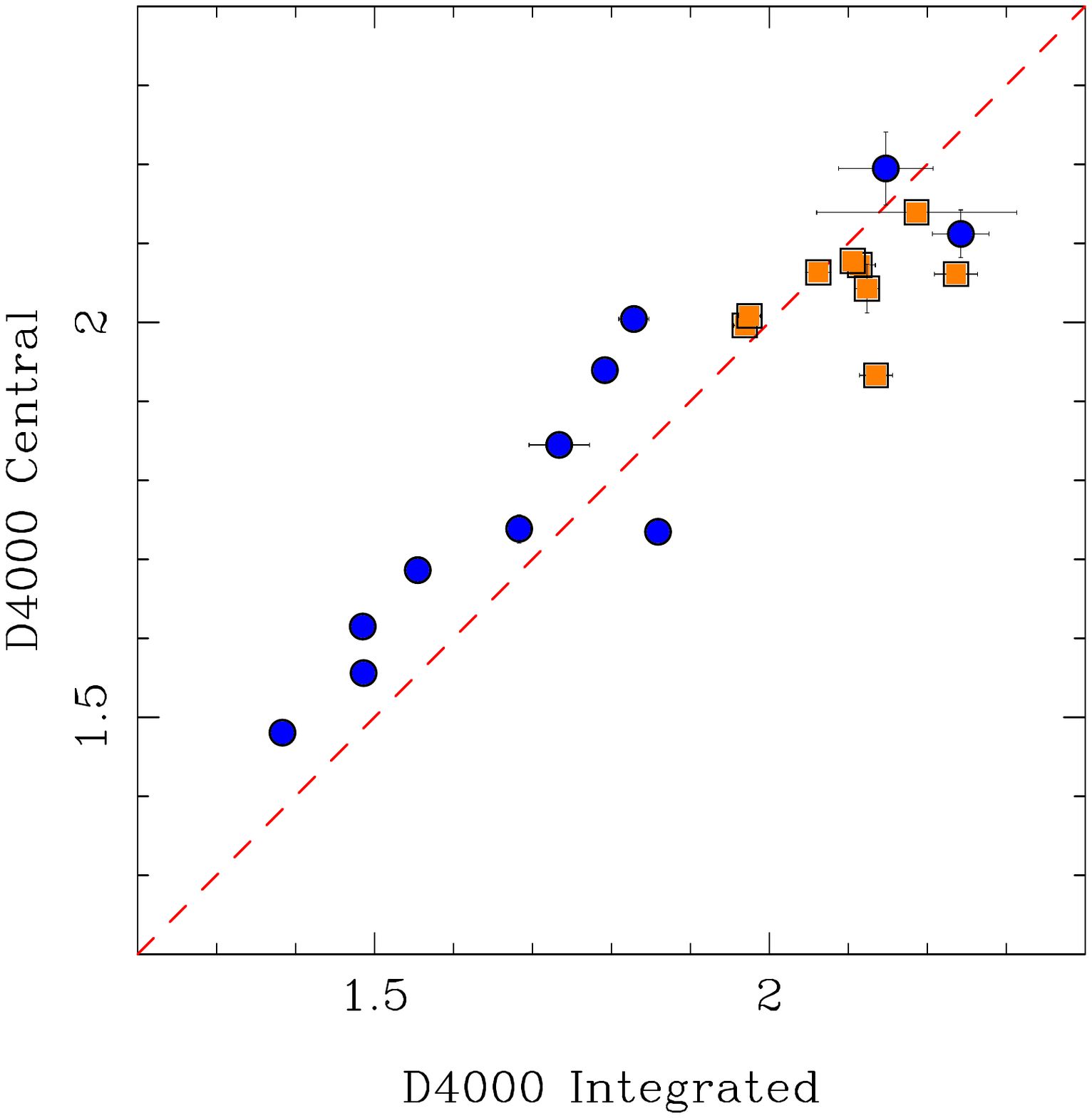}
 \includegraphics[width=6cm,angle=0,clip=true,clip=true,trim=110 20 170 70]{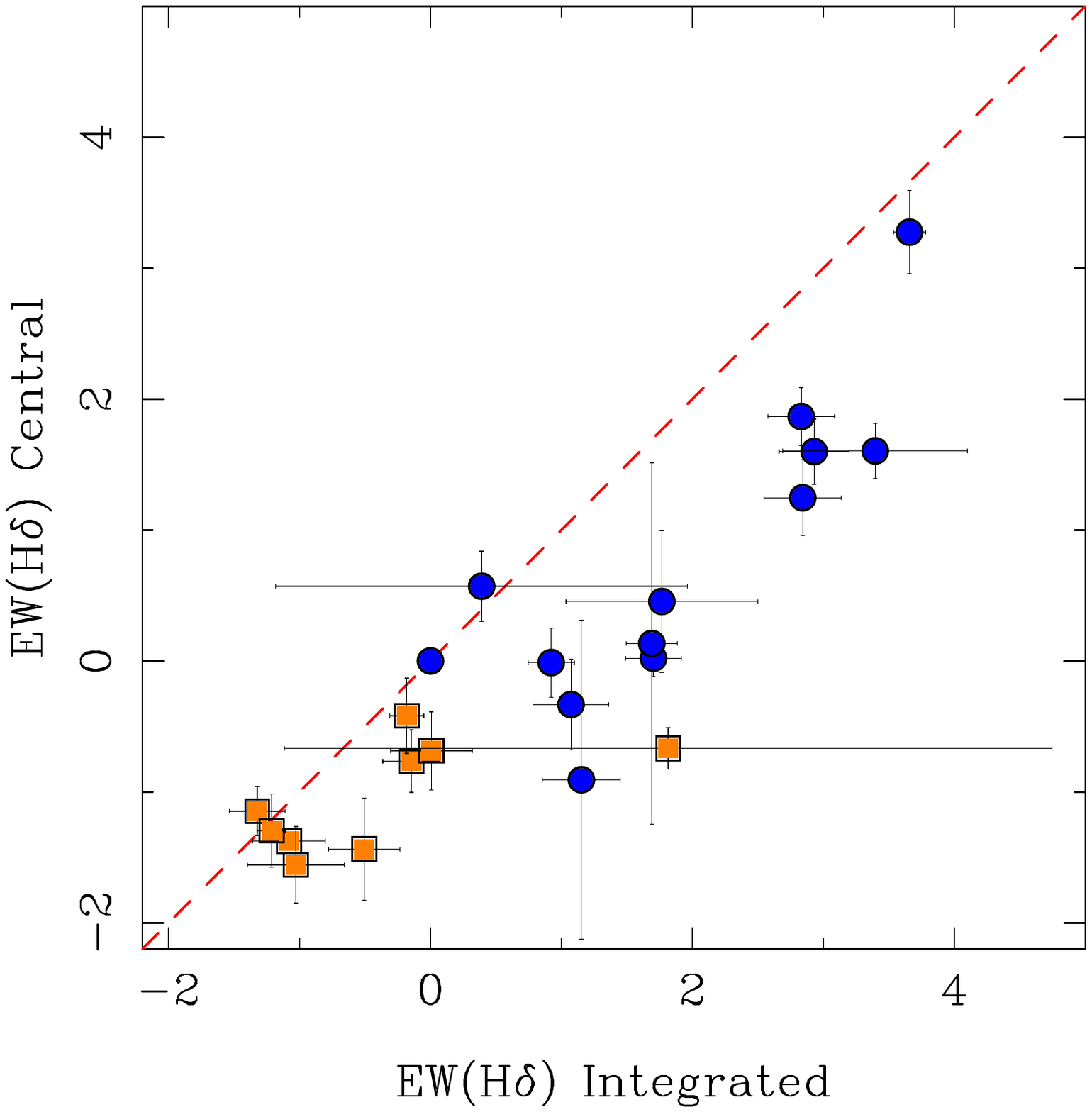}
}
\vspace*{5mm}
{
 \includegraphics[width=6cm,angle=0,clip=true,clip=true,trim=110 20 170 70]{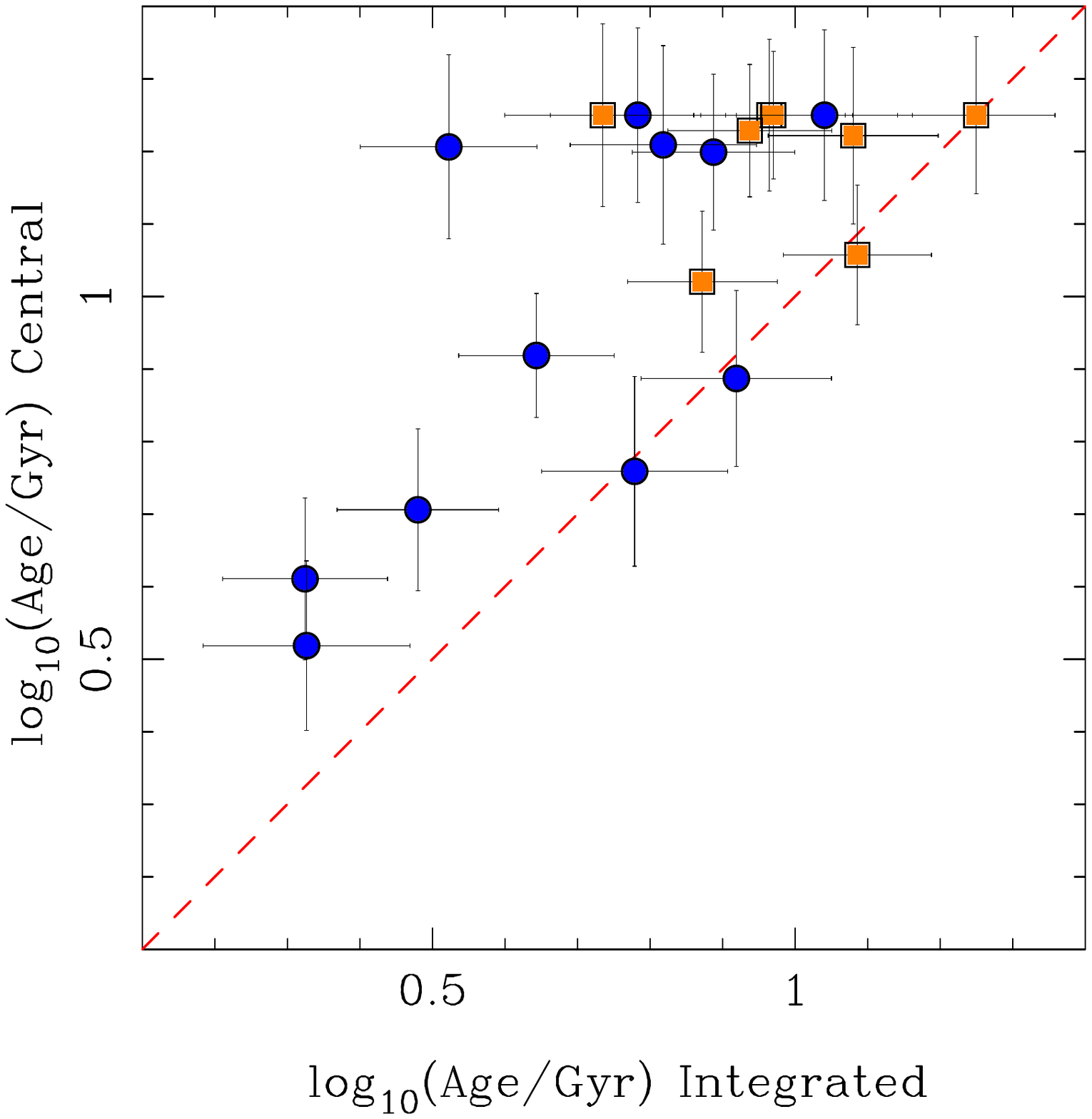}
 \includegraphics[width=6cm,angle=0,clip=true,clip=true,trim=110 20 170 70]{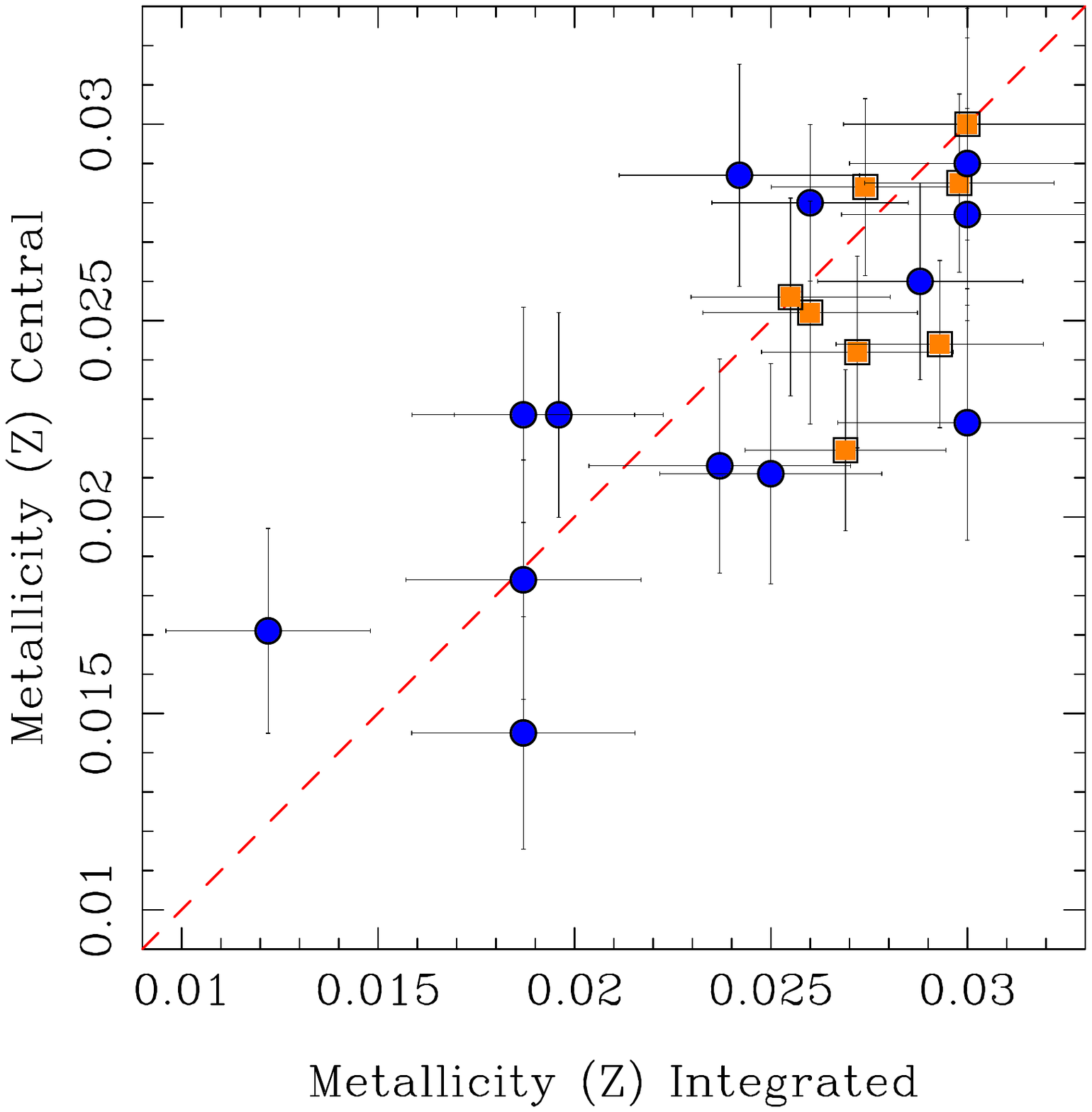}
 \includegraphics[width=6cm,angle=0,clip=true,clip=true,trim=110 20 170 70]{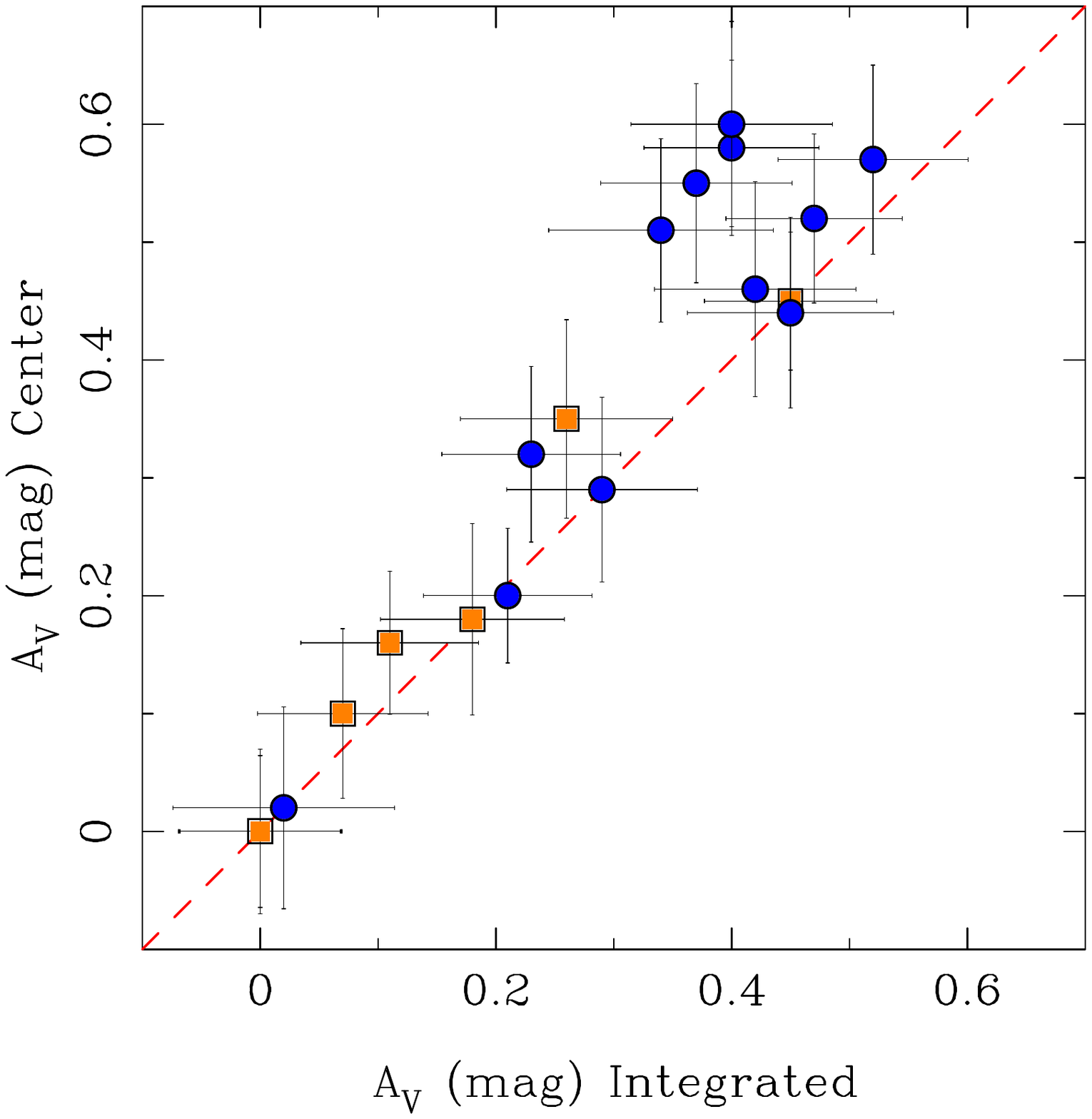}
}
\caption{Comparison between the properties derived for the stellar
  populations on the basis of the analysis of the stellar indices and
  the results of the multi-SSP fitting procedure, for the central and
  integrated spectra. From from left to right and top to bottom, each
  panel shows the comparison of the [MgFe] index,  the D4000
  parameter, the equivalent width of H$\delta$, the
  luminosity-weighted age, the luminosity-weighted metallicities, and the
  dust attenuation of the stellar component. The orange squares
  indicate the parameters for the early-type galaxies ($E$ or $S0$
  ones), while the blue circles indicate the parameters for the
  remaining galaxies.\label{fig:ssp_int}}
 \end{center}
 \end{figure*}

 The datacubes reached a 3$\sigma$ signal-to-noise ratio for an average
 surface brightness of $\sim$23.0 mag/arcsec$^2$ for the objects
 observed under good weather conditions with the V500 grating, and
 $\sim$22.7 mag/arcsec$^2$ for those observed with the V1200
 grating. This difference is expected since the difference in
 resolution between both setups is not completely compensated by the
 larger integration time for the latter grating. The area sampled with a S/N
 above 3$\sigma$ is $\sim$0.7 arcmin$^2$, for the V500 grating. This is
 clearly identified in Fig. \ref{fig:sn_map}, where the contours
 indicate the intensity level in the $V$-band, extracted from the
 datacube. The first contour, at $\sim$3$\sigma$ corresponds to
 $\sim$23.5 mag/arcsec$^2$ in this particular object. Both the limiting
 magnitude and the area sampled over the 3$\sigma$ limit agree with our
 expectations. 
 { The depth and area of the FoV sampled above the 3$\sigma$ limit are
   very similar for all the observed targets: the
   3$\sigma$ surface brightness depth has a range of $\pm$0.2 mag of
   the average values of each target and the area of
   the FoV above the 3$\sigma$ limiting surface-brightness ranges
   between a 50\% and a 90\%. This is because, due to the diameter
   selection of the sample, all the targets have similar surface
   brightness at a similar projected distance independently of their
   real physical size or light distribution. In addition, as all the
   targets are observed in dark time, there is no expected variation
   of the S/N or depth with night-sky brightness. 

  Obviously, this is only true for the objects observed on clear
  weather conditions. Under bad weather conditions the S/N is
  degraded. There are two cases of bad weather conditions considered
  in the design of the survey: (1) completely lost nights (i.e.,
  clouds, snow, rain...), and (2) nights with partial observations,
  i.e., nights with data taken under not-so-good weather
  conditions. For the first kind of nights, we have accounted an
  overhead of $\sim$30\%, based on published estimations of the
  fraction of useful time \citep{sanchez07a}, and they were considered
  in the total budget of nights. For the second case, our actual
  estimation is that only $\sim$10\% of the targets are observed under
  sub-optimal conditions. Of those, $\sim$60\% corresponds to the V1200
  setup, due to the larger fraction of time devoted to these
  observations. In summary, it is expected that we will obtain good
  quality data (i.e., with the described S/N and depth) for more than
  500 galaxies, once the survey is completed.}

 \subsection{Comparison with SDSS Data}
 \label{s:SDSScomp}

 As indicated above, the CALIFA mother sample has been selected from the SDSS
 imaging survey. This selection provides us with a large ancillary dataset, including 
 imaging in five bands for the complete sample and spectroscopic information for a considerable 
 subsample of the objects ($\sim$2/3). A simple quality test of the 
 obtained data is thus to compare them with the SDSS archive data. A collection of 
 archive auxiliary data and an analysis of the properties of the mother
 sample will be presented in a future paper.

 A first sanity check that we perform on all reduced data is to
 visually inspect how they look compared to the SDSS
 images. For each datacube we synthesize the corresponding $u$ (for
 both the V500 and V1200 dataset), $g$, and $r$-band (for the V500
 data) images, by convolving the cubes with the corresponding
 transmission curve, following the same procedure described previously for
 the photometric recalibration. These images are compared visually with
 the corresponding images in the SDSS imaging survey. Some simple
 problems with the data acquisition (e.g., off-centering, error in the
 scaling between pointings, error in the dithering pattern...), show up
 clearly in this visual inspection. Figure \ref{fig:img_sdss}
 illustrates the procedure, showing, for the same three objects shown
 in Figure \ref{fig:comp_r}, a comparison between the three-color
 images derived using both the SDSS and the CALIFA data. There is a
 clear morphological agreement between both image sets. Most of the
 structures seen in the SDSS images are clearly identified in the
 CALIFA data, in particular for the two brighter objects.  

 \begin{figure*}[tb]
 \begin{center}
 \includegraphics[width=6cm,angle=0,clip=true,clip=true,trim=50 20 100 70]{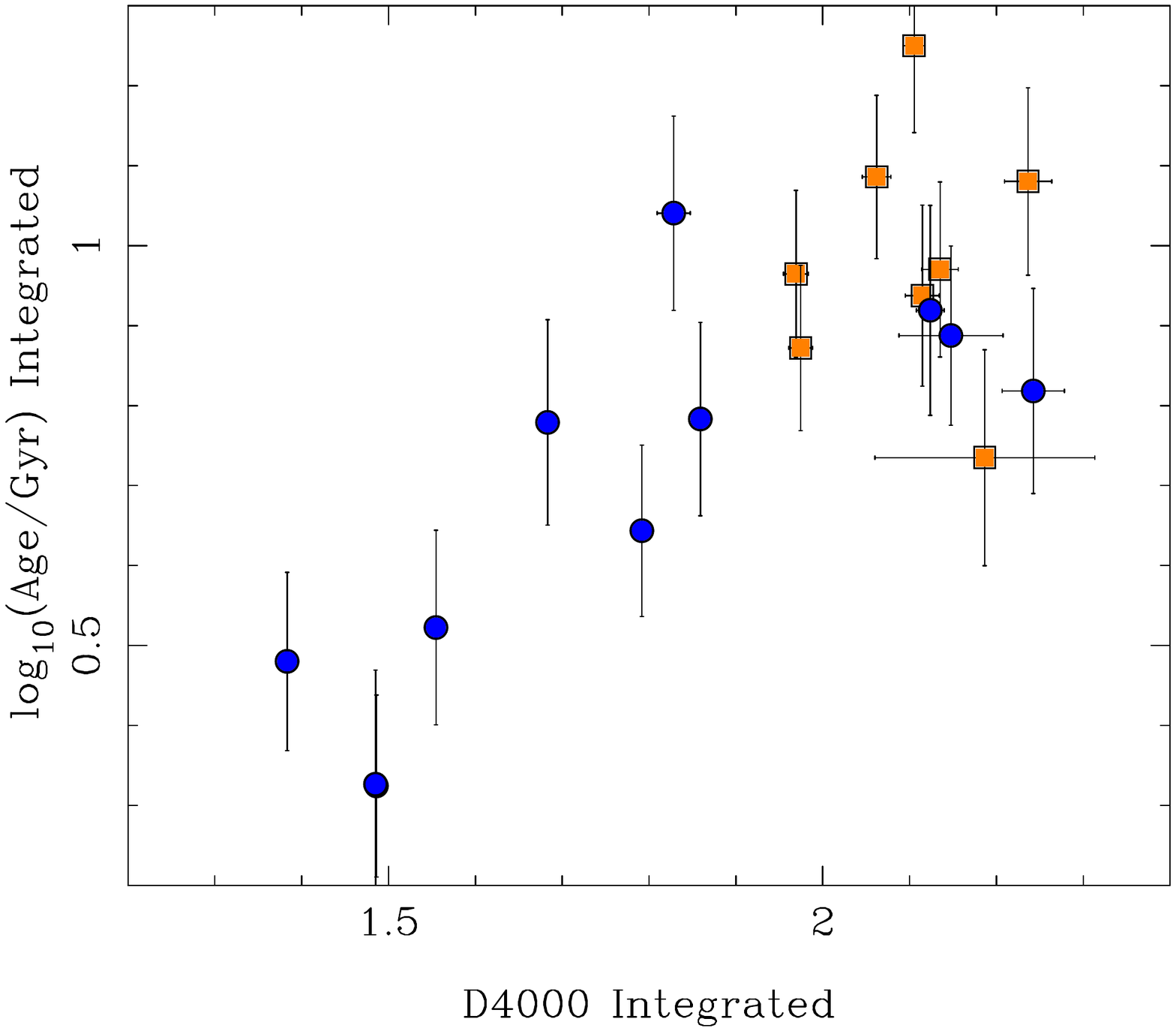}
 \includegraphics[width=6cm,angle=0,clip=true,clip=true,trim=50 20 100 70]{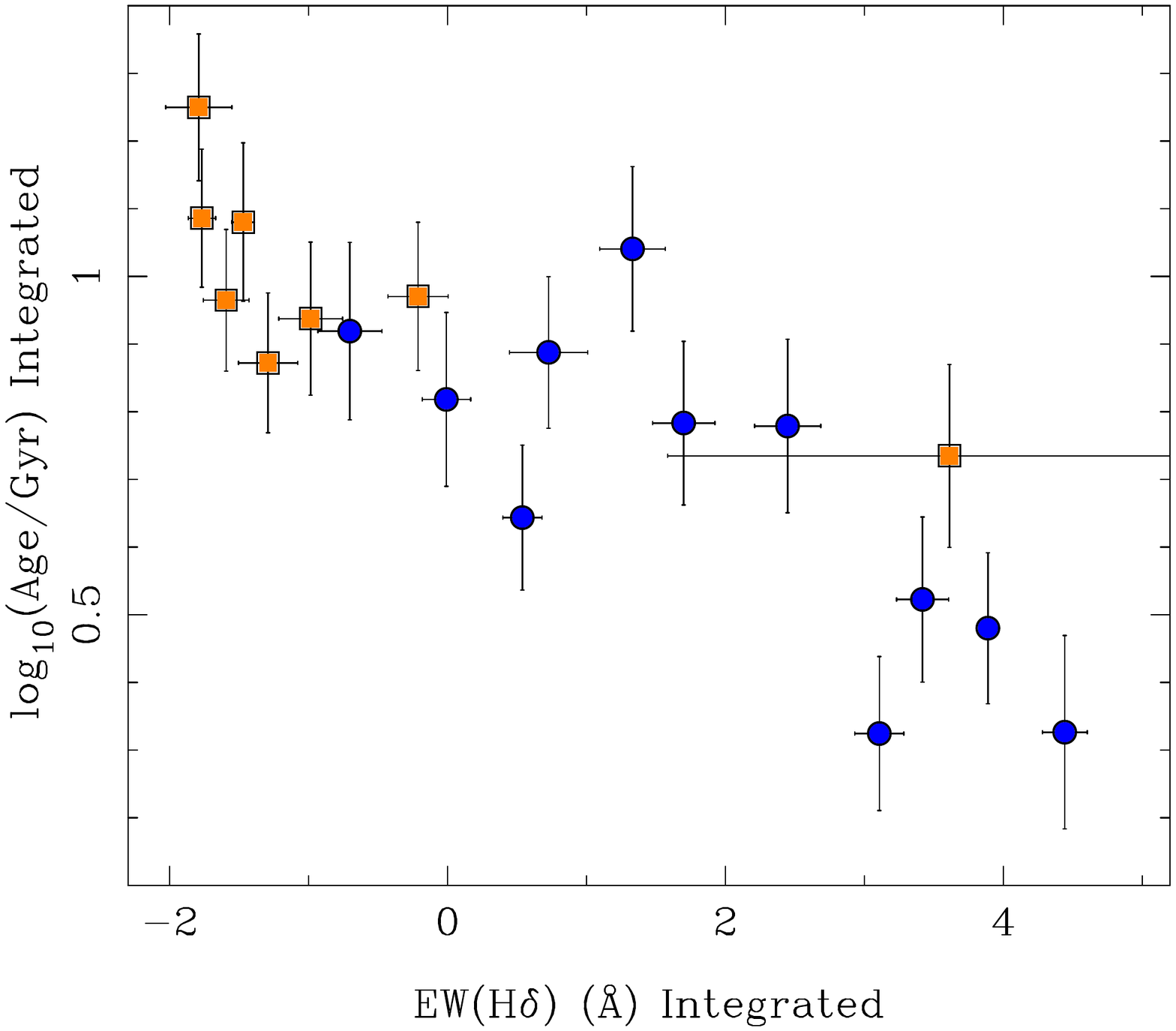}
 \includegraphics[width=6cm,angle=0,clip=true,clip=true,trim=50 20 100 70]{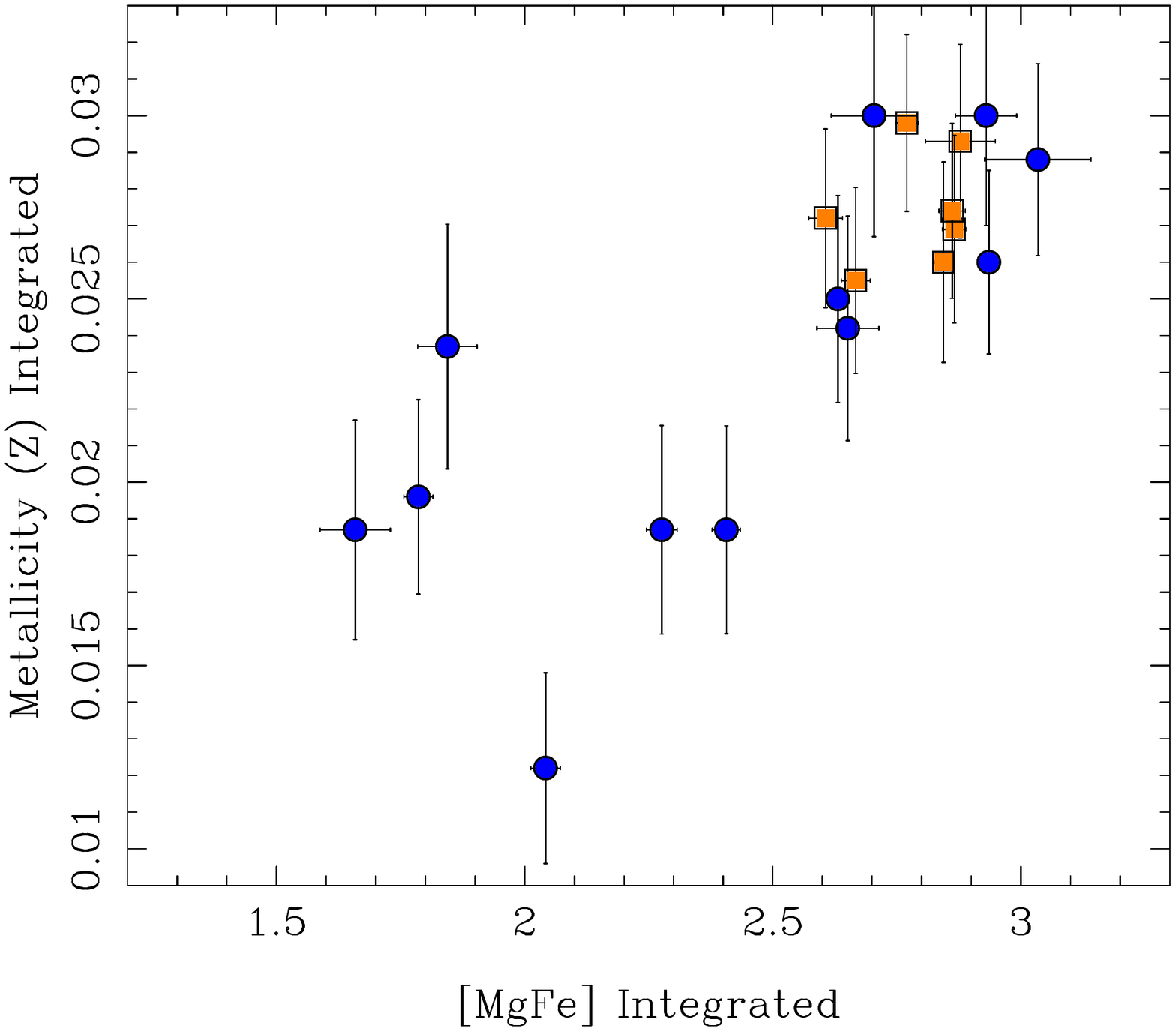}
 \caption{Comparison between the properties derived for the stellar populations
   on the basis of the multi-SSP fitting procedure (Age and Metallicity)  and the corresponding properties derived on the basis of the analysis of the stellar indices (D4000, EW(H$\delta$) and [MgFe]), for the integrated spectra. The symbols are defined in Fig. \ref{fig:ssp_int}.\label{fig:ind_int}}
 \end{center}
 \end{figure*}

 { Despite this qualitative agreement, there are more quantitative
   comparisons to illustrate the agreement with previously published
   broad-band images across the FoV \citep[e.g.,][]{sanchez11}. A
   simple one is to compare the surface brightness profiles. We
   extracted the surface brightness profiles for the considered
   $g$-band images, both synthesized from the V500 datacubes and
   obtained from the SDSS survey. The profiles were obtained by
   averaging the intensity within consecutive elliptical rings of
   3$\arcsec$ width, using the ellipticity and position angle listed
   in the SDSS-NYU catalogue for each object, and centred in the peak
   intensity of each image. The derived profiles are plotted in
   Fig. \ref{fig:img_sdss}, in the third panel. For each object the
   surface brightness profile derived from the $g$-band images derived
   from the datacubes is shown (blue solid circles), together with the
   same profile derived from the SDSS data (red solid squares). The
   error bars indicate the standard deviation around the average
   intensity. There is {\bf good} agreement between both profiles
   at all the radii. {\bf In addition to the surface brightness
     profiles, we include in the figure a comparison between the
     logarithm-scaled counter-plots derived from the SDSS $g$-band
     images and the corresponding ones synthetized from the CALIFA
     datacubes}. The largest differences are seen in the inner
   $\sim$3$\arcsec$ radius, where the datacubes have a flatter
   distribution than found using higher resolution
   images. This is expected since in some cases the datacubes have an
   incomplete coverage of the galaxy, due to the hexagonal pattern of
   the IFU. Within these two radii, the standard deviation of the
   differences between the surface brightness is $\sim$0.13 mag. This
   comparison illustrates the accuracy of the flux-calibration across
   the FoV. }

 We also perform a cross-check using the spectra. For each object with
 an SDSS spectrum, a combined spectrum within a 5$\arcsec$ diameter
 aperture is extracted from the V500 grating datacubes, centered on the
 intensity peak of the galaxy in the $V$-band. This aperture-extracted
 spectrum is selected to match the aperture of the SDSS fibers
 ($\sim3\arcsec$), taking into account possible errors in the
 centroiding accuracy of these fibers and seeing{ /resolution} effects. 
{ Slight differences due to seeing and resolution are expected, as illustrated in Fig. \ref{fig:img_sdss}, bottom panel.}
 In any case,
 no significant differences are found when slightly different apertures
 (3$-$6$\arcsec$) are selected, apart from the natural scaling between
 the spectra. Although the fraction of objects within the CALIFA mother
 sample with SDSS spectra is $\sim$60\%, for the particular dataset
 currently observed there were only 3 objects out of 21 with published
 SDSS spectra. The resolution of the SDSS spectra
 was degraded to match that of the V500 CALIFA data. Finally, a
 constant scaling factor was applied to the SDSS spectra to compensate
 for the differences in apertures.

 Figure \ref{fig:comp_sdss_spec} shows the comparison between the
 aperture-integrated spectra and the corresponding SDSS spectra for these
 three objects, including the residual after the subtraction of one from 
 the other. Both spectra agree to within $\sim$16\% ($rms$ of the
 residual) over the full wavelength range. Although the comparison
 sample is too small to derive statistical results, it is clear that
 the stronger differences are found at the wavelengths of the
 emission lines (e.g., NGC 6394). These differences are mostly due to a
 non-perfect matching of the spectral resolutions (which change along
 the wavelength in both cases), rather than to a real difference in the
 intensity of each emission line (as we will show later). A slight
 difference is also found in the blue-to-red spectral shape of
 UGC11649, which is clearly seen in the distribution of the residual
 spectra, although both spectra match within the expected errors (taking
 into account our blue-to-red spectrophotometric accuracy).


 { In summary, these preliminary quality checks show that:

 \begin{itemize}

 \item  The accuracy of the wavelength calibration corresponds to $\sim$8 km~s$^{-1}$ for the V500 ($\sim$3 km~s$^{-1}$ for the V1200), along the covered wavelength range.


 \item  The final spectral resolution in FWHM is $\sim$6.5\AA, ie.,
   $\sigma\sim$150 km~s$^{-1}$ for the V500 ($\sim$2.7\AA\, ie.,
   $\sigma\sim$85 km~s$^{-1}$ for the V1200). It has been homogenized
   across the FoV and along the wavelength range, taking into account
   the spectral and spatial variations.

 \item  The sky subtraction has a typical residual of $\sim$2\%\ of the original flux of the night-sky spectrum.

 \item  The flux calibration is anchored to the SDSS photometry. The final photometry is accurate to $\sim$8\% with respect to SDSS, with variation across the FoV of $\sim$0.13 mag/arcsec$^2$.

 \item  The datacubes reach a 3$\sigma$ S/N depth at an average
   surface brightness of $\sim$23.0 mag/arcsec$^2$ for the V500 data
   ($\sim$22.7 mag/arcsec$^2$ for the V1200 data), with $\sim$50-90\%\
   of the FoV above this S/N limit (with an average of $\sim$70\% of
   the FoV covered). It is estimated that $\sim$90\%\ of the survey
   datacubes will reach this depth.

 \end{itemize}

  In summary, the data have the quality foreseen in the
  proposal\footnote{http://www.caha.es/CALIFA/Accepted\_Proposal.pdf},
  which will allow us to study the spatial distribution of different
  spectroscopic properties, including: (1) the ionized gas emission
  lines from [OII]$\lambda$3727 to [SII]$\lambda$6731, (2) the stellar
  population features from D4000 to Fe5335 and (3) the gas and stellar
  kinematics, with a resolution of $\sim$85 km~s$^{-1}$. We conclude
  that we are on track for providing high-quality, well-calibrated and
  well-characterized reduced data to the users of the CALIFA survey.  
 }


 \section{Verification of scientific usability}
 \label{explore}

 In this section we describe a showcase analysis performed on the CALIFA
 data available to date, with the double purpose of verifying that we will
 be able to reach our science goals and of illustrating the information
 contained in the data.  It is beyond the scope of this article to
 perform a detailed analysis of these data to achieve any of the goals of
 the survey, which in any case require a statistical sample of
 objects to be observed.

 \begin{figure*}[tb]
 \begin{center}
{ \includegraphics[width=6.0cm,angle=0,clip=true,trim=70 20 140 70]{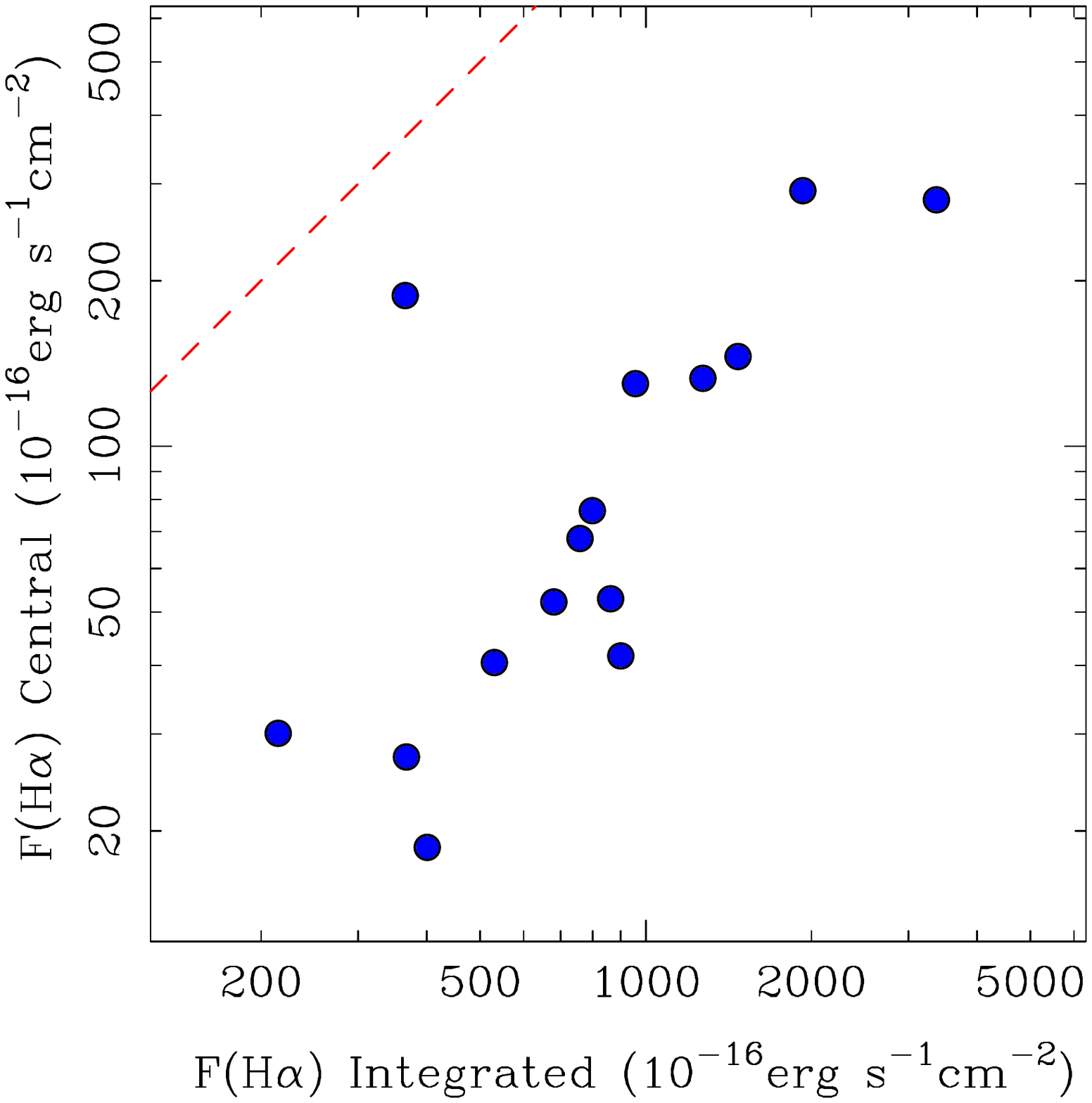}
 \includegraphics[width=6.0cm,angle=0,clip=true,trim=70 20 140 70]{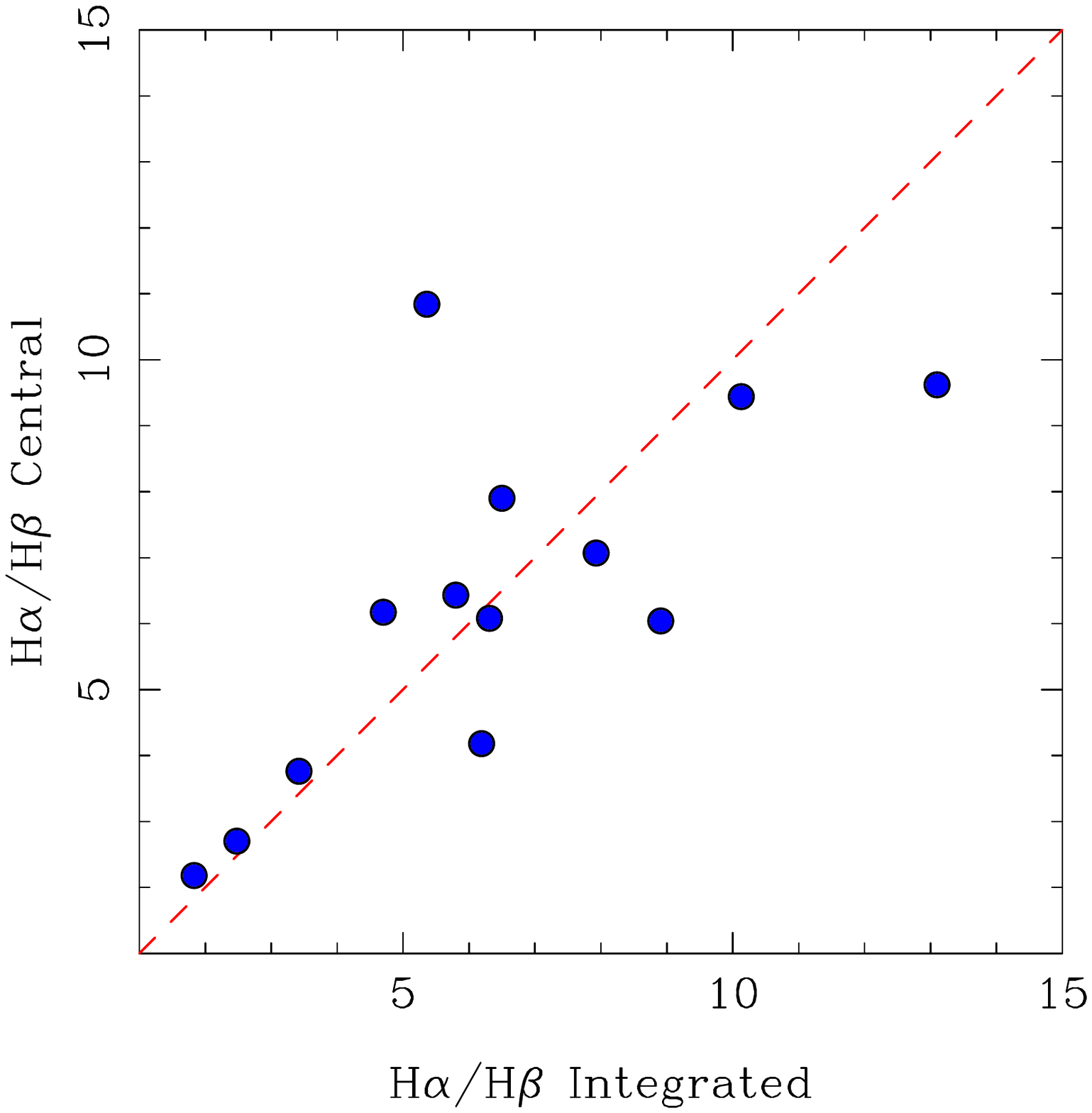}
 \includegraphics[width=6.0cm,angle=0,clip=true,trim=70 20 140 70]{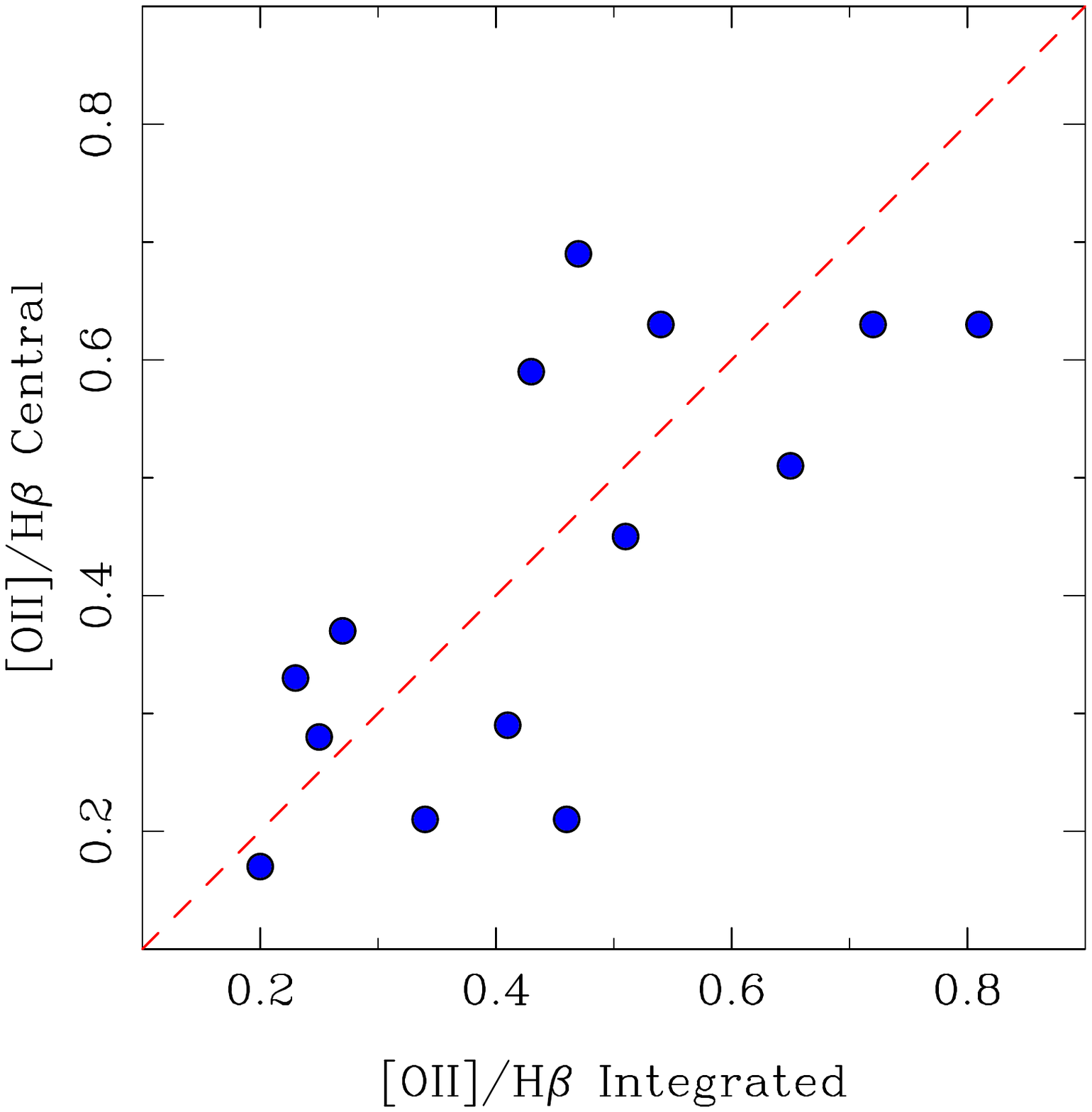}
}
\vspace*{5mm}
{ \includegraphics[width=6.0cm,angle=0,clip=true,trim=70 20 140 70]{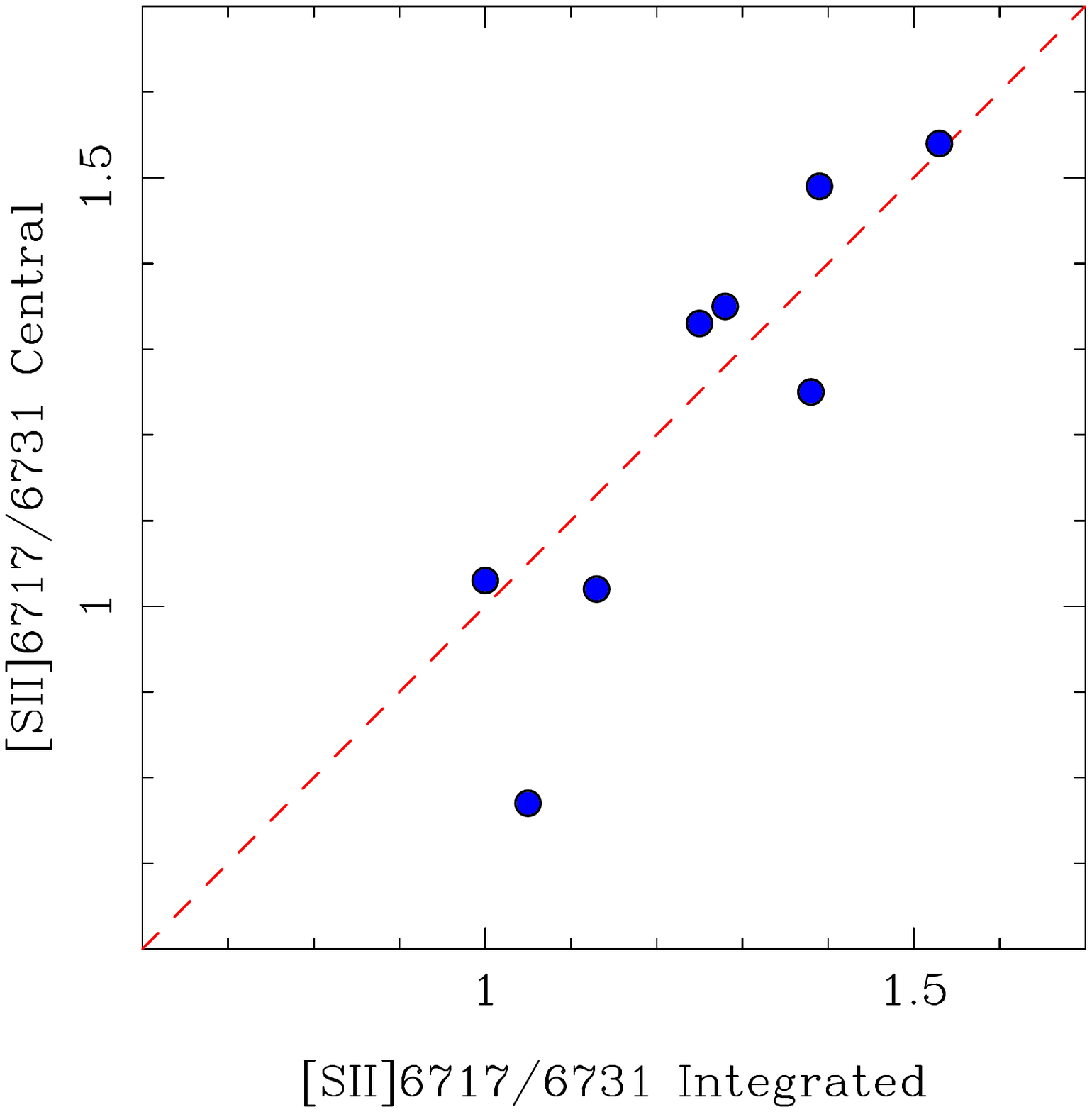}
 \includegraphics[width=6.0cm,angle=0,clip=true,trim=70 20 140 70]{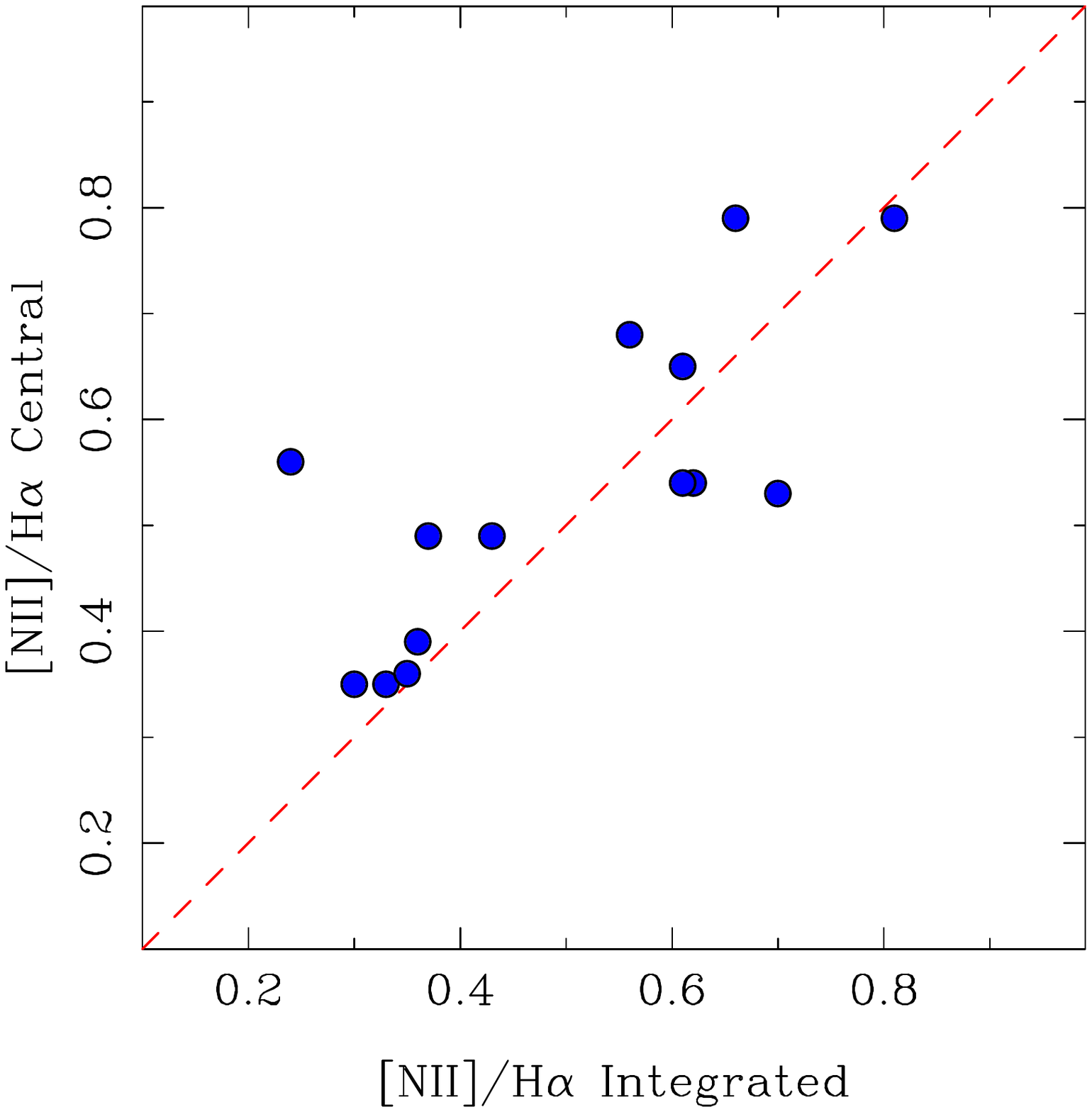}
 \includegraphics[width=6.0cm,angle=0,clip=true,trim=70 20 140 70]{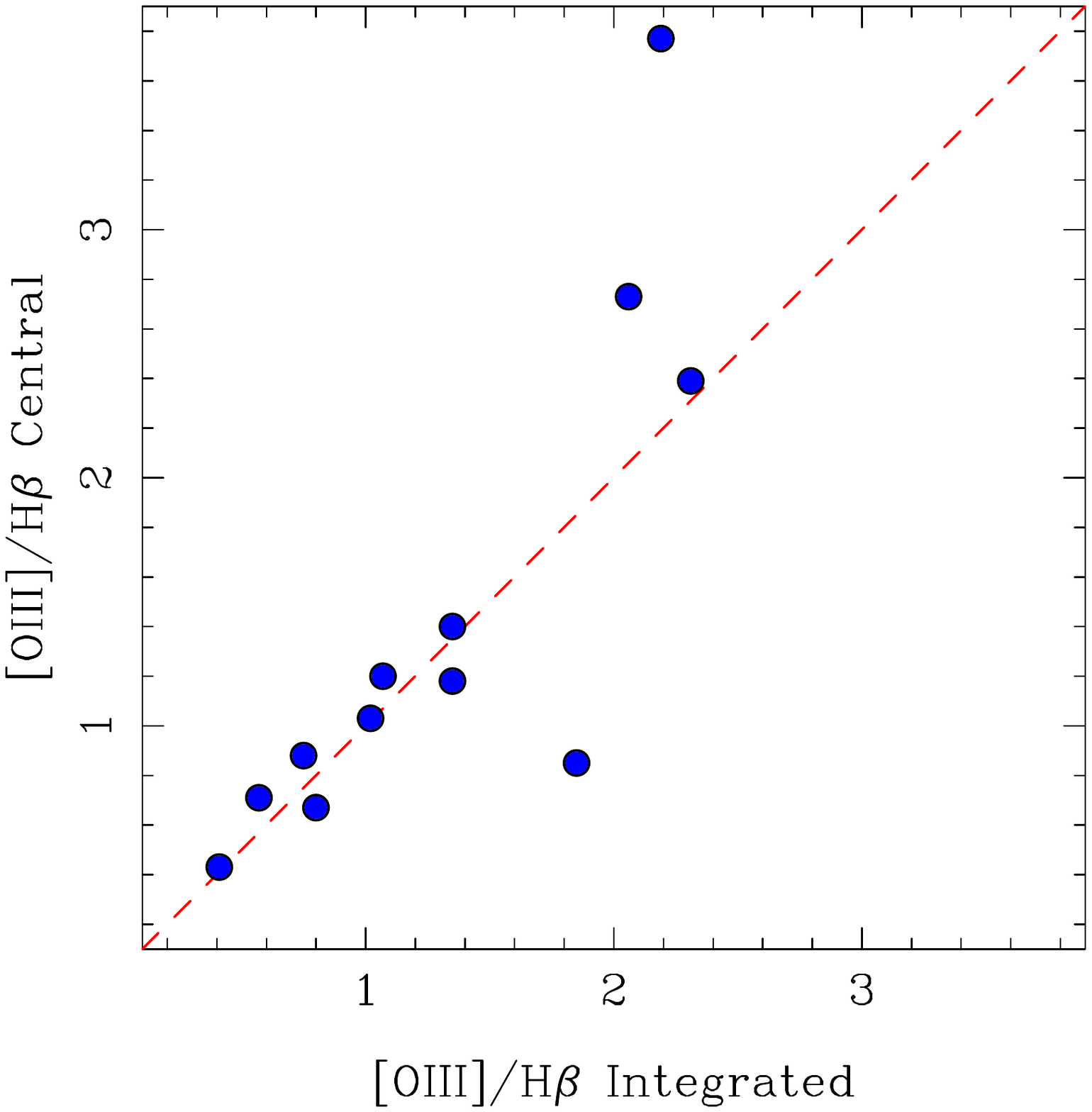}
}
 \caption{Comparison between the gas parameters derived for the
   central and integrated spectra using the fitting procedure
   described in the text. From from left to right and top to bottom
   each panel shows the comparison between the H$\alpha$ flux, and the H$\alpha$/H$\beta$,
   [OII]$\lambda$3727/H$\beta$, [SII]$\lambda$6717/[SII]$\lambda$6731, [NII]$\lambda$6584/H$\alpha$
   and [OIII]$\lambda$5007/H$\beta$ line ratios. Only values with a
   S/N$>$3 for both spectra have been included in each plot.
   \label{fig:comp_gas}}
 \end{center}
 \end{figure*}

 \subsection{Continuum modeling and gas decoupling}
 \label{s:continuum}

 To extract the information contained in the datacubes, the absorption
 spectra must be separated from the emission lines for each of the
 analyzed spectra \citep[e.g.,][]{sanchez11}.  Several different tools
 have been developed to model the underlying stellar population,
 effectively decoupling it from the emission lines
 \citep[e.g.,][]{cappe04,cid-fernandes05,ocvrik2006,sarzi2006,walch06,sanchez07a,koleva2009,macarthur2009,sanchez11}.
 Most of these tools are based on the same principles. It is assumed
 that the stellar emission is the result of the combination of
 different (or a single) single-stellar populations (SSP), and/or the
 result of a particular star-formation history. The stellar continuum
 is redshifted by a certain systemic velocity, broadened and smoothed
 due a certain velocity dispersion and attenuated due to a certain
 dust content.  Once the stellar continuum is determined, it is
 subtracted from each spectrum.  This provides a pure emission line
 spectrum, including the information of the ionized gas. The
 intensity, systemic velocity and velocity dispersion of each emission
 line can be derived by, for example, fitting with a set of Gaussian
 functions. In general it is known that the information derived from
 emission lines is much more accurate and stable than that derived for
 stellar populations \citep[see][for a review of the state of the
 art]{walcher11}. However, an accurate decontamination of the stellar
 continuum is required to measure emission line ratios, such as the
 [OIII]/H$\beta$ ratio, which allow us to interpret the data in terms
 of physical processes e.g. ionization source.

 \begin{table*}
 \caption{Summary of the physical properties derived from the analysis of the ionized
   gas for the integrated and central spectrum of each galaxy.}      
 \label{tab:eline2}     
 \begin{center}
 \begin{tabular}{rrrrrrrr}  
 \hline\hline                
    \multicolumn{8}{c}{Integrated spectrum}\\
 \hline
 NAME & Spec. Type & A$_V$  & SFR$_{H\alpha}$ & SFR$_{[OII]}$ & log(U) &
 12+log(O/H) & Electron Density \\
      & &  (mag) & M$_{\odot}$ yr$^{-1}$& M$_{\odot}$ yr$^{-1}$ & & & cm$^{-3}$ \\
 \hline               
 NGC 5947      & SF     & 1.2    $\pm$ 0.1   &  3.8    $\pm$ 0.1    & 3.8   $\pm$ 0.8    &  -3.5  $\pm$ 0.2    &  8.7   $\pm$ 0.1    &          ----        \\
 UGC 09892     & SF     & 1.3    $\pm$ 0.2   &  2.3    $\pm$ 0.1    & 3.9   $\pm$ 0.8    &  -3.7  $\pm$ 0.4    &  8.7   $\pm$ 0.3    &          ----        \\
 NGC 6394      & SF/AGN & 2.1    $\pm$ 0.2   &  12.1   $\pm$ 0.1    & 14.1  $\pm$ 2.8    &  -3.2  $\pm$ 0.3    &  8.6   $\pm$ 0.1    &  214 $\pm$ 64  \\
 NGC 6497      & SF/AGN & 1.5    $\pm$ 0.2   &  3.7    $\pm$ 0.1    & 3.4   $\pm$ 0.7    &  -3.5  $\pm$ 0.4    &  8.7   $\pm$ 0.1    &          ----        \\
 UGC 11262     & SF     & 1.5    $\pm$ 0.2   &  1.8    $\pm$ 0.1    & 1.4   $\pm$ 0.3    &  -3.3  $\pm$ 0.7    &  8.6   $\pm$ 0.2    &  145 $\pm$ 43  \\
 NGC 6762      & AGN    & 1.5    $\pm$ 0.2   &          ----          &         ----         &          ----         &          ----         &          ----        \\
 UGC 11680     & AGN    & 3.3    $\pm$ 0.2   &          ----          &         ----         &  -3.0  $\pm$ 0.5    &          ----         &          ----        \\
 UGC 11694     & SF     & 0.0    $\pm$ 0.5   &  0.3    $\pm$ 0.1    & 0.2   $\pm$ 0.1    &  -2.9  $\pm$ 0.7    &  8.6   $\pm$ 0.4    &          ----        \\
 UGC 11717     & AGN    & 3.3    $\pm$ 0.4   &          ----          &         ----         &  -3.6  $\pm$ 0.2    &          ----         &  4571$\pm$ 1371\\
 UGC 11740     & SF     & 1.8    $\pm$ 0.2   &  3.5    $\pm$ 0.1    & 5.1   $\pm$ 1.0    &  -3.5  $\pm$ 0.5    &  8.6   $\pm$ 0.2    &  525 $\pm$ 157\\
 UGC 12185     & SF/AGN & 2.8    $\pm$ 0.4   &  6.2    $\pm$ 0.1    &22.1  $\pm$15.0  &  -3.6  $\pm$ 0.6    &  8.6   $\pm$ 0.2    &          ----        \\
 NGC 7549      & SF     & 2.4    $\pm$ 0.1   &  10.8   $\pm$ 0.1    & 10.3  $\pm$ 0.1    &  -3.7  $\pm$ 0.2    &  8.7   $\pm$ 0.1    &          ----        \\
 \hline\hline
 \multicolumn{8}{c}{Central Spectrum}\\
 \hline
 NAME & Spec. Type & A$_V$  & SFR$_{H\alpha}$ & SFR$_{[OII]}$ & log(U) &
 12+log(O/H) & Electron Density \\
      & &  (mag) & M$_{\odot}$ yr$^{-1}$& M$_{\odot}$ yr$^{-1}$ & & & cm$^{-3}$ \\
 \hline
 NGC 5947      &SF/AGN& 1.5    $\pm$ 0.1   &  0.4    $\pm$ 0.2    & 0.2   $\pm$ 0.1    &  -3.2  $\pm$ 0.3    &  8.7   $\pm$ 0.1    &          ----         \\
 UGC 09892     &SF/AGN& 2.6    $\pm$ 0.3   &  0.2    $\pm$ 0.2    & 0.4   $\pm$ 0.2    &  -3.8  $\pm$ 0.6    &  8.7   $\pm$ 0.3    &  2512$\pm$ 754  \\
 UGC 10710     &SF/AGN& 4.5    $\pm$ 0.9  & 3.9    $\pm$ 0.1     & 23.2  $\pm$13.9     &  -3.8  $\pm$ 0.6    &  8.6   $\pm$ 0.5    &          ----        \\
 NGC 6394      & AGN& 2.9    $\pm$ 0.3    &          ----          &         ----         &  -2.8  $\pm$ 0.1    &          ----         &  309 $\pm$ 93   \\
 NGC 6497      & AGN& 1.8    $\pm$ 0.9    &          ----          &         ----         &  -3.4  $\pm$ 0.4    &          ----         &          ----         \\
 UGC 11262     &SF/AGN& 2.7    $\pm$ 0.4   &  0.1    $\pm$ 0.2    & 0.2   $\pm$ 0.2    &  -3.7  $\pm$ 1.1    &  8.7   $\pm$ 0.4    &  575 $\pm$ 173  \\
 NGC 6762      & AGN& 2.4    $\pm$ 0.5    &          ----          &         ----         &  -3.3  $\pm$ 0.1    &          ----         &  234 $\pm$ 70   \\
 UGC 11680     & AGN& 2.3    $\pm$ 0.2    &          ----          &         ----         &  -2.5  $\pm$ 0.1    &          ----         &          ----         \\
 UGC 11694     & AGN & 3.1    $\pm$ 0.5   &          ----          &         ----         &  -3.3  $\pm$ 0.2    &          ----         &  316 $\pm$ 95   \\
 UGC 11717     & AGN& 4.9    $\pm$ 0.7    &          ----          &         ----         &  -3.4  $\pm$ 0.1    &          ----         &          ----         \\
 UGC 11740     &SF/AGN& 2.8    $\pm$ 0.6  &  0.1    $\pm$ 0.2    & 0.5   $\pm$ 0.2    &  -3.7  $\pm$ 0.5    &  8.6   $\pm$ 0.3    &  34  $\pm$ 10   \\
 NGC 7549      &SF/AGN& 2.6    $\pm$ 0.2  &  0.4    $\pm$ 0.2    & 0.3   $\pm$ 0.2    &  -3.4  $\pm$ 0.3    &  8.6   $\pm$ 0.2    &          ----         \\
 \hline                
 \hline                
 \end{tabular}
 \end{center}
 \end{table*}

 It is beyond the scope of this article to analyze in detail the nature
 of stellar populations in the considered objects, since this will be
 one of the major goals of the full CALIFA project. Therefore, for the
 current demonstration we perform a simple modelling of the continuum
 emission. We use the routines described in \cite{sanchez11} and
 \cite{rosales-ortega10}.  These routines fit the underlying stellar
 population combining linearly a set of stellar templates within
 a multi-SSP model. They provide us with a number of parameters
 describing the physical components of the stellar populations (e.g.,
 luminosity-weighted ages, metallicities and stellar dust attenuation,
 together with the systemic velocity and velocity dispersion), and the
 properties of the analyzed emission lines (intensity, velocity and
 velocity dispersion). No detailed comparison between the different
 available tools or a detailed error analysis is considered here. A
 simple SSP template grid was adopted, consisting of three ages (0.09,
 1.00 and 17.78 Gyr) and two metallicities ($Z\sim$ 0.0004 and
 0.03). The models were extracted from the SSP template library
 provided by the MILES project \citep{vazdekis10}. This library is most
 probably too simple to describe in detail the nature of all the stellar
 populations included in the current dataset. However, it covers the
 space of possible stellar populations, and it allows us to obtain
 reliable information on the ionized gas.  This fitting procedure has
 been adopted to analyze different aperture-extracted and single
 spectra within the datacubes, as we will describe in the following
 sections.

 Figure \ref{fig:ssp} illustrates the results of the fitting procedure for the six 
 galaxies also shown in Figures
 \ref{fig:comp_r} and \ref{fig:comp_sdss_spec}. In each panel the
 solid black line represents the integrated spectrum of the
 corresponding galaxy (i.e., 30$\arcsec$ diameter aperture spectra),
 together with the best multi-SSP model for the stellar population
 (green line), and the gas emission recovered (red line). 


 \subsection{Comparing the results derived by CALIFA and SDSS}

 The SDSS has become a well-appreciated standard for spectroscopic 
 data and also has been used to select the mother sample of the CALIFA survey. It is therefore 
 of interest to make a quantitative comparison of the
 information provided by both datasets. To this end we analyze both spectra (see Section 
 \ref{s:SDSScomp}) using the fitting algorithm described in Sec. \ref{s:continuum}. 
 Table \ref{tab:comp_sdss} summarizes the
 results of this analysis. For each galaxy and each spectral dataset it lists
 the luminosity-weighted age, metallicity, dust attenuation and velocity
 dispersion, derived from the analysis of the stellar
 population. It also lists diagnostic emission line ratios, such as
  H$\alpha$/H$\beta$, [OIII]/H$\beta$, [NII]/H$\alpha$ and [SII]6717/6731
 line ratios. 
 The parameters
 describing the stellar populations match reasonably well, considering that a 
 typical error of $\sim$20-30\% is expected in all of them \citep{sanchez11}. The properties of 
 the ionized gas match each other within the reported errors.

 Despite the small number of compared spectra, the good agreement in all
 derived spectroscopic parameters indicates that the current data provided
 by CALIFA have, on average and per fiber, a similar information content to that
 provided by the SDSS. The major advantage of CALIFA is the
 spatial coverage of the spectral information.

 \subsection{Central vs. total spectra}
 \label{cen_total}

 One of the science goals of CALIFA is to allow  the extraction of 
 observed spectra of particularly interesting regions or the full galaxy to
 produce integrated spectra using the IFU as an adaptive aperture
 spectrograph. This technique has been used by \citet{rosales-ortega10} to
 produce aperture selected spectra of HII regions within the
 galaxies of the PINGS survey, or in \citet{sanchez11} 
 to derive the integrated spectrum of the complete galaxies. 

 The full-aperture integrated spectra can be used to derive, for the
 first time, the real integrated spectroscopic properties of these
 galaxies, as opposed to previous studies that attempted to describe
 average properties by the analysis of individual spectra taken in
 different regions. The most similar approach would be the
 drift-scanning technique \citep[e.g.][and part of the ancillary data
   of the SINGS survey]{moustakas06}, although in those studies the
 fraction of the galaxy covered by the spectra was smaller than that of
 CALIFA. Another advantage of an IFU with respect to the
 drift-scan technique is that the former allows a comparison between
 the integrated and the spatially resolved properties of the galaxy, or
 the integrated spectra of particular regions.

 In this section we analyze two spectra derived for each galaxy for
 different apertures, using the V500-grating dataset. One is a
 5$\arcsec$ diameter aperture spectrum, centered on the peak intensity
 in the $V$-band. This spectrum is representative of the central
 region of each galaxy, and is similar to the one obtained by the SDSS
 survey (see Section \ref{s:SDSScomp}). The other is a 30$\arcsec$
 diameter aperture spectrum, an aperture that contains most fibers
 above the 5$\sigma$ detection limit for each galaxy. The latter
 spectrum is representative of the integrated properties of each
 galaxy. Both spectra have similar S/N ($\sim$40, at $\sim$5000~ \AA).
 This is due to the fact that the central spectrum samples areas of much
 higher S/N, despite its smaller aperture, while the total spectrum
 samples areas of lower S/N, but with a larger aperture. On average,
 the 5$\arcsec$ aperture contains ∼10\% of the flux encircled by the
 30$\arcsec$ aperture. We applied the above fitting technique
 described in Sec. \ref{s:continuum} to both spectra, deriving the
 main properties of both the stellar population and the ionized
 gas. An example of typical integrated spectra, and their modelling,
 was already shown in Fig. \ref{fig:ssp}.

 \subsubsection{Stellar Populations}

 Gradients and, more generally, spatial variations of stellar
 population properties have been observed in galaxies, both early types
 \citep[e.g.][]{kuntschner+06} and late types
 \citep[e.g][]{macarthur+04}. CALIFA will allow the mapping of these
 variations for the first time over a statistically representative sample of
 galaxies, unbiased in terms of morphology and mass/luminosity. The
 detailed modeling required to this goal is beyond the scope of this
 paper and will be presented in forthcoming dedicated works. Here we
 just illustrate the potential of the dataset using either
 model-independent measurements, such as the spectral indices, or the
 output of the simple fitting technique described in Sec. 7.1.

 It is known \citep[e.g.][]{trager+00,gallazzi+05} that spectral
 indices can be used to infer stellar population parameters such as
 age, metallicity and $\alpha$ enhancement. They provide robust,
 model-independent, information, complementary to that provided by
 fitting the full spectrum with SSP templates as described in Sec. \ref{s:continuum}. For the current analysis,
 we have explored the D4000 index (i.e. the 4000~\AA~ break) and
 the H$\delta$ index (which represents the
 equivalent width of the H$\delta$ line), as tracers of stellar age,
 and [MgFe] as a tracer of stellar metallicity. For this later index,
 we adopted the formula:

 $$[{\rm MgFe}]~=~ \sqrt{{\rm Mg}b ~ (0.72 {\rm Fe}_{5270}~+~0.28 {\rm Fe}_{5335})}$$

 The indices were measured for both the central and integrated
 spectra, once decontaminated by the ionized gas emission { and
   normalized to the standard Lick/IDS resolution}, using the scripts
 included in FIT3D, described in \cite{sanchez07b}.

\begin{figure*}[tb]
\begin{center}
\includegraphics[width=9cm,angle=0,clip=true,trim=30 30 100 30]{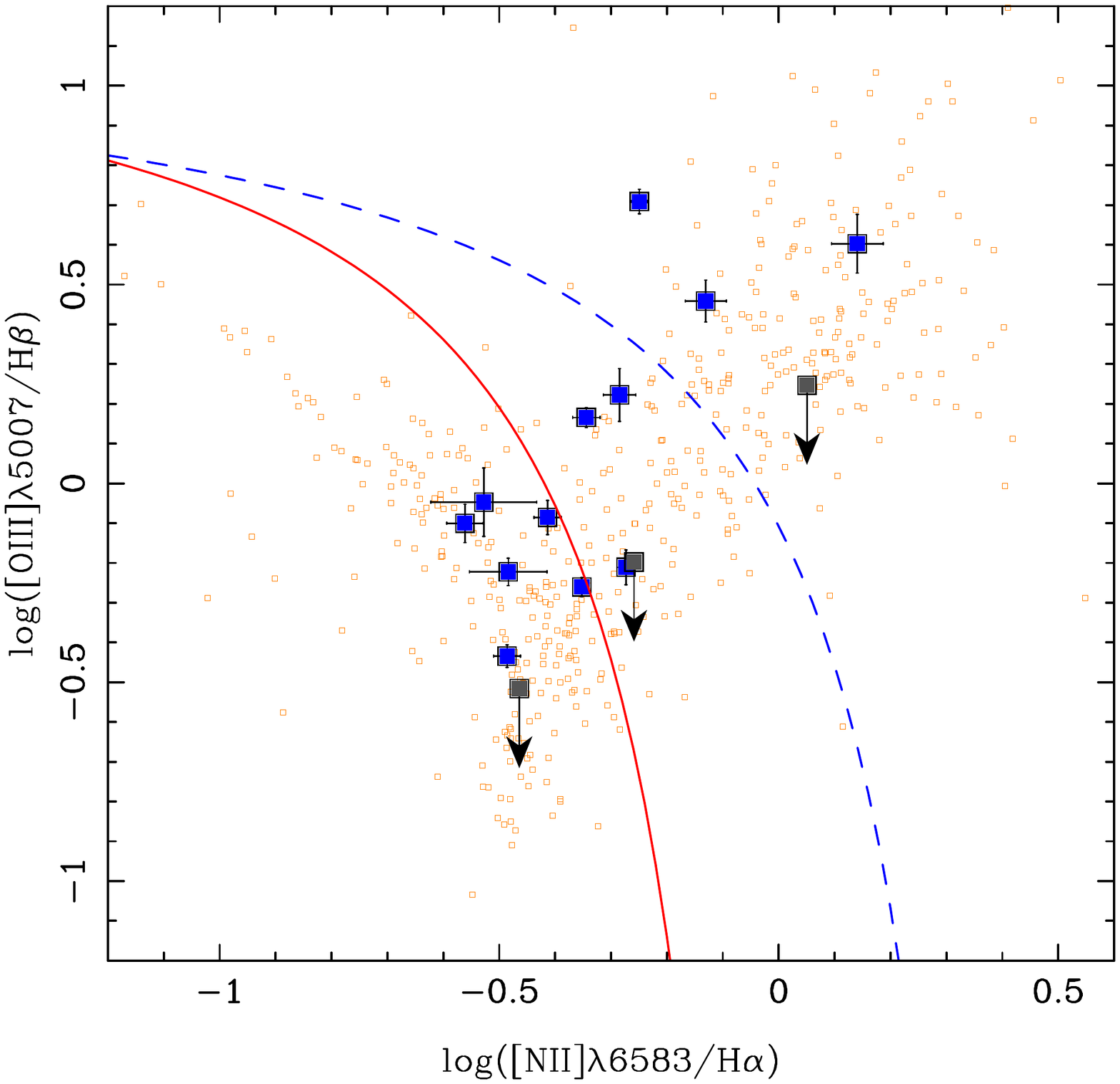}
\includegraphics[width=9cm,angle=0,clip=true,trim=30 30 100 30]{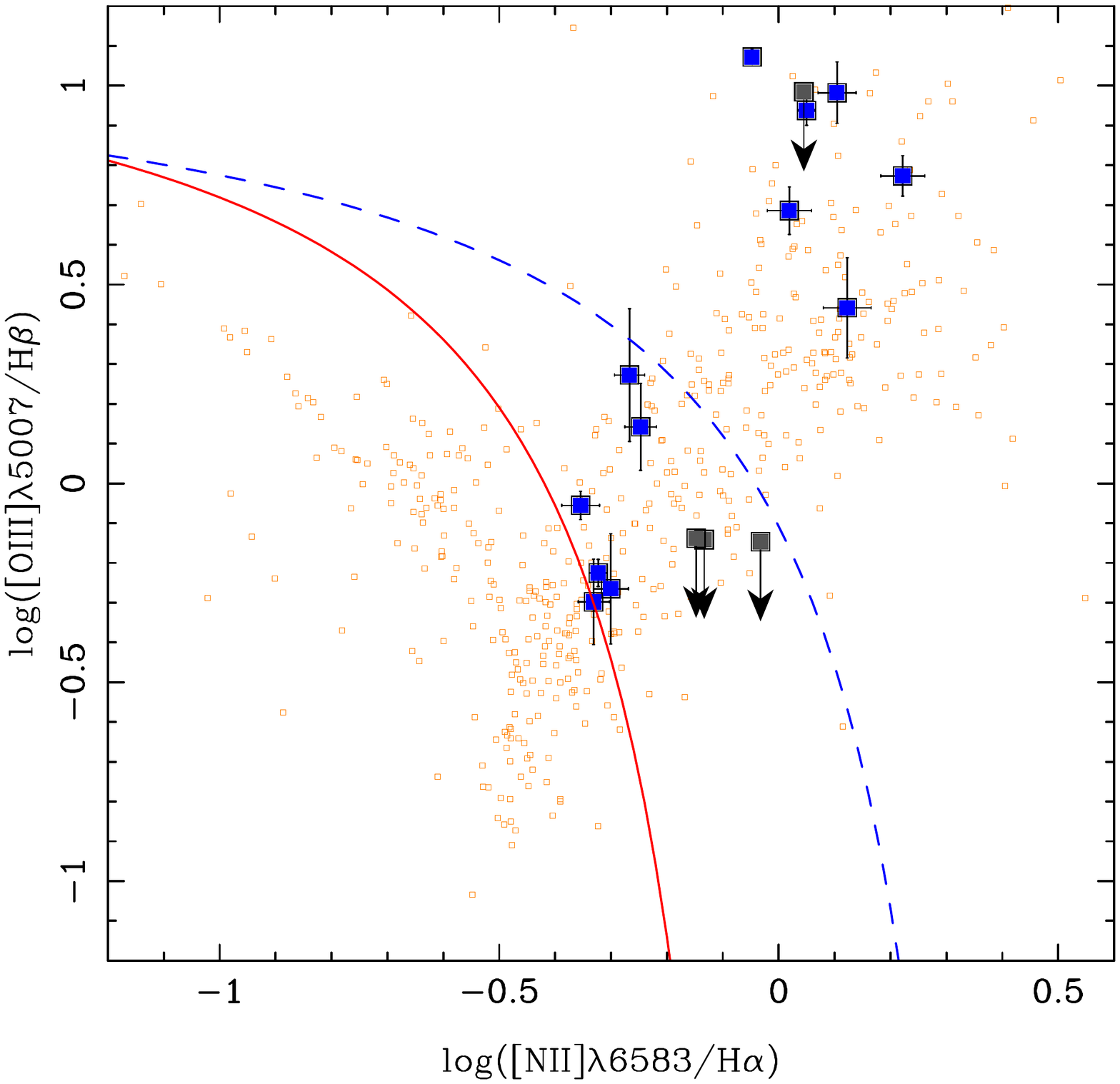}
\caption{[OIII]$\lambda$5007/H$\beta$
  vs. [NII]$\lambda$6583/H$\alpha$ diagnostic diagram for  integrated (left
  panel) and central (right panel)  spectra of the currently observed CALIFA sample. 
  Upper limits are indicated with an arrow, if only one of the two ratios could be 
  accurately determined. We also plot data derived from
  the SDSS spectra of the CALIFA mother sample (orange open squares). The
  Kauffmann et al. (2003) (blue dashed-line) and Kewley et al. (2001) (red
  solid-line) demarcation curves are usually invoked to distinguish between
  SF galaxies and AGN with an intermediate region in between.   
  \label{fig:diag}}
\end{center}
\end{figure*}

 Figure \ref{fig:ssp_int} shows a comparison of the properties derived
 for the underlying stellar populations for both spectral apertures. It
 includes the results from the analysis of the indices, i.e., the age-sensitive D4000 and
 H$\delta$ indices and the metal-sensitive [MgFe] index, 
 and the physical parameters luminosity-weighted
 age, metallicity and dust attenuation, estimated by fitting SSP templates to
 the full spectrum. The galaxy UGC11262 has been
 excluded due to its lower S/N level in the continuum due to technical
 problems during the observations. These problems affect the shape of
 the continuum and therefore the parameters derived by the fitting
 procedure, although they affect neither the stellar indices nor
 the parameters of the emission lines.

 Fig. \ref{fig:ssp_int} shows the correlations between central (within 5 arcsec)
 and total integrated indices. Red dashed lines are used to display the
 one-to-one correlation and guide the eye. For all indices we find
 tight correlations between central and total values. However, we find
 systematic offsets with respect to the one-to-one line, which indicate
 gradients: in particular, for the stellar population indices,
 \textit{i)} [MgFe] central is always significantly higher than total,
 thus indicating negative metallicity gradients in all galaxies;
 \textit{ii)} H$\delta$ is typically stronger in the integrated spectra
 than in the central ones, thus indicating that young stellar
 populations are mainly found in the outskirts of galaxies;
 \textit{iii)} D4000 has typically flatter radial distribution, with
 some indication for positive (age) gradients for the oldest galaxies
 (those with the strongest D4000), and vice versa for the younger ones.

 Similar trends are appreciated in the parameters derived from the SSP
 fitting procedure.  On average the central stellar populations are
 slightly older than the integrated ones, and they suffer a stronger or
 equal dust attenuation. The metallicity shows no clear trend, with
 values randomly distributed around the one-to-one relation. Maybe the
 selected library template, with only two extreme metallicities in the
 grid, is not good enough to sample this parameter properly, which is
 better represented by the [MgFe] index. Differences are also
 appreciated among the different galaxy types. The stellar component of
 the early-type galaxies ({\it E} or {\it S0}) is, in all the cases,
 dominated by old ($>$7 Gyr, high D4000 and low H$\delta$ values)
 and metal rich ($Z>$0.025, hight [MgFe] values) populations. In
 addition, very low dust attenuation is found in both the integrated
 and central spectra ($A_v<$0.5 mag, with a mean value of $\sim$0.2
 mag). On the other hand, the late-type galaxies show a wider variety
 of properties in their stellar populations.

 The information provided by the indices and the SSP analysis is
 complementary, but leads to similar conclusions, in general. Figure
 \ref{fig:ind_int} shows a comparison between each stellar population
 index and its corresponding parameter derived by the SSP analysis, for
 the integrated spectra (similar results are found for the central
 spectra).  As expected there is a clear correlation between the
 age-sensitive indices (D4000 and H$\delta$) and the luminosity
 weighted ages, on one hand, and a weaker trend between the [MgFe] and
 the luminosity weighted metallicity, on the other hand. The larger
 scatter in this relation is most probably due to an incorrect sampling
 of the metallicities in the over-simplistic stellar population library
 grid adopted for the current analysis.

 Although a more detailed analysis is required to understand the
 described trends, these preliminary results seem very encouraging,
 illustrating the kind of studies that can be done when applying more
 refined analysis techniques over the full CALIFA sample.

\begin{figure*}
\begin{center}
\includegraphics[angle=0,width=18.5cm,clip=true,trim=60 0 0 0]{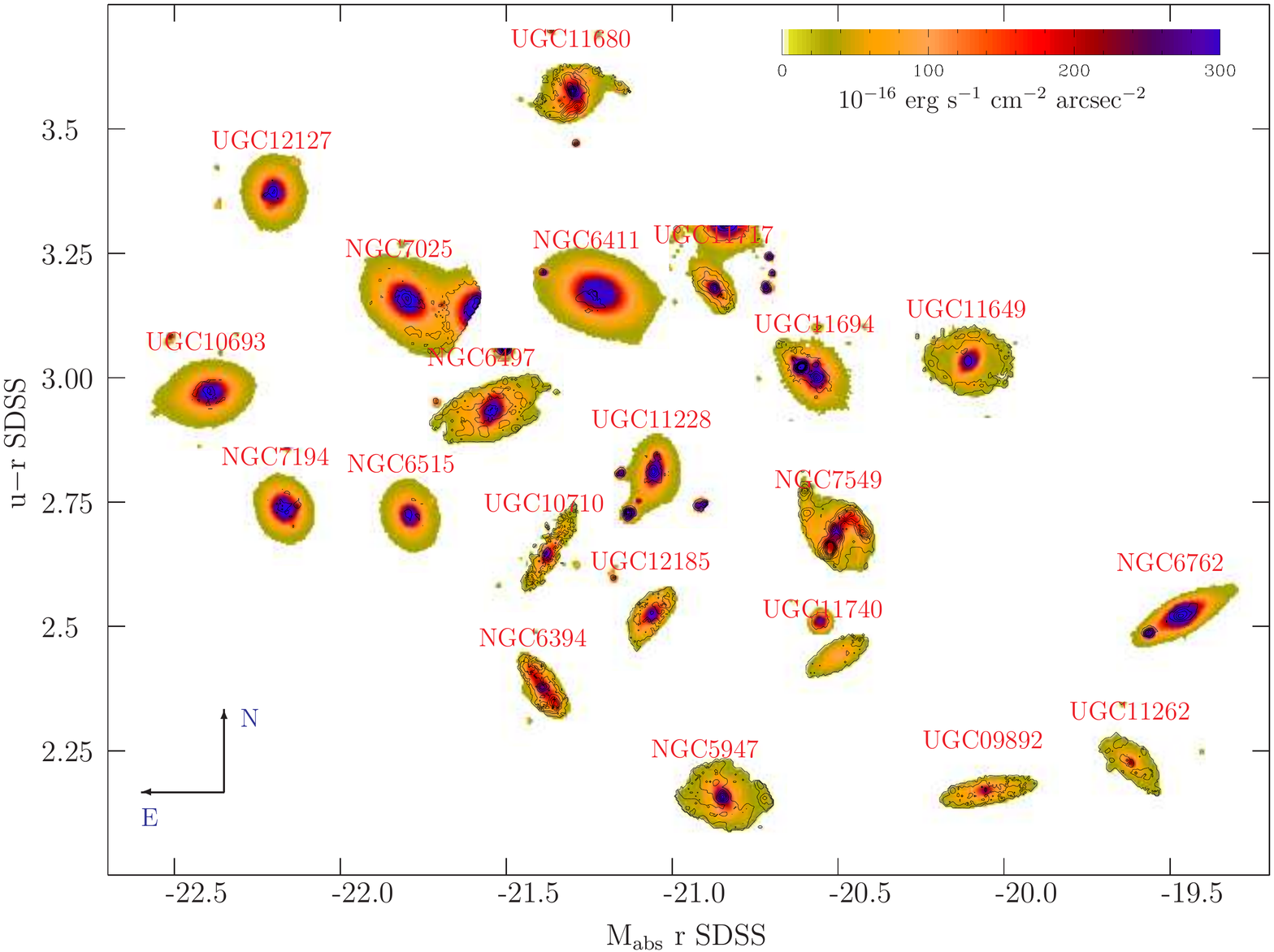}
\caption{\label{fig:fig_gas_cont} Color magnitude diagram of the galaxies
  currently observed within CALIFA, for the SDSS $u-r$ observed colors versus the
  $r$-band absolute magnitude. For each galaxy we plot in color a map for
  the continuum intensity at $\sim$6550~\AA\ \ down to a surface brightness of
  $\sim$23 mag/arcsec$^2$ AB ($>$3$\sigma$ detection limit per spaxel), together with a
  contour plot of the H$\alpha$ emission (if detected), derived by the fitting
  procedure described in the text (applied to the V500 data). The
  contours start at 0.3~\FunitsAREA, and are spaced on a logarithmic scale,
  following the equation $f=0.3+N^{1.5}$ (0.3,1.3,3.1,5.5...). The NED name of
  the galaxies has been included, together with the orientation of the map on the sky. 
  Note that some images have been shifted by up to $\pm$0.5 mag around the 
  nominal color-magnitude coordinate to avoid overlap. Nearby (projected) companions (e.g., NGC 7025) and foreground stars have not been masked (e.g., UGC 11694, UGC 11717). }
\end{center}
\end{figure*}

\subsubsection{Ionized gas properties}

{ 

Figure \ref{fig:comp_gas} shows a comparison between the emission line
fluxes and ratios of the ionized gas in the integrated and central
spectra, derived using the previously described fitting procedure
(Sec. 7.1). The figure includes the observed flux intensity of the
H$\alpha$ emission line, and different line ratios between some of the
strongest emission lines, including H$\alpha$/H$\beta$,
[OII]/H$\alpha$, [SII]6717/6731, [NII]/H$\alpha$ and
[OIII]/H$\beta$. None of the plotted values have been corrected for
dust attenuation, and only those with a S/N$>$3 have been included in
each plot. H$\alpha$ gas emission is detected in the integrated
spectra of 15 galaxies, and in the central spectra of 16. As expected,
most of the galaxies without detected ionized gas emission (i.e.{}
F$(H\alpha)<3\sigma$), either in the central or in the integrated
spectra are early-type galaxies (E and S0).

The comparison illustrates the importance of mapping the full optical
size of galaxies to derive the average properties of the ionized gas. For
example, the H$\alpha$ intensity, as an additive property, is much larger in
the integrated spectra than in the central aperture. The differences in other
non-additive properties (e.g., the H$\alpha$/H$\beta$ line ratio), illustrate
changes in the physical conditions of the gas from the inner to the outer
regions.

The plotted emission line ratios can be used to distinguish the ionizing
source of the gas associated with each region. In particular, the
[OIII]/H$\beta$ and NII/H$\alpha$ can be used as a diagnostic probe
which is almost insensitive to the dust attenuation. Figure \ref{fig:diag}
shows the classical BPT diagram \citep{baldwin81,veilleux87} for both the
integrated (left-panel) and central (right-panel) spectra for the galaxies
with detected ionized gas. We included in the figure the corresponding line
ratios estimated from the 496 SDSS spectra of the CALIFA mother sample with
clearly detected ionized gas. The diagram shows the typical butterfly
structure, with a left branch corresponding to H{\sc II} regions powered by
the ionizing radiation of young OB stars (i.e., star forming regions), while
the branch on the right side is usually attributed to ionization by an AGN. We
include  the theoretical boundary between ionization from stars (SF) and from
an AGN from \citet{kewley01} as well as the one empirically derived by
\citet{kauffmann03}. Objects are thus classified as star forming (SF),
AGN-ionized (AGN), or intermediate (SF/AGN) if they lie between the empirical
boundary and the theoretical one. Table \ref{tab:eline2} includes the result
of this classification. Note that the current classification scheme has not
considered the possibility of ionization due to post-AGB stars as suggested by
several authors  \citep[e.g.,][]{trin91,binn94,stas08,cid-fernandes10,sarzi10}. It is beyond the scope
of the current study to analyze that possibility in detail, although it may
lead to the reclassification of the ionized sources presented here. 

\begin{figure*}
\begin{center}
\includegraphics[angle=0,width=18.5cm,clip=true,trim=60 0 0 0]{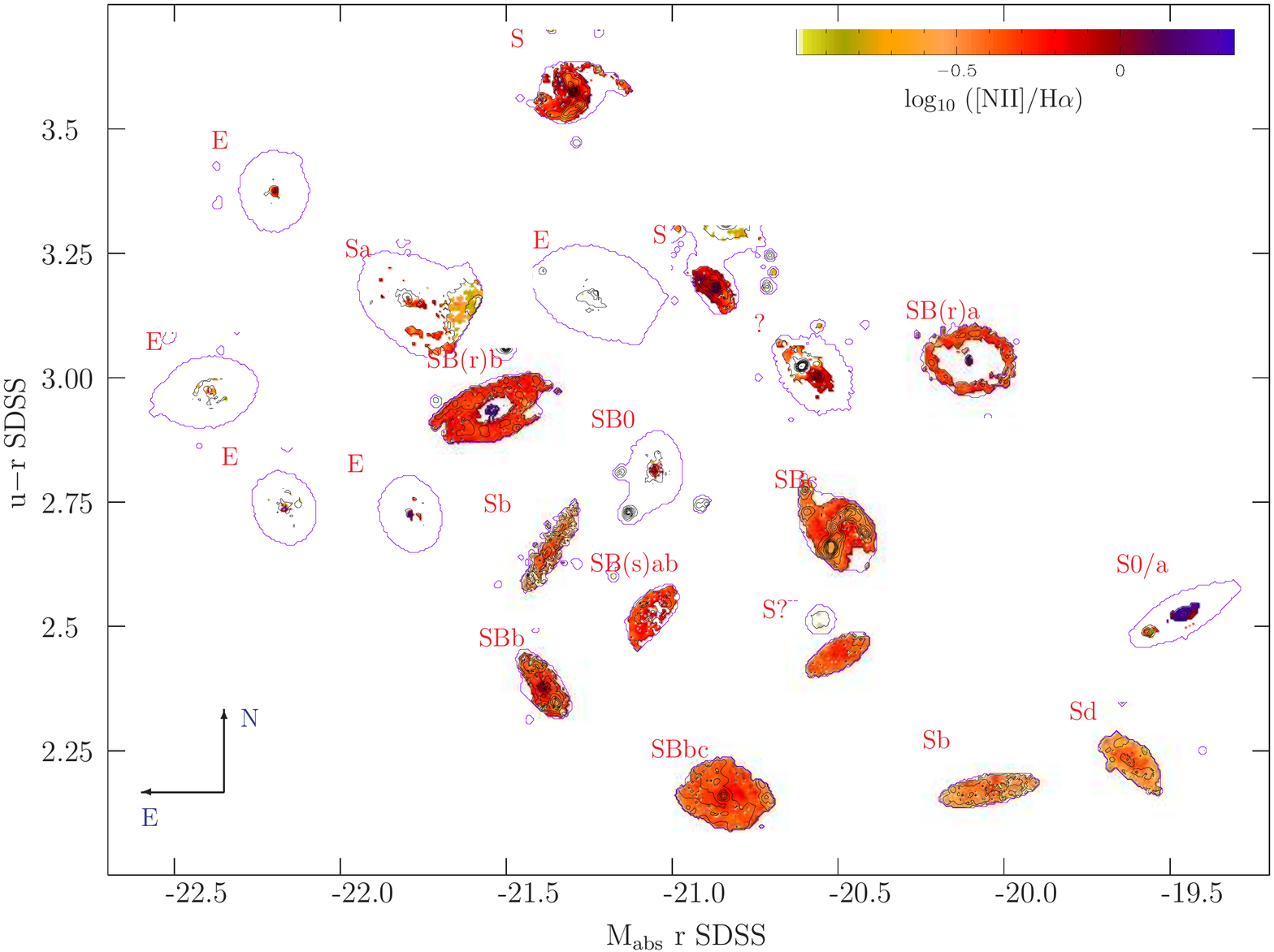}
\caption{\label{fig:fig_gas_NII} Color maps showing the distribution of the
  emission line ratios between [NII]$\lambda$6583 and H$\alpha$, in a 
  color-magnitude diagram similar to that of Figure \ref{fig:fig_gas_cont}.
  In the present case the solid contours show the intensity of
  the H$\alpha$ emission, adopting the same levels as in Figure
  \ref{fig:fig_gas_cont}. The dashed blue contours show 3 intensity levels of
  the continuum emission at $\sim$6550~\AA\  (starting at 3~\FunitsAREA). They have
  been included to indicate the physical extent of the continuum emission
  in the galaxies.}
\end{center}
\end{figure*}

For the integrated spectra with both line ratios detected (+upper limits) we find 
that 6(+1) galaxies are classified as SF, 3(+1) as AGN and 3(+1) as intermediate. 
On the other hand, based on the properties of the central spectra, 6(+1)
galaxies are classified as pure AGN and 6(+3) as intermediate. The fraction of
AGN detected in this first reduced sub-sample observed so far by CALIFA is much
larger than the one expected on the basis of the analysis of the SDSS spectra
(i.e. $\sim$1/3 of the objects), as is clearly seen in the figure. Targets
were selected from the mother sample purely based on
observability. Furthermore, we have so far not found significant differences
between the properties of the SDSS and CALIFA central spectra. Therefore,
most probably this effect is a result of small number statistics.

That the fraction of galaxies classified as SF is larger for integrated spectra than 
for central spectra is of course expected as an aperture effect. It is one of the design 
goals of CALIFA to quantify this effect.

In addition to the determination of the main ionization source, the
emission line spectrum can be used to derive properties of the ionized
gas. The dust attenuation ($A_V$) of emission lines along the line of
sight through a galaxy can be derived from the Balmer decrement,
assuming an extinction law \citep[e.g.][]{cardelli89} and comparing
the theoretical vs. observed I(H$\alpha$)/I(H$\beta$) ratio. Table
\ref{tab:eline2} lists the derived dust attenuation for each object,
for both the integrated and central spectra\footnote{Assuming
  $R_V=3.1$ and case B recombination: temperature $10^4$~K, density
  $10^2$~cm$^{-3}$; \citep{osterbrock89}}.

Extinction-corrected line fluxes can be used to derive the star
formation rate (SFR), based either on the H$\alpha$ or [OII]$3727$
line intensities. The SFRs can be derived adopting the classical
relation between this parameter and the luminosity of both emission
lines \citep{kennicutt98}.  Note that the absolute luminosities at the
distances involved in CALIFA are insensitive to the assumed cosmology
\footnote{A H$_o$=70 km s$^{-1}$ Mpc$^{-1}$ and q$_o$=0.5 was
  adopted.}.

\begin{figure*}[tb]
\begin{center}
\includegraphics[width=8cm,angle=0,clip=true,clip=true,trim=30 30 100 30]{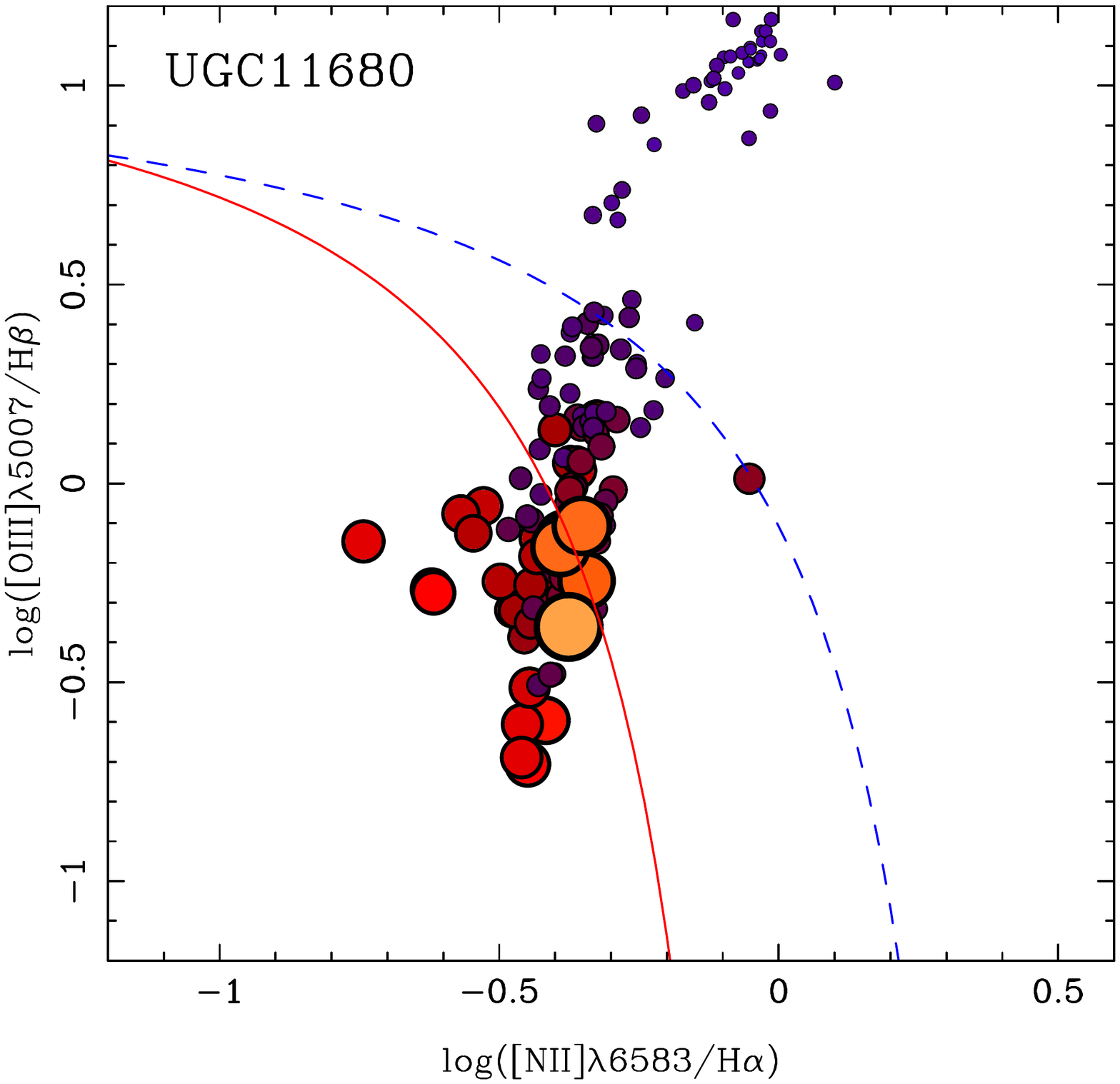}
\includegraphics[width=8cm,angle=0,clip=true,clip=true,trim=30 30 100 30]{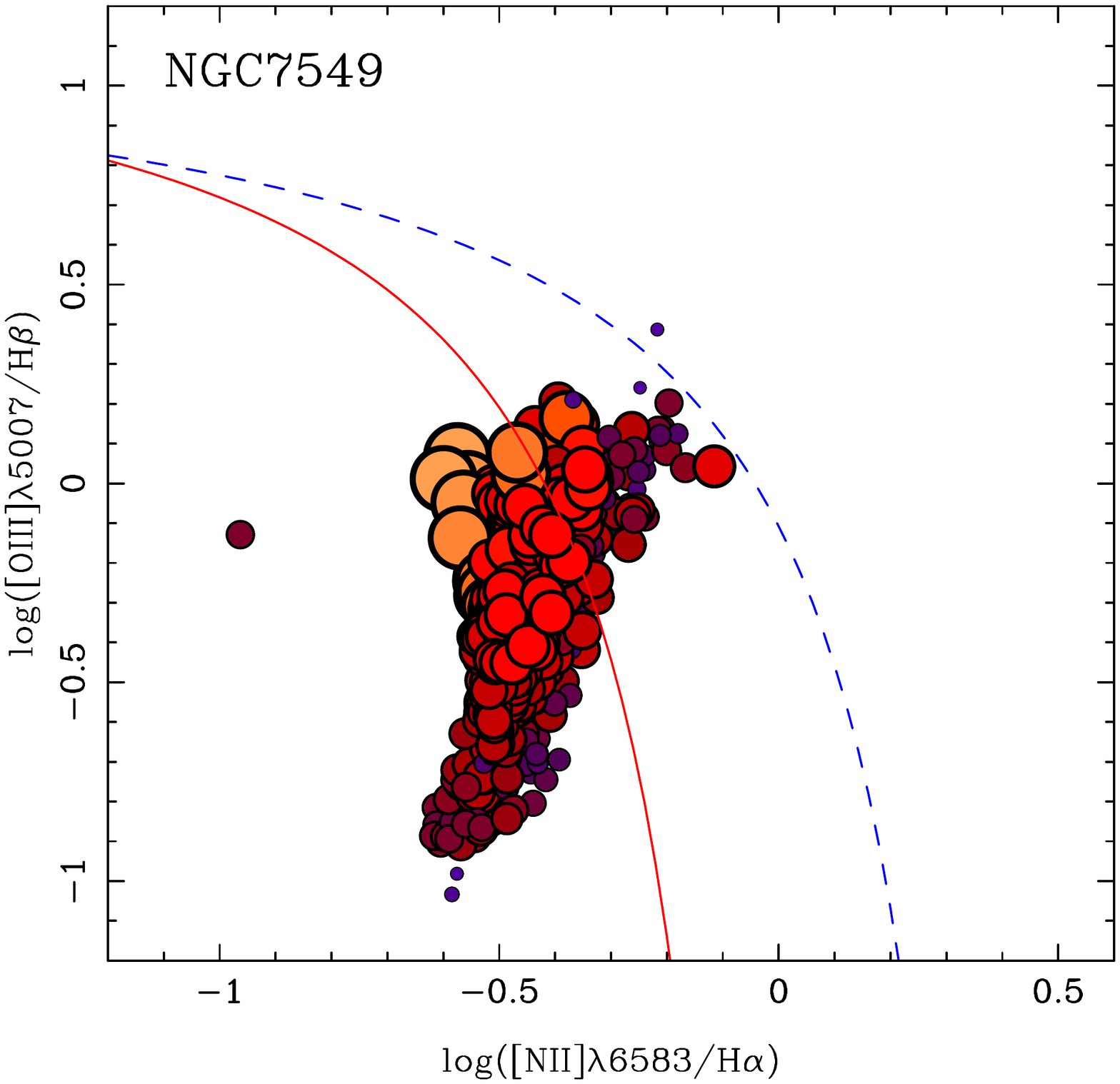}
\caption{[OIII]$\lambda$5007/H$\beta$ vs. [NII]$\lambda$6583/H$\alpha$
  diagnostic diagram for the individual spectra within the V500
  datacubes corresponding to two different objects, UGC 11680 (left
  panel) and NGC 7549 (right panel) in the currently observed CALIFA
  sample.  The solid and dashed-lines are similar to those shown in
  Figure \ref{fig:diag}. The size and colors of the symbols indicate
  the distance to the center of the galaxy, with bluer and smaller
  circles located in the inner regions and orange and larger ones
  located in the outer parts. \label{fig:diag_maps}}
\end{center}
\end{figure*}

Table \ref{tab:eline2} includes the results of
this calculation for those galaxies in which the ionization is not
dominated by an AGN. In most of the cases the SFRs derived using both
lines are consistent within the errors. In a few cases ($\sim$10\%), a
much higher SFR is found from [OII] than using H$\alpha$
(e.g. UGC12185 for the integrated spectrum, and UGC10710 for the
central one). In these cases the observed [OII]/H$\alpha$ ratio does
not seem to be abnormally high. An inspection of these spectra shows
that for such cases the correction of H$\beta$ by the underlying
stellar absorption is particularly high, which may affect the
derivation of the dust attenuation and most probably produces an
over-correction of the [OII] flux. Galaxies for which the
[OII]/H$\alpha$ ratio is higher than the average are located in the
intermediate region of the BTP diagram, between the pure SF and pure AGN
ionization regimes.

The emission lines present within the spectral range of CALIFA allow us to
examine other properties of the ionized gas, such as the hardness of the
ionization, typically characterized by the ionization parameter $U$, defined as
the ratio of the density of ionizing photons to the particle density.
This parameter is best determined using the ratios of emission lines of the
same element originating from different ionization stages. The lines available
in our spectra allow us to derive the ionization parameter from the ratio
[O\,{\sc ii}]/[O\,{\sc iii}]=$\lambda 3727/(\lambda4959+\lambda5007$), using
the relation $\log u = - 0.80 \log([O II]/[O III]) - 3.02$, after
\cite{diaz00}.
The derived ionization parameters are listed in Table \ref{tab:eline2}. The
reported values are in the higher range compared to most of the known 
HII regions \citep[e.g. the Orion Nebula,][]{sanchez07c}. The main
reason for these high values is most probably the presence of an AGN in a
substantial fraction of the galaxies of the current sample.

Gas-phase oxygen abundances can also be obtained from the emission lines
present in the CALIFA data. However, a direct oxygen abundance determination
requires the presence of temperature-sensitive auroral lines such as [O\,{\sc
  iii}]4363~\AA~ and/or [N\,{\sc ii}]5755~\AA, which are only present in the
low-to-intermediate metallicity regime and are generally much fainter than
the corresponding nebular ones, our spectra are not deep enough to detect
them, in most of the cases.
Nevertheless, we can turn to strong-line methods to derive the gas metallicity
\citep[e.g.,][]{kewley08,rafa10}, bearing in mind that these
calibrations are only valid for gas ionized by star formation. In the case of
hard ionization due to AGN, the derivation of the oxygen abundance is largely
unexplored, and only detailed photo-ionization models can be used to
quantitatively estimate it \citep[e.g.][]{storchi-bergmann98}. For the purpose
of the current exploratory study, we restrict ourselves to objects which are
not AGN-dominated and we adopt the simple O3N2 indicator \citet{pettini04}, 
$12+{\rm log (OH)}=8.73+0.32~ {\rm log_{10}}\frac{[{\rm O}\,{\sc III}]\lambda5007/{\rm H}\beta}{[{\rm N}\,{\sc II}]\lambda6583/{\rm H}\alpha}$.
The derived abundances are listed in Table \ref{tab:eline2}. They are in the
range of values expected for galaxies of this kind. No significant variation
is found between the values reported for the central and integrated spectrum.

Table \ref{tab:eline2} lists also the electron density of the ionized gas from
the [S\,{\sc ii}]6717~\AA/6731~\AA~ doublet ratio. This particular line ratio
is sensitive to the electron density only for a particular range of values
\citep{osterbrock89}. Only electron densities within this range are listed in
Table \ref{tab:eline2}. A wide range of values is found for the electron
densities, indicating different gas conditions for the different objects and
spatial regions.

We conclude from this section that the emission line spectra delivered
by CALIFA live up to the expectations in terms of their scientific
usability. Future analyses should be performed, improving the
subtraction of the underlying stellar population, the treatment of
dust attenuation and the derivation of the oxygen abundance.

}


\begin{figure*}
\begin{center}
\includegraphics[angle=0,width=18.5cm,clip=true,trim=60 0 0 0]{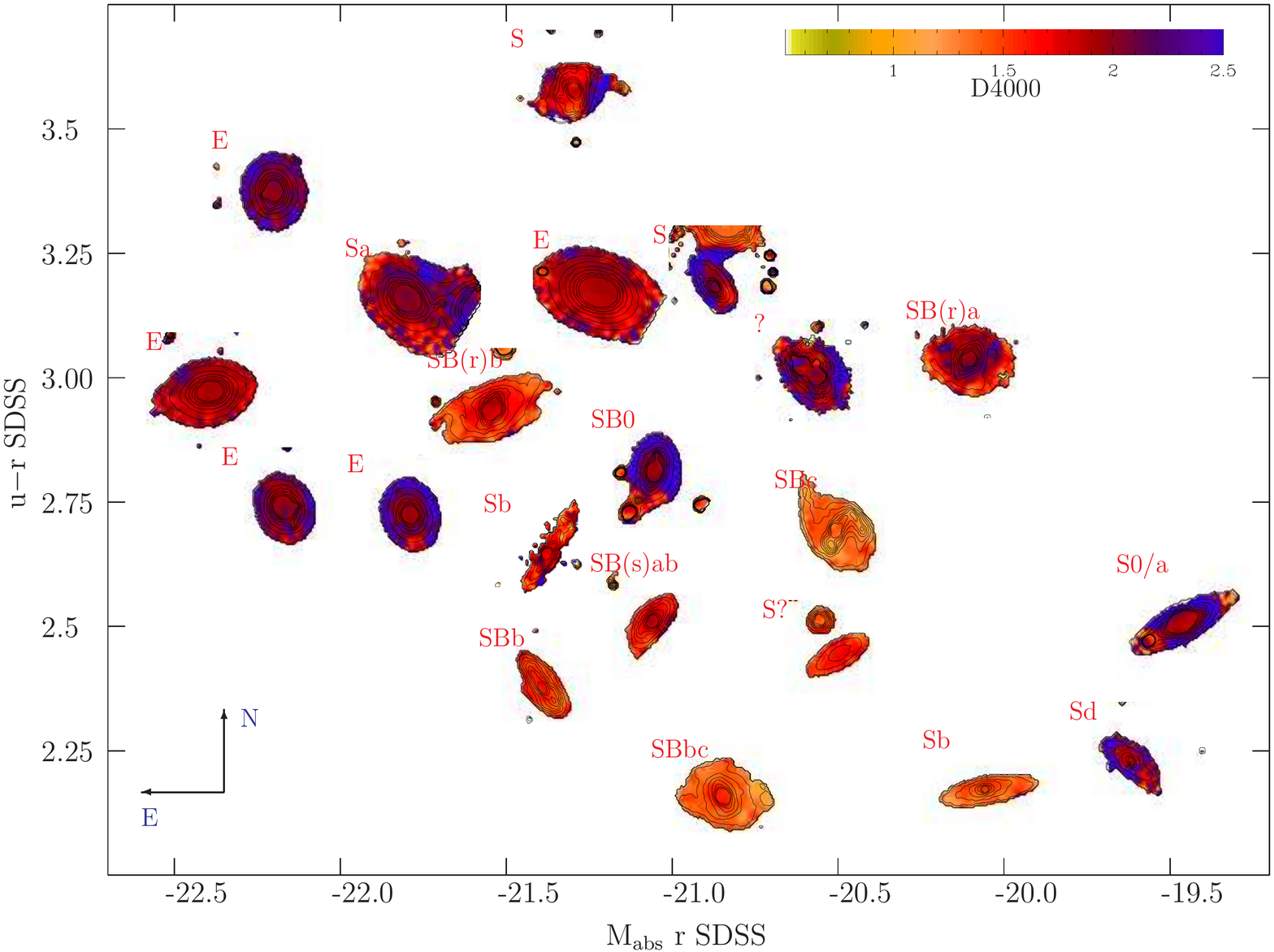}
\caption{\label{fig:fig_age} Color maps showing the distribution of
  the D4000 stellar population index in a color-magnitude diagram
  similar to that of Figure \ref{fig:fig_gas_cont}. }
\end{center}
\end{figure*}

\subsection{Spatial distribution of the spectroscopic properties}

In the previous section we have described the results of the analysis of single
spectra, selected from two different apertures. A further design goal of CALIFA 
is, however, the analysis and comparison of the spatial distribution of those
properties. We now turn towards showing the power of CALIFA in this domain.

We have applied the fitting procedure described in previous sections to each
individual spectrum of each datacube, for both the V500 and V1200
datasets.  For each spaxel of each datacube we thus recover the corresponding
spectral information, and it is therefore possible to create two-dimensional
maps for each of the derived parameters that describe the stellar populations
(luminosity weighted ages, metallicities, dust attenuation, and/or coefficients
of the mix of the adopted SSP templates), and the ionized gas (line 
intensities, dust attenuation, diagnostic line ratios, ionization strength,
oxygen abundance, electron density). In addition, the
two-dimensional kinematic structure for both the ionized gas and the stellar
populations are derived. We will illustrate the results of this analysis in the next figures.

\begin{figure*}[tb]
\begin{center}
\includegraphics[width=7.5cm,angle=0,clip=true,clip=true,trim=30 30 100 30]{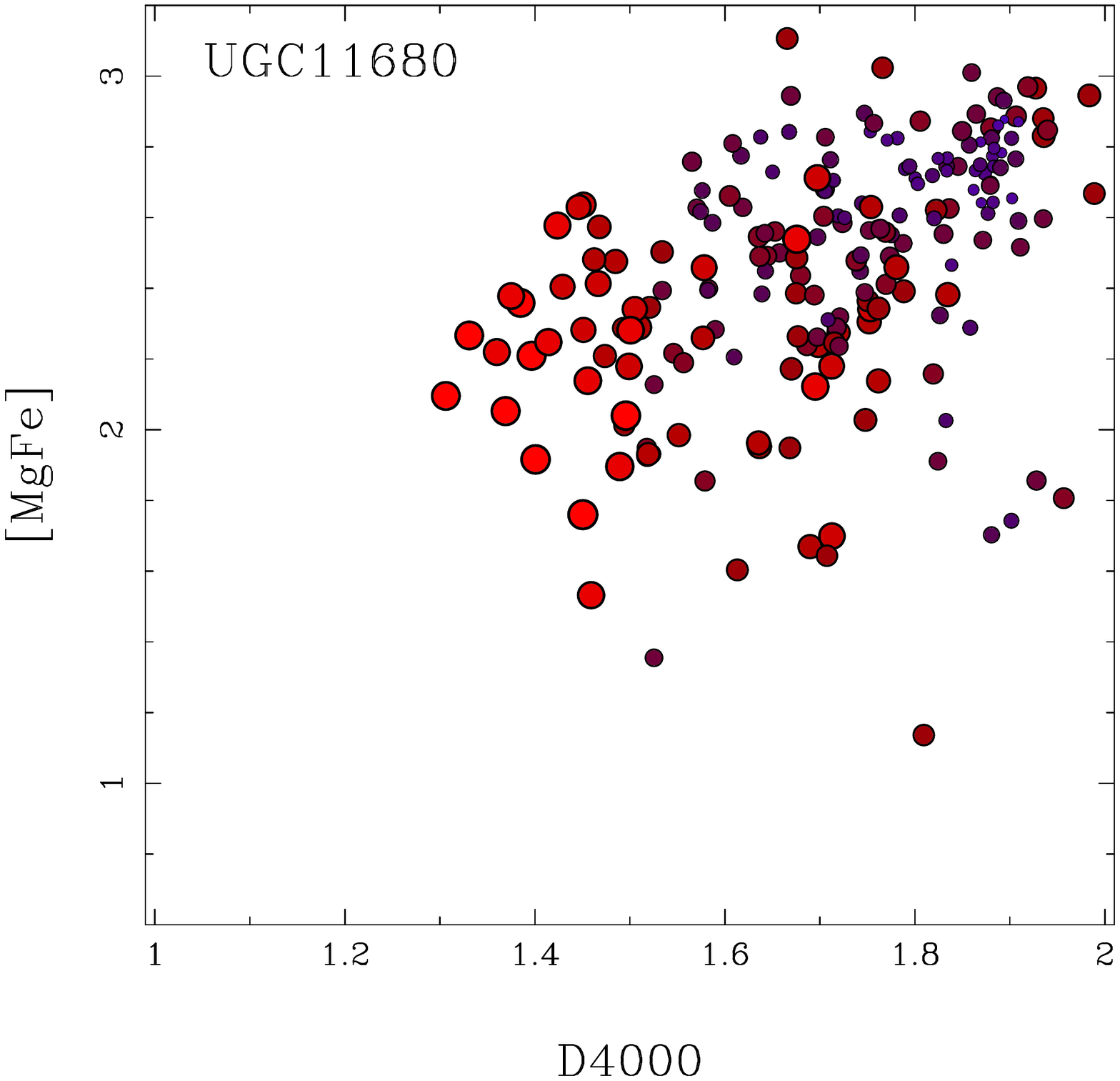}
\includegraphics[width=7.5cm,angle=0,clip=true,clip=true,trim=30 30 100 30]{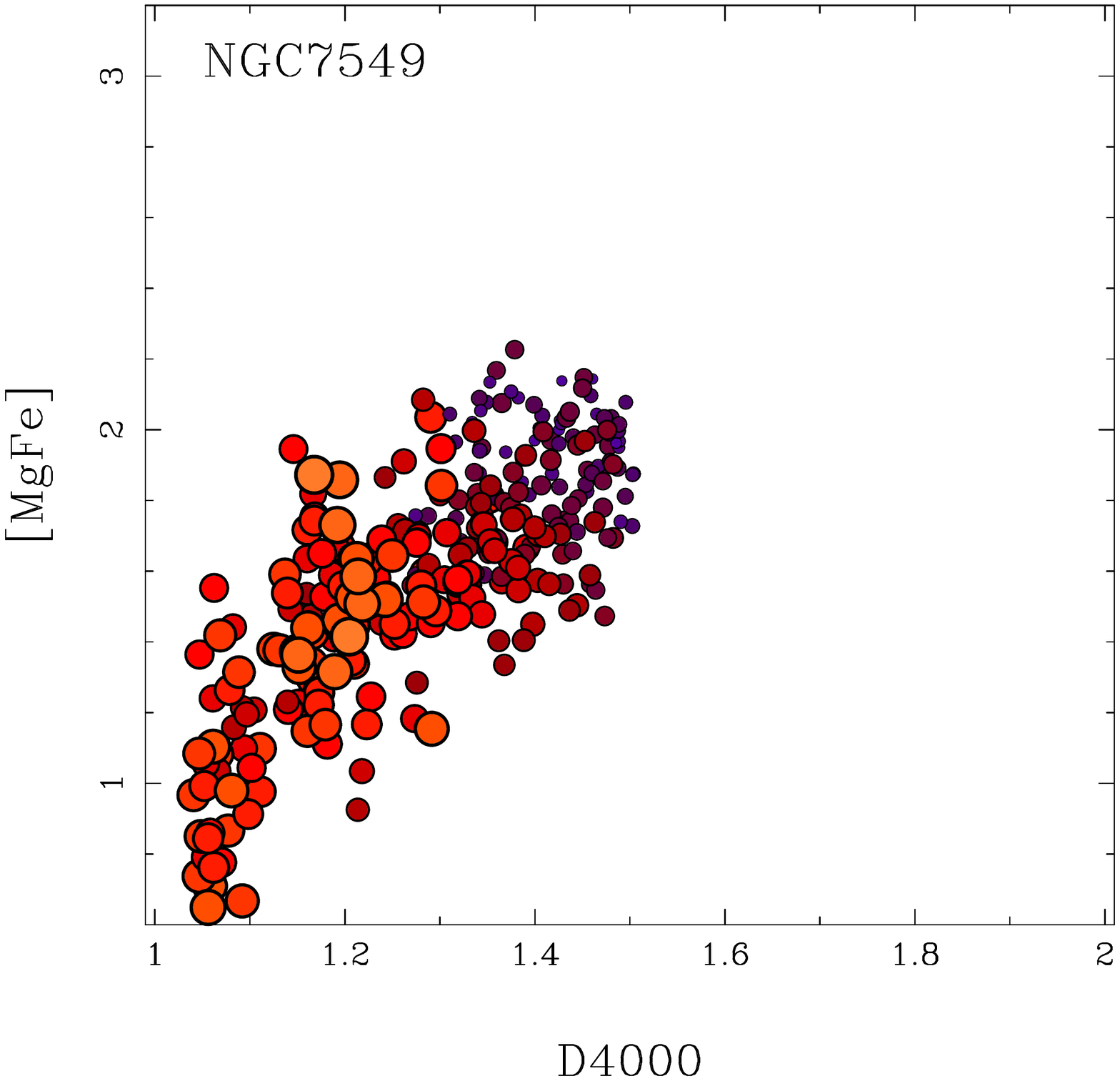}
\caption{Distribution of the metallicity sensitive [MgFe] stellar index and age sensitive D4000 index across the field of view for the two galaxies shown in Figure \ref{fig:diag_maps}. The colors and sizes of the circles indicate the distance from the center, defined as the intensity peak in the V-band. The bluer/smaller circles indicate the central regions and the orange/larger symbols the outer ones. The figure illustrates in more detail the change in the stellar populations described in Figure \ref{fig:fig_age}.  \label{fig:index}}
\end{center}
\end{figure*}

Figure \ref{fig:fig_gas_cont} shows the color-magnitude diagram for
the galaxies of the analyzed subsample, based on the SDSS $u-r$
observed colors and the $r$-band absolute magnitude. For each galaxy,
we plot a color-map for the continuum intensity at
$\sim$6550~\AA\ \ down to a surface brightness of $\sim$23
mag/arcsec$^2$ AB ($>$3$\sigma$ detection limit per spaxel), together
with a contour plot of the H$\alpha$ emission (if detected). The names
of the galaxies have been indetified. As expected, the brightest and
reddest galaxies show { the least} ionized gas emission. These dry or almost dry galaxies are mostly morphological
early-types (E and S0). Some of them show some emission lines in the
central regions. However, the detected emission in H$\alpha$ is below
the conservative detection limit adopted in Table \ref{tab:eline2}.  A
more detailed decoupling analysis between the emission lines and the
underlying stellar population is needed to ensure the reliability of
this detection, in particular of E and S0 galaxies (Kehrig et al., ).

{ A greater morphological diversity and a larger range of colors is
  found among the galaxies with ionized gas.}  There are the typical
face-on (e.g. NGC 5947, NGC 6497) and edge-on (e.g. UGC 10710, UGC
09892) spiral galaxies; galaxies with strong bars (e.g. NGC 6392); low
surface brightness galaxies (e.g. UGC 11740), and galaxies with
evidence of recent interactions (e.g.  UGC 11680 and NGC 7549). The
two reddest galaxies with gas (UGC 11680 and UGC 11717) both harbor an
AGN and have strong dust attenuation in their integrated spectrum (as
derived from the ionized gas, Table \ref{tab:eline2}). The spatial
distribution of the ionized gas also shows { a} wide diversity. In
some galaxies, the ionized gas is concentrated in the central regions
(e.g. NGC 6762, dominated by a ionization different than star
formation), while in other objects the ionized gas follows the spiral
arms (UGC 11680, NGC 5947 or NGC 7549), or is located in a ring (UGC
11649). The ionization source in most of these galaxies is purely star
formation, based on their integrated spectra. The most distorted
H$\alpha$ morphology is shown by NGC 7549, which is also the galaxy
with the highest SFR of all those dominated by pure star formation in
the integrated spectrum. The SDSS image of this object shows the
typical distorted morphology produced by an interaction and/or merging
process.

The ionization conditions change considerably not only from galaxy to
galaxy, but also within each galaxy. Figure \ref{fig:fig_gas_NII}
illustrates these changes, by showing the spatial distribution of the
[NII]$\lambda$6583/H$\alpha$ emission line ratio, for each galaxy. In
Figure \ref{fig:diag} it was shown how this line ratio, in combination
with [OIII]/H$\beta$, can be used to derive the main source of the
ionization. In most of the ionization models, values of
log([NII]/H$\alpha$)$>$1 cannot be produced by star formation, even in
the case of very strong starbursts.  Figure \ref{fig:fig_gas_NII}
shows that such values are only found in the central regions of some
objects: NGC 6762, NGC 6497 and UGC 11649. In all these cases the
ionization is clearly dominated by an AGN (Table
\ref{tab:eline2}). However, while in some the ionization is
concentrated in the central region (NGC 6762), in other cases there
are clear extended emission regions most problably ionized by the AGN
(NGC 6497 and UGC 11649).  Similar comparative analyses between
different galaxy types and within galaxy classes can be performed
using any of the parameters that characterize the ionized gas, like
those listed in Table \ref{tab:eline2}.

Figure \ref{fig:diag_maps} shows two diagnostic diagrams, similar to
the ones presented in Fig. \ref{fig:diag}, corresponding to the
individual spectra within the V500 datacubes of two objects, UGC 11680
and NGC 7549. These two diagrams illustrate how the ionization source
changes with location, as already illustrated in
Fig. \ref{fig:fig_gas_NII}. { The track along the diagnostic
  diagram due to pixel-to-pixel variations in a single galaxy is an
  important tool to understand the nature of the ionization
  \citep[e.g.,][]{sharp10}. }  In the case of UGC 11680, there are
regions corresponding to all three regimes, dominated by
star-formation, intermediate and dominated by an AGN. On the other
hand, the gas ionization in NGC 7549 is clearly dominated by
star-formation, with a tail of $\sim$10\%\ of the spectra showing
possible intermediate ionization \citep[e.g.,][]{alonso10}.  In future
studies these diagrams will be used to separate the different
ionization sources in the spatial domain, trying to correlate them
with other local properties of the galaxy.

Figure \ref{fig:fig_age} shows maps of the D4000 stellar
population index, sensitive to the age of the stellar component (e.g.,
Fig.\ref{fig:ind_int}), in a similar color-magnitude diagram to
Fig. \ref{fig:fig_gas_cont}. Radial gradients are readily observed,
which in some cases highlight abrupt transitions between different
structural components of the galaxies (e.g. bulge, bar, disc). As expected, the morphologically
early-type galaxies (E/S0), bright, red and dry, are dominated by old
stellar populations (D4000$>$2), with little variation within their
optical extent. Late-type galaxies, on the other hand, show a wider
variations of the age of the stellar population, both from galaxy to galaxy and
within the same galaxy.  In most cases there is a gradient in the
ages, such that the central region is older than the outer parts. This
is most evident for the face-on spirals.

A more detail analysis of the variation of the stellar populations
across the field requires the comparison of either the distribution of
the luminosity-weighted ages and metallicities or the corresponding
indices. Figure \ref{fig:index} shows the distribution of the
metallicity-sensitive [MgFe] index and the age-sensitive D4000 index,
derived spaxel-to-spaxel across the field of view of the two galaxies
shown in Fig. \ref{fig:diag_maps}. Only those spaxels with an intensity above 3~\FunitsSSAREA~
(S/N$\sim$10) at $\sim$5000\AA~ have been included in the plots. Both
figures illustrate in more detail the change in the stellar
populations from the inner to the outer regions.

It is interesting to note that the stellar populations of NGC 7549 show
a clear gradient, with the inner regions dominated by old and metal
rich stars, and the outer ones dominated by younger and metal poor
ones. This is the typical trend expected in an inside-out secular
evolution of a disk-dominated galaxy. On the other hand, this gradient
is less evident in UGC 11680, where there is a more diverse mix of
stellar populations at all radii. 
While both galaxies are face-on spirals with evidence of a recent
interaction, the gas ionization is dominated by 
star-formation in most of the extension of NGC 7549,
while UGC 11680 hosts an AGN.  Whether this is
related to the observed differences in the stellar populations is one
of the main goals of the CALIFA survey, i.e., to study the
inter-relation of the spatially resolved spectroscopic properties of
different types of galaxies.


The results presented in this subsection illustrate the potential of
the CALIFA data to study the spatially resolved properties of the
stellar population of galaxies. We will analyze the implication of
these results on our current understanding of the evolution of
galaxies in forthcoming articles.


\begin{figure*}
\begin{center}
\includegraphics[angle=0,width=18.5cm,clip=true,trim=60 0 0 0]{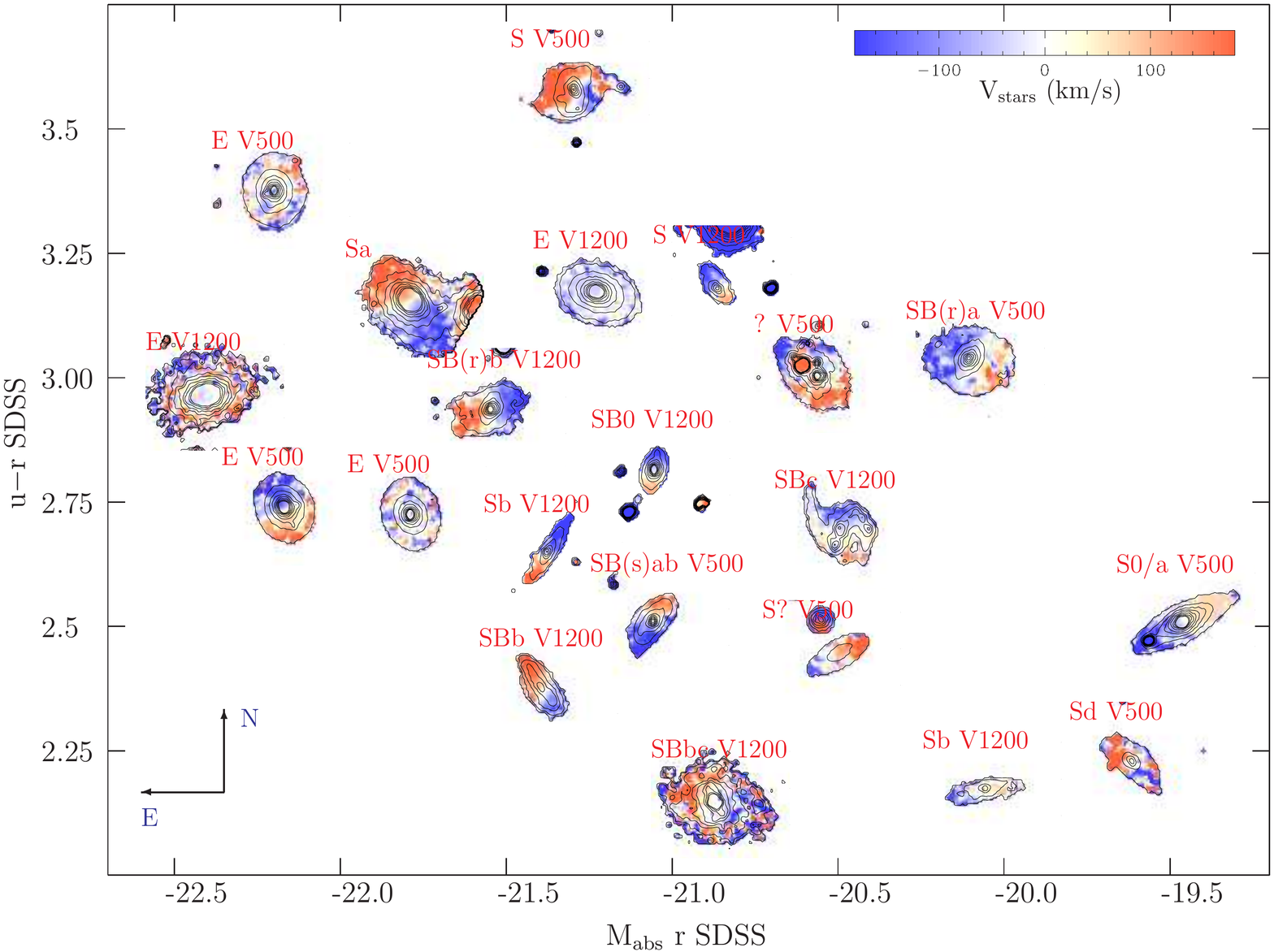}
\caption{\label{fig:fig_stars_kin} Color maps showing the velocity field of the
  stellar component from the V1200-grating dataset in a 
  color-magnitude diagram similar to that of Figure \ref{fig:fig_gas_cont}. 
  For those galaxies for which the 3$\sigma$ limiting
  surface brightness per spaxel was brighter than 22 mag/arcsec$^2$, the 
  velocity map was derived from V500 dataset. The contours show the intensity 
  level of the continuum emission at $\sim$4000~\AA, for the same levels as in 
  Figure \ref{fig:fig_gas_cont}. }
\end{center}
\end{figure*}

\begin{figure*}
\begin{center}
\includegraphics[angle=0,width=18.5cm,clip=true,trim=60 0 0 0]{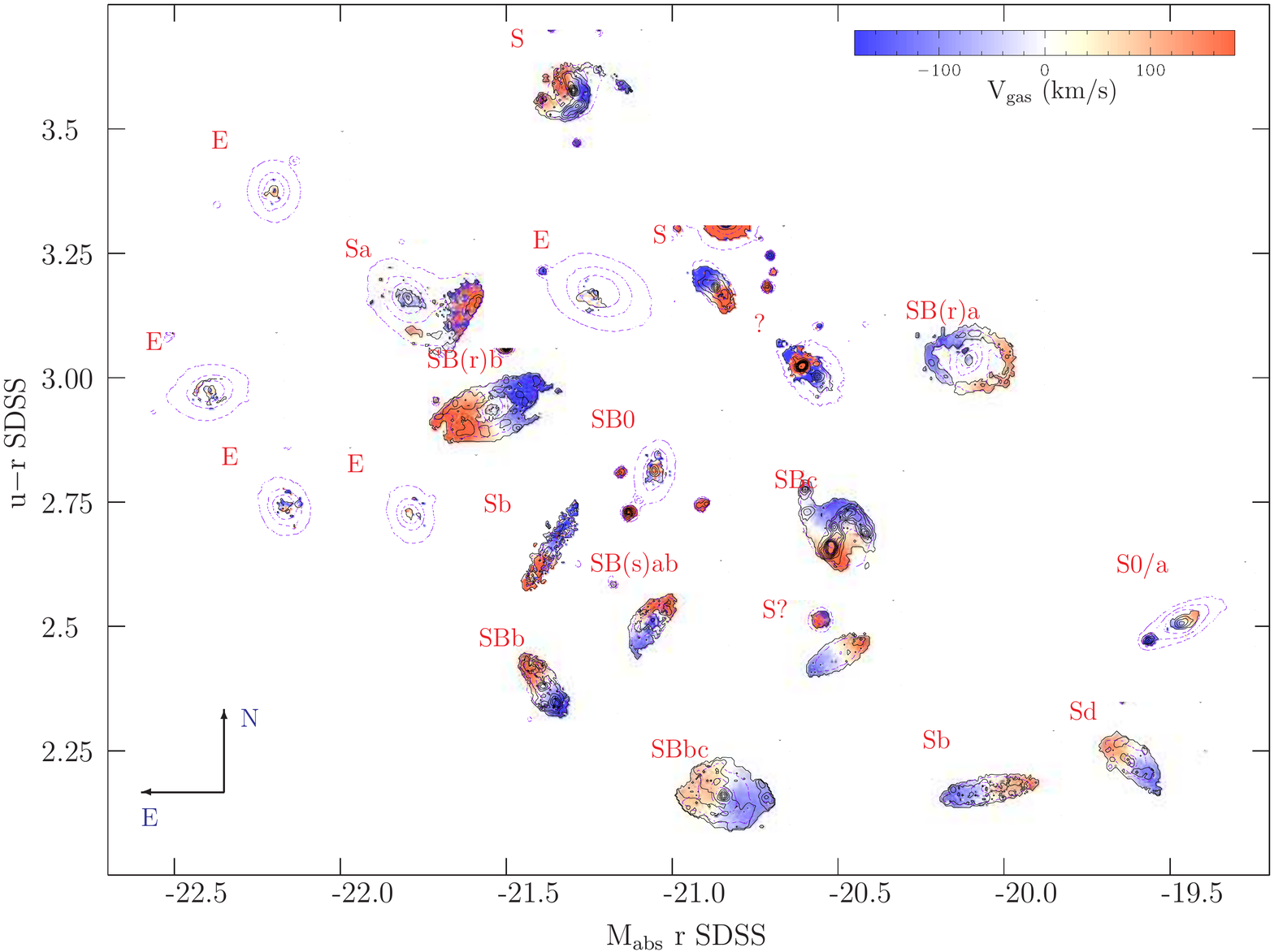}
\caption{\label{fig:fig_gas_kin} Color maps showing the velocity field  of the  ionized gas derived from 
the H$\alpha$ emission.  The contours are similar to those shown in Figure \ref{fig:fig_gas_NII}.}
\end{center}
\end{figure*}

\subsection{Kinematic maps}

The stellar kinematics were derived from
the V1200 dataset in the wavelength range between 3850 and
4600~\AA\ through the stellar population fitting procedure described in
Section \ref{s:continuum}. Only the V1200 datacubes with a 3$\sigma$
surface brightness detection limit above 22 mag/arcsec$^2$ are used. For the remaining objects, we performed a
similar analysis from the corresponding V500 data, restricted to a
similar wavelength range. The accuracy in the derivation of the
stellar kinematics is expected to be better for the V1200 data than
for the V500, in particular in the analysis of the velocity
dispersion.  

Figures \ref{fig:fig_stars_kin} and \ref{fig:fig_gas_kin} show the
velocity maps derived for the stellar population and the ionized gas,
respectively, for each individual galaxy within the current dataset.
{ These velocity maps were derived based on spaxel-to-spaxel analysis,
  without adopting any optimal binning. A proper binning will increase
  the accuracy of the kinematic analysis, as described below.}  The
velocity maps are labelled with the corresponding grating name of the
data used to derive them. Similar diagrams can be constructed for the
velocity dispersion maps. Combined these diagrams allow us to classify the
galaxies in terms of ordered vs. random motions by identifying
non-regular motions and disturbances.  For the gas kinematics we have
chosen here to present the H$\alpha$ kinematics, since this emission
line is one of the strongest. No significant differences are found
when using other strong emission lines.

Obviously, the gas kinematics can only be derived in those regions (or
objects) with detected emission lines. For those galaxies, mostly
late-type objects, the kinematics of both the gas and the stars are
dominated by rotation, with an amplitude of $\pm$150 km
s$^{-1}$. There are no significant differences between the gas and stellar
kinematics, at least for the objects observed so far, { for those galaxies for which the stellar and gas kinematics is sampled in the same physical regions: i.e., regions with high S/N ionized gas and stellar continuum. In a few cases (e.g., NGC6497 and UGC1164), we detect high S/N ionized gas in regions with low intensity continuum. For those galaxies we do not have information of the stellar kinematics derived spaxel-to-spaxel in those regions where we have derived the ionized gas one.}
In these cases the gas kinematics fits well with the
expected values from the extrapolation of the stellar kinematics to { these} outer
regions.  NGC 7025 is a special case because it is the only morphological late
galaxy without detected gas emission (so far).  This galaxy,
apparently dry, bright and red, shows a clear rotational pattern,
consistent with its morphological classification.

All galaxies without rotation are morphologically elliptical galaxies:
UGC 12127, UGC 10693, NGC 6411 and NGC 6515 (4 out of the { 5} elliptical
galaxies in the current sample). We have no gas kinematics for these,
since no emission lines were detected in these objects. On the other
hand, their stellar kinematics show a patchy structure, consistent
with a flat distribution, without significant gradients. Although
rotation may become apparent through a more detailed analysis {, making
  use of spatial binning approaches}, we detect none for the moment.

On the other hand, the remaining early-type galaxies, NGC 7194 (E),
UGC 11228(SB0) and NGC 6762(S0/a), show significant rotation. Emission
lines were detected only in NGC 6762 and only in the central region,
with line ratios consistent with the presence of an AGN. Particularly
interesting is the case of NGC 7194, an object that does not present
any morphological signature of a disk. When compared to other galaxies
classified as E, e.g. NGC 6515, both objects have similar morphologies,
colors, and luminosities. The continuum emission of both galaxies is
dominated by old ($>$8 Gyr) and super-solar ($Z\sim$0.028) stellar
populations, with small or no dust attenuation($A_V<$0.1
mag). Despite all these similarities, both objects clearly
show a different kinematic behavior. { However, a full analysis of the
  velocity and the velocity dispersion maps is required to disentangle
  the real nature of the kinematic structure of these objects.}

\begin{figure*}
\begin{center}
\includegraphics[angle=0,width=14cm,clip=true,trim=30 30 30 30]{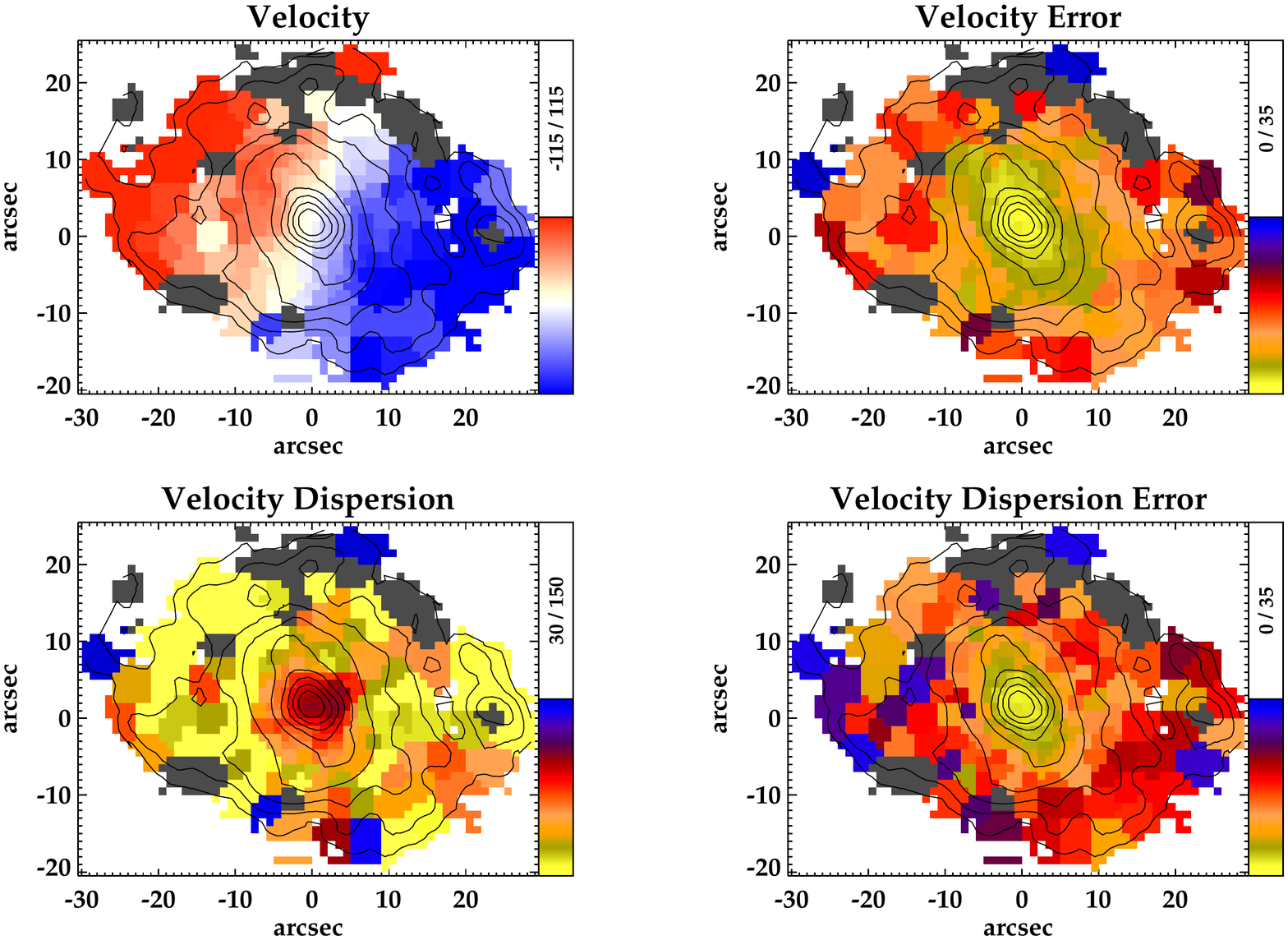}
\caption{ Results from the stellar kinematics analysis. From left-top
  to right-bottom the binned maps corresponding to the derived
  stellar velocity, its error, the stellar velocity dispersion and its
  corresponding error, all in units of km~s$^{-1}$. Values
  with errors larger than 35 km~s$^{-1}$ have been masked-out, and
  indicated with grey-colors.  \label{fig:kin_NGC5947}}
\end{center}
\end{figure*}

{ We use the high S/N integrated and central spectra described above
  (Sec. \ref{cen_total}, S/N$>$40) to estimate the accuracy in the
  derivation of the stellar velocity. First, each spectrum was
  split into consecutive, non overlapping wavelength ranges of
  200\AA. Then, each sub-spectrum was fit with a multi-SSP, adopting
  the procedure described in Sec. \ref{s:continuum}. The stellar
  systemic velocity derived for each wavelength range is recorded. The
  typical standard deviations around the central values are $\sim$5 km
  $^{-1}$ for the V1200 data, and $\sim$10 km $^{-1}$ for the V500
  one. Taking into account that 200\AA\ is only a short wavelength
  range, these values agree with the expectations based on the
  accuracy of the wavelength calibration (Sec. \ref{emp_res}). 

  This estimation of the accuracy of the kinematic properties may be
  misleading, since the S/N of the individual spaxels is lower than
  that of the integrated and central spectra. To derive maps of the
  stellar kinematics, including the mean velocity and velocity
  dispersion we need a minimal S/N per spectrum, which requires the
  adoption of a spatial binning technique \cite[e.g.][]{emsellem04}. First, we
  apply an {\it a priori}  cut threshold in the signal-to-noise
  per spaxel of $S/N>$3. Next, we perform a Voronoi binning following
  \cite{capp03}. Different experiments have been performed to derive
  the optimal limit of the S/N goal in the binned data, regarding the
  accuracy of the derived parameters. For the velocity and velocity
  dispersion a compromise has been adopted between maximizing the S/N
  per bin and minimizing the size of the final bins: i.e., keeping as
  much spatial information as possible.

  Our tests indicate that to achieve a typical accuracy of $\sim$5
  km~s$^{-1}$ in the mean velocity and a typical accuracy of $\sim$15
  km~s$^{-1}$ in the velocity dispersion, for the V1200 data, a final
  S/N per bin of $\sim$20 is required. With this S/N requirement, the
  typical bin has a projected size of a few arcseconds. Figure
  \ref{fig:kin_NGC5947} illustrates the result of these experiments,
  showing, for the velocity and velocity dispersion maps, together
  with the estimated errors for both parameters based on an extensive
  Monte Carlo simulation for a particular target. At
  $\sim$20-30$\arcsec$ from the center of the galaxy, there is a rise
  in the error to $\sim$30-40 km~s$^{-1}$. This effect is stronger in
  the distribution of velocity dispersions. This is due to the slight
  decrease of the S/N per bin and the fact that in the outer regions
  the velocity dispersion is smaller (in general), and therefore
  harder to measure. This effect is shown in the simulations presented
  by \cite{marmol-queralto11}, for the CALIFA feasibility studies.

  Similar results are derived for all the galaxies observed so-far, as
  expected due to the sample selection and survey strategy which
  ensures a similar S/N pattern across the field for the different
  targets. A more detailed report on the accuracy of the stellar
  kinematics will be described in forthcoming papers. }

\subsection{Kinematics Analysis}

As indicated in the introduction one of the goals of the current
survey is to use the stars and gas kinematics to derive physical
properties of the galaxies, such as dynamical mass. This requires a
detailed modelling of the velocity curves (and dispersion maps) of
each particular galaxy, which is outwith the scope of the current
article. To probe the quality of the current data to perform these
proposed studies a simple modelling of the velocity is performed.
We used the maps shown in Figure \ref{fig:fig_stars_kin}, trying to recover the
asymptotic rotation speed (which is related to the dynamical mass),
for the disk-galaxy NGC 5947. 

For this study the observed rotational curve was
de-projected. De-projection requires the position of the kinematics
axis and the inclination angle to be known. The kinematic center was
fit by eye using the continuum and H$\alpha$ intensity maps, and the
axes were adjusted to give the most accurate rotation curve, i.e.,
symmetrical and with singular velocity at the center. The inclination
angle was found from the 2MASS K-band axis ratios { assuming an infinitesimally thin circular disk.}

\begin{figure*}[tb]
\begin{center}
\includegraphics[width=6cm,angle=0,clip=true,clip=true,trim=40 10 80 30]{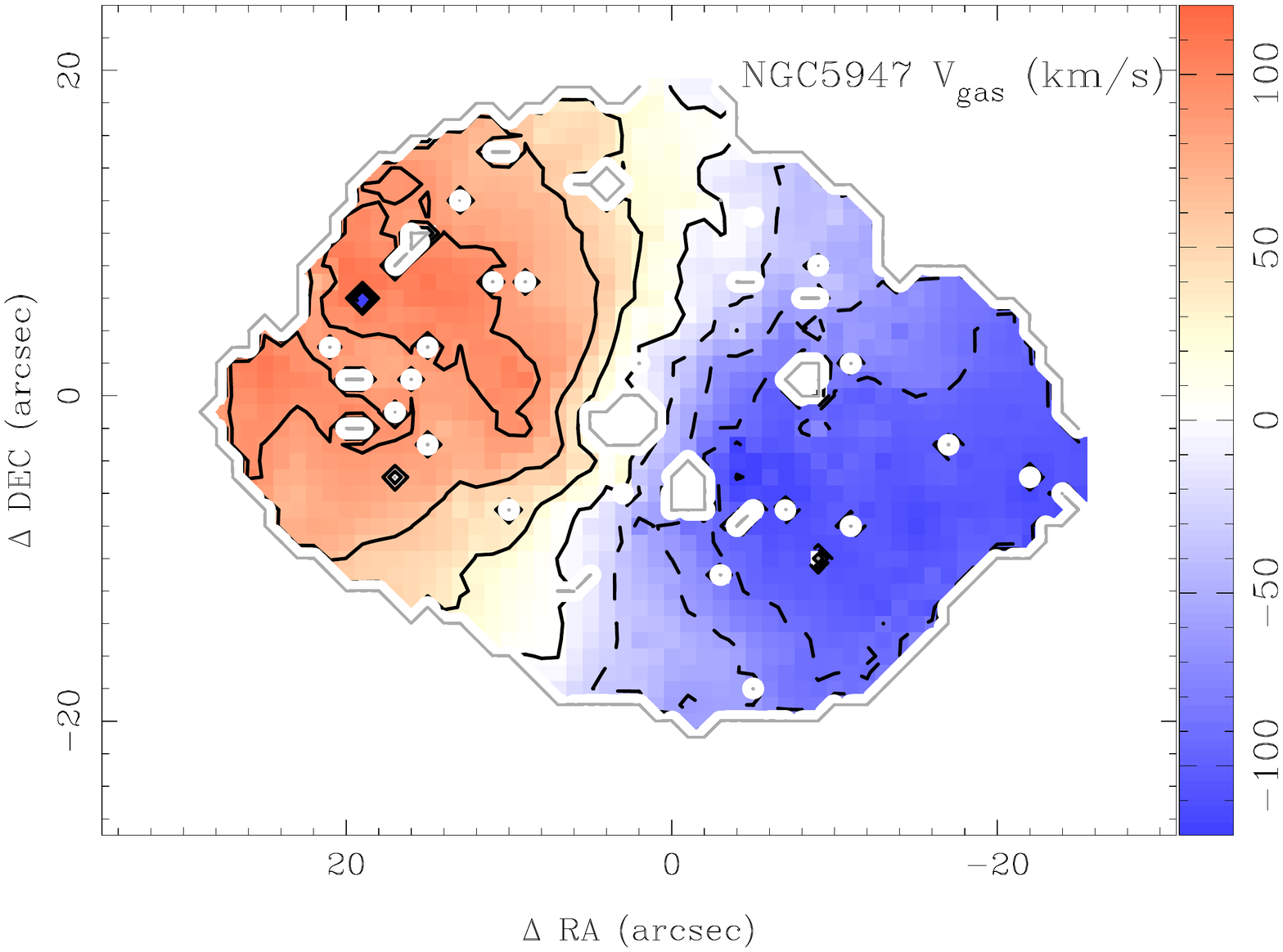}
\includegraphics[width=6cm,angle=0,clip=true,clip=true,trim=40 10 80 30]{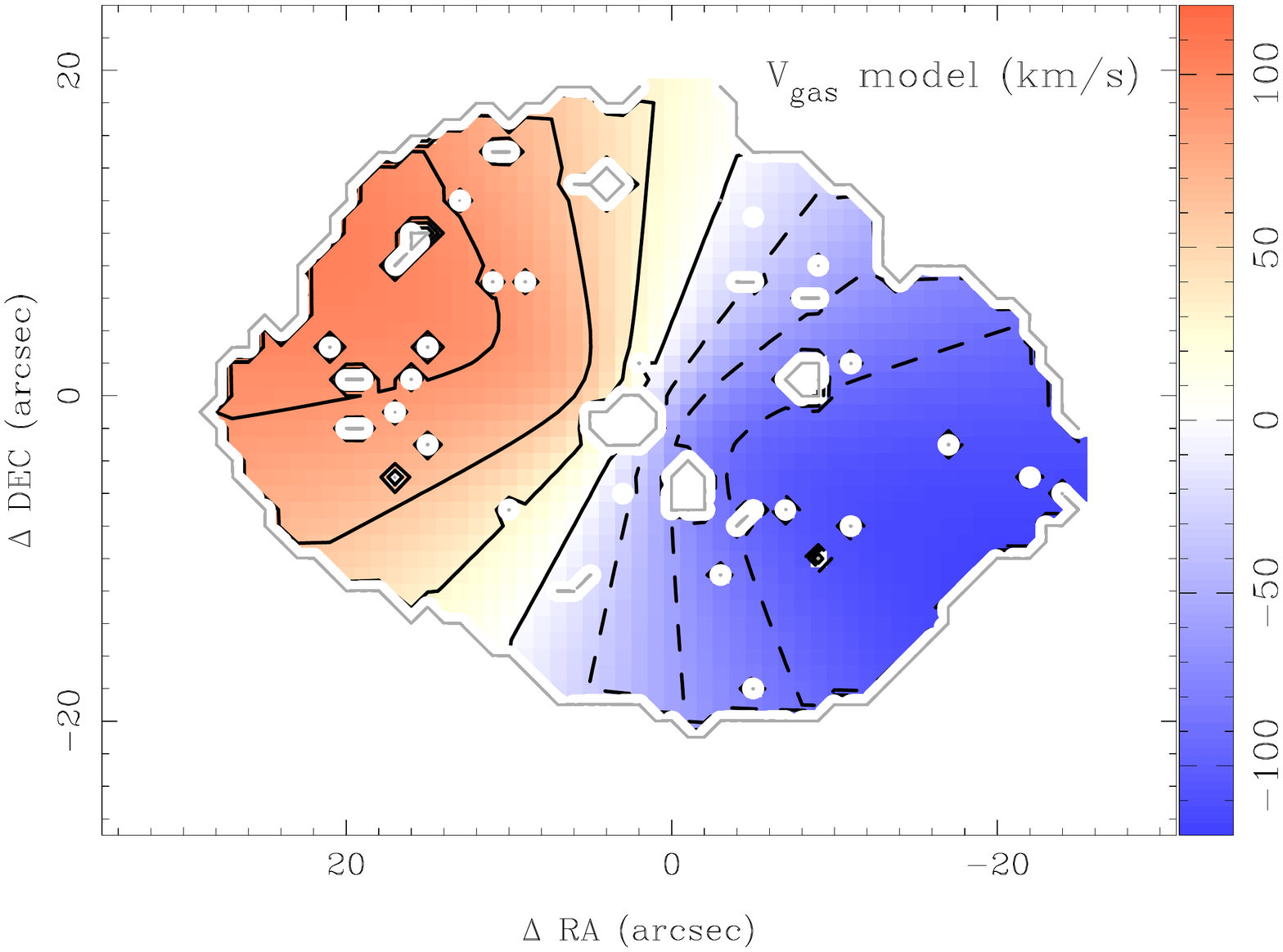}
\includegraphics[width=6cm,angle=0,clip=true,clip=true,trim=40 10 80 30]{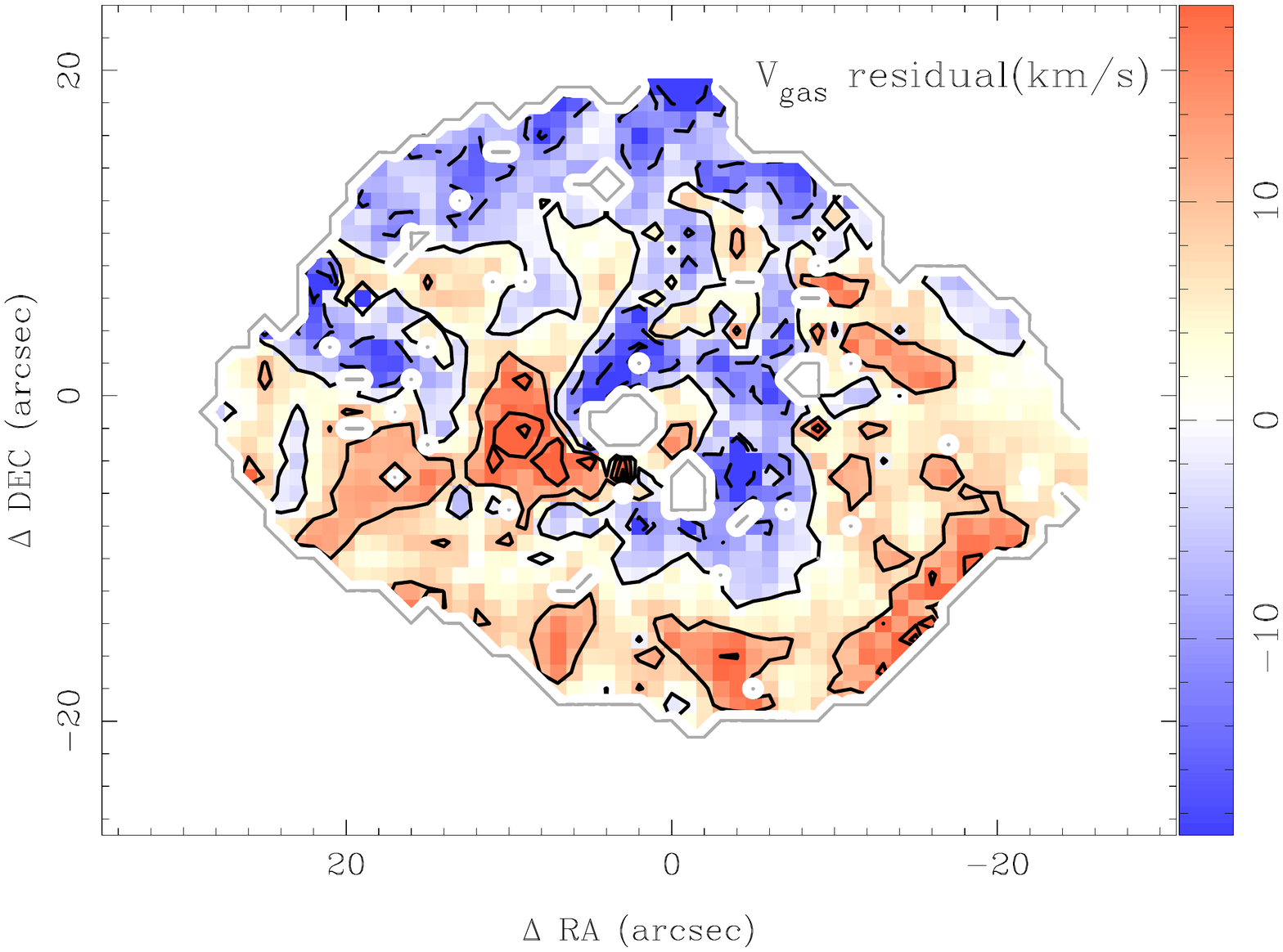}
\caption{Results from the { gas} kinematics analysis. The left panel
  shows the H$\alpha$ velocity map of NGC 5947 shown in Figure
  \ref{fig:fig_gas_kin}. The central panel shows the best-fit model
  assuming a simple arctan-rotational curve. Finally the right panel
  shows the residual after subtracting the model to the velocity
  map. The spaxels not used in the fitting procedure have been
  masked. Contours indicate different velocity levels, with constant
  step of 30 km s$^{-1}$ in the first two panels, and 10 km s$^{-1}$
  in the last one. \label{fig:ana_kin}}
\end{center}
\end{figure*}

This deprojected rotation curve was fit with a simple (but robust) arctan function of the form,

$$ v(r) = v_{sys} + \frac{2~v_r}{\pi} {\rm arctan} (sr-c)$$

\noindent where $v_{sys}$ is the systemic velocity of the gas and
$v_r$ is the asymptotic rotation speed of the disc, $s$ characterises
the slope of $v(r)$ in the inner part of the galaxy and $c$ is the
parameter that characterises any offset in the rotation axis of the
galaxy. A model of the velocity map was created by re-projecting the
best fitting arctan function. Figure \ref{fig:ana_kin} shows the
original velocity map, the re-projected model and the residual map, in
which the model is subtracted from the original map. The areas with
noisy gas emission, excluded from the fitting analysis, have been
masked from the figures. The residual map highlights the deviations
from purely circular velocity. { The gas velocity map of NGC5947 is
  well described by a purely rotating disc, assuming an asymptotic
  velocity ($v_r \sim$230 km~s$^{-1}$) and an inner regions slope
  ($s\sim$0.3 arcsec$^-1$). The residuals are in general lower than 30
  km s$^{-1}$ across the field.}

The conclusion from our exploratory kinematic analysis is that the
global information that CALIFA will provide will allow us  not only
to classify galaxies through their kinematics, but also to
build rotation curves and dispersion profiles { (as shown
  in Fig. \ref{fig:kin_NGC5947})}, as well as detailed mass models
for a large sample of galaxies of all morphological types and over a
wide range of stellar mass.

\section{Conclusions}
\label{sum}

The Calar Alto Legacy Integral Field Area survey (CALIFA
survey), has been designed to be the first survey to provide integral
field spectroscopy data for a statistical sample of all galaxy types in the
local Universe.  On completion it will be the largest and the most
comprehensive wide-field IFU survey of galaxies carried out to
date. It will thus provide a valuable bridge between large single
aperture surveys and more detailed studies of individual galaxies.

In the first part of this paper we have presented the survey design, the data reduction 
pipeline and a number of quality control procedures. We have reached the following 
conclusions: 

\begin{itemize}

\item At completion the survey will yield a statistically significant sample of 
$\sim$600 galaxies, which are representative of the galaxy population in the 
local Universe. 

\item Already at this early stage in the survey we have a working 
data reduction pipeline, which operates without human intervention. We will 
 be able to improve it regularly and provide consistently re-reduced 
complete datasets to the community. 

\item To give the community a feeling for the quality of data that CALIFA will deliver, 
we have compared the CALIFA spectra to the more generally known SDSS spectra 
for equivalently sized central regions of those galaxies that are in common between both 
datasets. From our preliminary analysis we find that the information content of the CALIFA 
spectra is at least equivalent to those of the SDSS (for a matched aperture).

\item We have presented the main quality parameters of the dataset: depth, spectrophotometric 
accuracy and stability, wavelength calibration and final spectral resolution. 
The analysis of the quality of the currently observed objects indicates that we
reach the target depth for the survey in 75\%\ of the datacubes and that the 
loss of depth in the remaining cases can be clearly attributed to adverse observing conditions. 
From all our quality checks we conclude that we are on track for providing 
high-quality, well-calibrated and well-characterized reduced data to the 
users of the CALIFA legacy survey.

\end{itemize}

In the second part of the paper we have presented an exploratory analysis of the 
first set of 21 galaxies observed through 2010 in order to verify the scientific usability of the 
data. We reached the following conclusions:

\begin{itemize}

\item We have shown that the absorption line spectra yield information on ages (and 
therefore mass-to-light ratios) and metallicities of the stellar population. We expect 
that with a more stringent analysis we will also be able to recover abundance ratios 
for the $\alpha$ elements. 

\item We have shown that the CALIFA emission line spectra will yield dust attenuation,
star formation rates, excitation mechanisms and ionization parameters, 
metallicities and electron densities. 

\item The conclusion from our exploratory kinematic analysis is that the global information 
from CALIFA will allow us to not only classify galaxies through their 
kinematics, but will also allow us to build rotation curves and dispersion profiles, as 
well as detailed mass models for a large sample of galaxies of all morphological 
types and over a wide range of stellar mass. 

\item CALIFA will allow us to quantify the effects of sampling different physical aperture 
sizes. As an example we have shown that the 
fraction of galaxies classified as star forming (vs. AGN) is larger for integrated 
spectra than for central spectra. This is of course expected. CALIFA, however, will 
allow us to \emph{quantify} these effects.

\item Finally, we have shown how comparative analyses between
  different galaxy types and within individual galaxies can be
  performed using any of the derived parameters.  CALIFA will thus be
  the first survey allowing comparative studies of the 2D distribution
  of galaxy properties in a statistically meaningful way{, including galaxies of many types}.

\end{itemize}

With two-dimensional maps of all these observables in hand, we expect that the data from 
the CALIFA survey will 1) characterize the local galaxy population in a way that will constitute  
a benchmark for models and other datasets alike and 2) will produce qualitatively new results on many 
topics of current active research, such as the build-up of galaxy disks, the influence of 
environment, galaxy bi-modality (morphology, kinematics, stellar populations) and 
possible transition objects (green valley), comparative studies of AGN and non-AGN 
galaxies, kinematic galaxy classification, stellar vs. dark mass, the origin and evolution of the warm ISM,  to name but a few. 

CALIFA data will become public in regular data releases and we very much hope that 
the community will actively use the data for their own projects - only then will CALIFA 
have fulfilled its main goal, namely to be a \emph{legacy survey}.

\begin{acknowledgements}

We thank the referee Eric Emsellem for his detailed comments which helped to improve the content and presentation of the article.

We thank the director of CEFCA, Dr. M. Moles, for his sincere support to this project.

We thank the {\it Viabilidad , Diseno , Acceso y Mejora } funding program,
ICTS-2009-10 , and the {\it Plan Nacional de Investigaci\'on y Desarrollo}
funding program, AYA2010-22111-C03-03, of the Spanish {\it Ministerio de
  Ciencia e Innovacion}, for the support given to this project.

I.M. and J.M. acknowledge financial support from the Spanish grant
AYA2010-15169 and Junta de Andaluc\'{\i}a TIC114 and Excellence Project P08-TIC-03531.

CK, as a Humboldt Fellow, acknowledges support from the Alexander von Humboldt Foundation, Germany.

B. Jungwiert acknowledges support by the grants
AV0Z10030501 (Academy of Sciences of the Czech Republic) and
LC06014 (Center for Theoretical Astrophysics, Czech Ministry of Education).

T. Bart\'akov\'a acknowledges support by the grants No. 205/08/H005 (Czech
Science Foundation) and MUNI/A/0968/2009 (Masaryk University in Brno).

This paper makes use of the Sloan Digital Sky Survey data. Funding for the
SDSS and SDSS-II has been provided by the Alfred P. Sloan Foundation,  the
Participating Institutions,  the National Science Foundation,  the
U.S. Department of Energy,  the National Aeronautics and Space Administration, 
the Japanese Monbukagakusho,  the Max Planck Society,  and the Higher Education
Funding Council for England. The SDSS Web Site is http://www.sdss.org/.

The SDSS is managed by the Astrophysical Research Consortium for the
Participating Institutions. The Participating Institutions are the
American Museum of Natural History,  Astrophysical Institute Potsdam, 
University of Basel,  University of Cambridge,  Case Western Reserve
University,  University of Chicago,  Drexel University,  Fermilab,  the
Institute for Advanced Study,  the Japan Participation Group,  Johns Hopkins
University,  the Joint Institute for Nuclear Astrophysics,  the Kavli
Institute for Particle Astrophysics and Cosmology,  the Korean Scientist
Group,  the Chinese Academy of Sciences (LAMOST),  Los Alamos National
Laboratory,  the Max-Planck-Institute for Astronomy (MPIA),  the
Max-Planck-Institute for Astrophysics (MPA),  New Mexico State University, 
Ohio State University,  University of Pittsburgh,  University of Portsmouth, 
Princeton University,  the United States Naval Observatory,  and the
University of Washington.

\end{acknowledgements}

\bibliographystyle{aa}

\end{document}